\documentclass[journal]{IEEEtran}
\usepackage{dsfont}
\usepackage{graphicx}
\usepackage{subfigure}
\usepackage{amsfonts,amssymb}
\usepackage{bm}
\usepackage{lipsum}
\usepackage{cite}
\usepackage{bm}
\usepackage{subeqnarray}
\usepackage{algorithm}
\usepackage{algorithmic}
\usepackage{caption}
\usepackage{lipsum}
\usepackage{multicol}
\graphicspath{{figures/}}
\usepackage[cmex10]{amsmath}
\usepackage{amsfonts,amssymb}
\usepackage{epstopdf}
\usepackage{amsmath}
\usepackage{bm}
\usepackage{algorithmic}
\begin{document}
%\title{Online Max Weight Learning for Utility Optimal Wireless Networks Scheduling under Reactive Jamming Attack}
%\title{Utility Optimal Max Weight Online Learning in Wireless Networks under Reactive Jamming Attack}

\title{Almost Optimal Energy-Efficient Cognitive Communications in Unknown Environments}
%Towards Optimal {Adaptive} Wireless Communications in Unknown Environments
%\author{\IEEEauthorblockN{Pan Zhou and Tao Jiang\\}
%\IEEEauthorblockA{Department of EI, Huazhong University of Science and Technology, Wuhan, Hubei, P.R. China\\
%Department of Engineering, Trinity College, Hartford, CT, USA\\
%%School of ECE, Georgia Institute of Technology, Atlanta, Georgia, USA$^3$\\
%Email:panzhou@hust.edu.cn, tao.jiang@ieee.org}
%}

\author{\normalsize Pan Zhou, \emph{Member, IEEE}, Chenhui Hu,
\emph{Student Member, IEEE}, Tao Jiang, \emph{Senior Member, IEEE}, Dapeng Wu, \emph{Fellow, IEEE}}%and Yevgeny Seldin
%\thanks{This work was supported by the National Science Foundation of China with Grant 60872008,
%Program for New Century Excellent Talents in University of China
%under Grant NCET-08-0217, and the Research Fund for the Doctoral
%Program of Higher Education of the Ministry of Education of China
%under Grant 200804871142.}

%»áÒé
%\thanks{%Pan Zhou and Tao Jiang are from the School
%of Electronic Information and Communications, Huazhong University of
%Science and Technology, Wuhan 430074, China. This research is supported by National Science Foundation of China with Grant 61401169.

%Dapeng Wu is with Department of Electrical
%and Computer Engineering, University of Florida, Gainesville, FL
%32611, USA. E-mails: zhoupannewton@gmail.com, Tao.Jiang@ieee.org, wu@ece.ufl.edu

%The authors appreciate Prof. Yevgeny Seldin, University of Copenhagen, Denmark, for useful discussions.

%\dag Corresponding author of this paper.

%Yevgeny Seldin is from Department of Computer Science, University of Copenhagen, Denmark,
%E-mail: Yevgeny.Seldin@gmail.com
%} \\}
%»áÒé
%\underline{{$^{^\dagger}$}Corresponding Author's Address:}\\
%$\mbox{Tao Jiang}$\\
%Wuhan National Laboratory For Optoelectronics\\
%Department of Electronics and Information Engineering\\
%Huazhong University of Science and Technology\\
%Wuhan 430074, P.R.China\\
%Tel: +86-27-87793073\\
%Fax: +86-27-87795842\\
%Email: Tao.Jiang@ieee.org\\
%http://ei.hust.edu.cn/teacher/jiangtao

\maketitle
\thispagestyle{empty}
\pagestyle{plain}

\begin{abstract}
 Cognitive (Radio) (CR)  Communications (CC) are mainly
 deployed within the environments of primary (user) communications,  where
 the channel  states  and accessibility are usually  stochastically distributed (benign or IID). However,  many practical
 CC are also exposed to disturbing events (contaminated) and vulnerable jamming attacks (adversarial or non-IID).  Thus,
 the channel state distribution
  of spectrum could be stochastic, contaminated or adversarial
at different temporal and spatial locations.  Without any \emph{a priori}, facilitating optimal CC is a very challenging issue.  In this paper,
 we propose an online learning algorithm that performs
the joint channel sensing, probing and adaptive channel access for multi-channel CC
 in general unknown environments. We take \emph{energy-efficient} CC (EECC)
 into our special attention, which is highly desirable for  green wireless communications and demanding to combat
 with potential jamming attack who could greatly mar the \emph{energy} and \emph{spectrum}  efficiency of CC.
 The EECC is formulated as a \emph{constrained regret minimization} problem with power budget constraints. By
 tuning a novel exploration parameter, our algorithms could adaptively find the optimal channel access strategies and achieve the almost optimal
  learning performance of EECC   in different scenarios provided with the vanishing long-term power budget violations. We also consider the important scenario that
  cooperative learning and information
  sharing among multiple CR users to see further performance improvements. The proposed algorithms are resilient to both
  oblivious and adaptive jamming attacks with different intelligence and attacking strength. Extensive
  numerical results are conducted to validate our theory.
\end{abstract}

\begin{keywords}
Energy Efficiency, Cognitive Radio, Online learning, Jamming attack and  Multi-armed bandits
\end{keywords}

% \fi
\IEEEpeerreviewmaketitle

%Wireless channel access is a fundamental issue in wireless networks.
\section{Introduction}%wide facilitate the wide application of the advantage
Cognitive (Radio) (CR)  Communications (CC) are widely recognized as one of the promising Information and
 Communication technology (ICT) to release the tension of current spectrum-scarcity issue. Meanwhile,
as  growing explosively, ICT is playing a more and more important role in global greenhouse gas emissions,
the energy-consumption of which contributes to $3$ percent of the worldwide electric energy consumption nowadays \cite{Fettweis}. Thus, Energy-Efficient
(EE) CC (EECC) has received great attention from the research community in recent years \cite{CommMag14}. Admittedly, the joint
design of channel sensing, probing, and  accessing (SPA)  scheme with the consideration of energy efficiency (EE) is pivotal for CC. Stimulated by the recent appearance of smart CR devices with adaptive and learning abilities, modern CCs have  raised very high requirements to its solutions, especially in complex environments, where
 accurate channel  distributions and states can barely be modeled and acquired due to unpredictable
 Primary User Activity (PUA) \cite{CRAHN09} in Primary Communications (PC), behaviors of other CC,  potential jamming attacks, and other distributing events frequently
seen in practice. Thus, it is critical
for CR devices  to learn from the environments and keep a good  balance of allocating its transmission power wisely to achieve the goal of \emph{energy-efficiency}  and of designing almost optimal channel access schemes to reach the goal of  \emph{spectrum-efficiency} in EECC.

%As noticed,
Undoubtedly, the communication model has a great impact on the performance of CC. A great amount of works assume
priorly known statistical information and have proposed \emph{deterministic}  channel states  models, e.g., POMDP \cite{JSAC07},  and accessibility models, e.g.,    Poisson Modeling of PUA \cite{TON11} to make good approximations in benign
wireless environments. Clearly, they are not suitable for complex or even
unknown environments.  To cope with the problem, a fairly reasonable and realistic line of studies assume
  no statistical prior information about the channel states and accessibility. Thus, online learning based methods (e.g.,
  reinforcement learning (RL) \cite{Ref98}) are desirable to be adopted, e.g., \cite{PanJSAC12} \cite{ICNP11}\cite{XYinfocom11}\cite{ZhaoInfocom10}.  Within this context, the use of the Multi-armed bandit (MAB) theory \cite{Bubeck12} is highly identified over other learning approaches.

In summary, these works assume that
the \emph{nature} of CC environments is either \emph{stochastic} ({benign}), where the
channel state  and accessibility are stochastically distributed \cite{ZhaoInfocom10} (IID), or \emph{adversarial}\cite{ICNP11} \cite{XYinfocom11}, where they   can vary arbitrarily (adversarial or non-IID) by jamming attackers or adversarial UPA, etc.  Respectively, these works
are mainly categorized into two MAB models, namely, stochastic MAB \cite{Anandkumar10,Dys10,TSP10}
with IID assumption and adversarial MAB \cite{ICNP11}\cite{XYinfocom11} with non-IID assumption.   Accordingly, the analytical approaches and results for the two
models are distinctively different. Note that the learning performance  is qualified by the classic term ``\emph{regret}", i.e., the performance difference between the
 proposed learning algorithm and the optimal one known in hindsight.  A well-known fact is that
stochastic MAB and adversarial MAB have the respective optimal regrets
$O(log(n))$ [18] and $O(\sqrt{n})$ [19] over time $n$. Obviously, the stochastic MAB
highly outperform that of the adversarial MAB in learning of convergence to the optimal strategies.

 However, all related works \cite{Anandkumar10,Dys10,TSP10} \cite{ICNP11,XYinfocom11,XYTMC13,Yi11,QianJSAC12,ZhaoInfocom10,Lai07,ICASSP10}  \emph{still} rely on the priori of
 either the stochastic or the adversarial assumption, which is limited in describing practical CC environments.  Because, the nature of the practical CC environments are not restricted to these two types and it usually can not be known in advance. On the one hand, consider a CC under potential
 jamming attack. Since the number and locations of jamming regions  are often unrevealed, it is uncertain which regions may (or may not) suffer
 from the attack. Thus, the usual mind of applying adversarial MABs models \cite{ICNP11}\cite{XYinfocom11} on all channels will lead to
 large values of regret, since a great portion of channels can still be stochastically distributed, while applying the stochastic MABs models is not
 feasible due to the existence of adversaries.

 On the other hand,  the stochastic MAB model \cite{XYTMC13}\cite{Yi11,ZhaoInfocom10,ICASSP10}\cite{Lai07}  will face practical implementation issues. In almost all CC systems, the commonly seen occasionally disturbing events would make the stochastic channel distributions contaminated. These include the burst  movements of individuals, the spectrums handoff and mobility \cite{CRAHN09} among users of PC and CC, and the
 jitter effects of electronmagnetic waves, etc. In this case, the channel
 distribution  will not follow an IID process for a small portion of time. Thus, it is not clear to us whether the stochastic MAB is still applicable, how the contamination affects the learning performance
 and to what extent the contamination is negligible. Therefore, the design of a unified SPA scheme without any prior knowledge of the operating environment is very challenging. It is highly desirable and bears great theoretical value.

  In this paper, we propose a novel adaptive multi-channel SPA algorithm for EECC that achieves almost optimal learning performance without \emph{any} a priori
   of the CC environments. Importantly, we take EE into our special consideration with power budget constraints on each of the multi-channel access strategy.  As such, our work can be
  regarded as the first work for the EECC in unknown environments, where optimal
   strategies can be gradually learned.
    Our innovative SPA   scheme is based on the famous EXP3 \cite{non_MAB02}  algorithm in the non-stochastic MAB with three main features: 1) We introduce a new control
parameter into the exploration probability for each channel to facilitate \emph{automatically}  detection of the feature of environments; 2) we
  use and design the Lagrangian method delicately to model the the power budget constraints for our own EECC problem;  3) By joint control of learning rate and exploration probability, the proposed algorithm achieves almost optimal learning performance in different regimes with
 vanishing (sublinear) long-term power budget violations. Our main contributions are summarized as follows.
%
%  When the
%environment happens to be adversarial,  our proposed algorithm enjoys the same behavior as classic adversarial MABs-based algorithms and has the optimal regret``root-n" bound in the adversarial regime. When the environment happens to be stochastic, we indicate a problem-dependent
%``polylogarithmic-n" regret bound, which is slightly worse than the optimal ``logarithmic" bound in \cite{Robbins1985}. Furthermore, we prove that
%the proposed  algorithm retains the ``polylogarithmic-n" regret bound  when the stochastic regime is only \emph{lightly}\footnote{See Section III
%for strict definition} contaminated. Note that all regret bounds of EE of EECC and power budgets violation are \emph{sublinear} to time horizon, which indicates the optimal  strategy is achievable.

%\begin{itemize}

\emph{1)}  We define an appropriate EE model that is suitable for SPA scheme-based EECC  over
large spectrum pools and with fairness considerations.  We   categorize the features of the EECC environments mainly into four typical regimes, each of which
 are proved to achieve the almost optimal regret bounds with sublinear long-term power budget violations.  Our proposed algorithm neither need to distinguish the type of PC, other CC and adversarial (jamming) behaviors, nor need to know the channel accessibility and quality within
all the different features of the complex environments. Thus, it provides a complete solution for practical CC in general unknown environment.
%We believe that our proposed concepts of these regimes are not restricted only to wireless communications, it can be used in other engineering disciplines, e.g.,
% the virus contaminations of internet information dissemination in social and computer networks,
%and the events of long term roadwork and occasionally appearance of  obstacles in traffic planning in civil engineering, etc.

\emph{2)}  The proposed AOEECC-EXP3++  algorithm considers  information sharing of   channels that belong to
     different channel access strategies, which can be regarded as
    a special type of combinatorial semi-bandit\footnote{The term first appears in \cite{OR2014}, which is
    the combinatorial version of the classic MAB problems.  } problem. In this case,
  given the size of all channels $K$ and the number of transmitting  channels $k$, the AOEECC-EXP3++ algorithm has the optimal tight regret bounds in
both the adversarial settings  \cite{OR2014} and the stochastic settings \cite{Branislav2015}, which indicates the good
scalability for different size of CC systems.

%Compared with the deterministic solutions in the previous work on EECC \cite{Hasan09}\cite{TVT15} with
%the present of  computational complexity  for multi-channel CC that is hard to
%resolve in practice for iterative (gradient) algorithms \cite{TVT2014},   our  multi-channel combinatorial semi-bandit framework  using the
%idea of try-and-learn that avoids the problem.  However, with the increasing of the size of CC, our original algorithms scale poorly£®Fortunely,

\emph{3)} This work is also the \emph{first} MAB-based \emph{constrained regret minimization} (optimization) framework for CC in unknown
environments in the online learning setting.  Our proposed algorithms have polynomial time implementations, which result in
 good computational efficiency in practice.

 \emph{4)} We propose a novel \emph{cooperative} learning algorithm that considers information sharing among multiple users of the CC systems to accelerating the learning speed of each individual users, which is desirable for the widely acknowledged feature of
  CC systems with cooperative spectrum sensing and sharing schemes \cite{CRAHN09}. It further improves the
    {energy-efficiency}  and  {spectrum-efficiency} of the EECC within a fixed time period.
%and conduct plenty of diversified numerical experiments.

\emph{5)} We conduct plenty of diversified  experiments based on
 real experimental datasets. Numerical results demonstrate that all  advantages of the proposed algorithms are real and
  can be implemented easily in practice.
%\end{itemize}

The rest of this paper is organized as follows: Section II discusses Related works. Section III describes the problem formulation. Section IV introduces the distributed learning algorithm, i.e, AOEECC. The performance results are presented in Section V. The multi-user cooperative learning algorithm is discussed
in Section VI. Proofs of previous sections are in Section VII and Section VIII. Simulation results are available in Section IX. The paper is concluded in Section X.
%\vspace{-0.3cm}

\section{Related Works}
Recently, online learning approach to address the dynamic channel access (DSA) problem  in CC with less prior
channel statistical information have received more and more attention than
classic deterministic model approaches, e.g. channel states \cite{JSAC07} and accessibility modeling \cite{TON11}.  The characteristics of repeated interactions with environments are  usually categorized
 into the domain of RL \cite{Ref98}, e.g. DSA by RL \cite{PanJSAC12}, anti-jamming
 CC by RL \cite{Wang11,Wu12,Dog09,Oksanen12}. It is worth pointing out that there exists extensive literature in RL, which is generally
targeted at a broader set of learning problems in Markov
Decision Processes (MDPs). The RL approach guarantees  the performance  asymptotically
to infinite. Hence, it is not quite suitable for mission-critical  advanced applications of CC, which
is commonly seen in next generation wireless communications.  By contrast, MAB problems
constitute a special class of MDPs, for which the no-regret
learning framework is generally viewed as more effective
in terms of fast convergence time, finite-time optimality guarantee \cite{Finite02}, and low computational complexity.
Moreover, it has the inherent capability in keeping
a good balance between ``exploitation" and ``exploration".  Thus, the use of MAB models is
 highly identified.

The works based on the stochastic MAB model often consider about the stochastically distributed channels in benign  environments, such as
\cite{XYTMC13}\cite{Yi11}\cite{ZhaoInfocom10,Lai07,ICASSP10}\cite{ZhaoTSP10}.   The adversarial  MAB model is applied
 to adversarial channel conditions, such as the anti-jamming CC \cite{ICNP11}\cite{XYinfocom11}.  In the machine
 learning society, the  stochastic and adversarial MABs have co-existed in parallel for almost two decades. Only until recently, the first practical
 algorithm for both stochastic and adversarial bandits is proposed in \cite{Seldin14} for the classic MAB problem. The current work uses the idea of introducing a novel exploration parameter \cite{Seldin14}. But our focus is on the much harder
  \emph{combinatorial semi-bandit} problem that needs to exploit the channel dependency among different SPA strategies, which is a nontrivial task. Moreover, our introducing of  the Lagrangian method into
 the online EECC problem leads to an important finding that we need to  set the learning rate and exploration probability together and
 the same for all regimes (as we defined) rather than could be  adjusted separately for stochastic and adversarial regimes in \cite{Seldin14}. This
 phenomenon indicates
 that the online learning for the EECC in unknown environments is a harder problem  than classic regret minimization without constraints \cite{ICNP11}\cite{XYinfocom11} \cite{Seldin14}.

The topic of EECC has recently received great attention in wireless communications society
\cite{CommMag14} due to the stimulation of green communications for ICT.  The \emph{spectrum efficiency} and \emph{energy efficiency} are the
two critical concerns.  Almost all of them  consider about \emph{deterministic} channel state and accessibility models
\cite{PirmICC12,TVT15,InfoES12,TWCES12,Tao14,Hasan09} for DSA in CC. Some of the works try to achieve the {spectrum efficiency}  \cite{TVT15},  {energy efficiency}  \cite{PirmICC12}, while others
try to achieve both goals \cite{InfoES12,TWCES12,Tao14,Hasan09}. Being worthy of mention, there are a small
amount of works focus only on optimization of the EE for spectrum sensing part\cite{ZhanginfoWkSP10} \cite{BayTVT13} within the
whole CC circle. This part of energy cost is comparatively minor in scales when compare to circuit and  transmission
energy cost \cite{CRAHN09}, which can be categorized into the circuit and processing energy cost as in classic wireless communications \cite{TVT2014}.

  Recently works \cite{Merti15_1} \cite{Merti15_2} have used the exponential weights (similar to EXP3) MAB model to
study the \emph{no-regret} (sublinear) online learning for the \textup{EE} of OFDM and MIMO-OFDM wireless communications. However, the problems are different from the  EECC, and
the dynamic channel evolution process is only assumed to be \emph{adversarial}.  Thus, our work is the first SPA scheme for EECC
in general unknown environments that targets on both {spectrum efficiency} and {energy efficiency} without
any \emph{deterministic} channel model assumption.

%Nevertheless, our proposed algorithms are order optimal with respect to $K$ and ${k}$ for all different regimes, which indicate  the good  scalability for general CC. (See \cite{TechEEECC15} for detailed discussions.)

\section{Problem Formulation}
\subsection{Cognitive Communication Model}
We first focus on EECC from the perspective of a single CR user (or called ``secondary user" (SU)), which is \emph{distributed} or \emph{uncoordinated} with other CR users. It is consisted by
  a pair of transmitter and receiver within the region of PC.  The transmitter sends data packets to the
receiver synchronically  over time with classic slotted model. The wireless environment is
 highly flexible in dynamics, i.e., besides the most influential PUA of a number of $M$ PUs that affect the CC's channels'
   qualities (states) and accessibility, there are interference from other SU transceiver pairs, potential jamming attack and channel fading, etc, would make the environment to be
  generally unknown. During each timeslot, the SU transmitter selects multiple channels $k$ to transmit data to the receiver
   over a  set $[K]= \{c_1, c_2,..., c_f,...,c_K\}$ of $K$ available orthogonal channels with
  possibly different data rates across them.  When a channel $c_f$ is occupied by primary user (PU), it is called as \emph{busy}, otherwise,
  it is called  \emph{idle}. However, the busy (or idle) probability of PU is not unknown. There are a set $[L]=\{S_1, S_2,...,S_u,..., S_L\}$ of $L$
 SU transceiver pairs  making contention or interfering power among each other. However, the
  behavior is transparent to a  single SU $S_u$.  W.l.o.g, if there are adversarial events, we ascribe them to be launched by one
  jammer who attacks the set or a subset of $K$ channels, where its attacking strategies are
  unrevealed. At each time $n$, the  \textup{EE} calculated from the allocated power and received data rate of the SU $u$
   on channel $f$ is denoted by $g_{n, u}(f)$, $g_{n, u}(f) \in [0,  M]$.  We omit subscript $u$ if there is no confusion from context. Here constant $ M$ is
   the maximum value of \textup{EE} for all channels. W.l.o.g., we normalize $M=1$ as usual in the  regret analysis.

%  $g_{n}(f)=0$ denoting that the channel $c_f$ is busy at time $n$ or severely been attacked by jamming attacker with
%   no data package reception.

 We employ the classic energy detection method \cite{CRAHN09} for spectrum sensing, i.e., if the transmitting  single strength is above a threshold, we regard the
 channel is busy or attacked, i.e., $g_{n}(f)=0$. Otherwise, CC are allowed and $g_{n}(f)$ is
 released for the frequency $f$ even though there are other potential PUs, jammers and CR users transmitting with
 low interfering power. Thus, our model is suitable for both spectrum \emph{overlay}  and spectrum \emph{underlay} \cite{CRAHN09} schemes. We assume that each radio out of the $k$ radios on the SU transmitter needs time $n_s$ for sensing the status of a channel and time $n_p$ for probing its quality.
  The actual time depends on the technology and device: the typical values of $t_s$ is about 10$ms$ and $t_p$ is from 10$ms$ to 133$ms$ \cite{Arbaugh09}.  Let
  $t_{sp}=t_s + t_p$. When a channel $f \in [K]$ is idle, transmitter/reciever can only access it for at most $t_a$ time at most, so it can detect
  the return of a PU. In practice, $t_a$ has the typical value of 2$s$.  Let ${\mathds{1}_n}(f)$
  be the indictor function  that denote whether the SU decides to transmit data using the probed channel $c_{{\mathds{1}_n}(f)}$.

%Noticed that, although the
%  channel quality often has a coherence time $t_c > t_a$, to avoid the possible collisions with PUs, other CR users and jammers.

%
%We consider the  AOEECC problem as a sequential
%decision problem, where the choice of transmitting  channels at each timeslot is a decision. Denote $\{0,1\}^K$
% as the vector space of all $K$ channels. The strategy space for the transmitter
% is  denoted as $S \subseteq \{0,1\}^K$ of size $N=\binom{K}{k}$. If the $f^{th}$-channel is selected for
% transmitting and transmitting  data, the value of the $f$-th entry of a vector (channel access strategy) is
% $1$, and 0 otherwise. In the case of the existence of jamming attack on a subset of $k_j$ channels, the
%  strategy space for the jammer is denoted as $S_j \subseteq \{0,1\}^K$ of size $\binom{K}{k_j}$. For convenience,
%  we say that the $f$-th channel is \emph{jammed} if the value of $f$-th entry is 0 and otherwise is 1.  At each timeslot, after choosing a strategy $s_r$, the value of the data rate (or called ``reward") $g_n(f)$ is revealed
%to the receiver if and only if $f$ is chosen as a transmitting  channel.

\subsection{Preliminary in EECC with Deterministic CSI}
% For the PC, after every primary data
%transmission, primary receiver will feedback the interference level to its transmitter in terms of an interference value that
%includes the background noise and the collection of the sum-interference
%made from CR users. This information is used
%by PUs to monitor  the environment
%and notify the violation of interference power
%constraint messages to CR networks or other anchor benchmarks.   to conduct their
%own transmit power control.

%Before going into the online learning problem formulation of the EECC problem, let us first
%review the classic EECC with deterministic channel accessibility and states.

Before the discussion  of our own problem, let us first
review the classic EECC with deterministic channel accessibility and states. For the multi-channel CC, SUs only know its own payoff and strategy for each channel
$f$ at timslot $t$, i.e., the realized transmission rate ${r_t}(f)$ and
  transmission power $\textsf{P}_t(f)$.  At each timeslot $t$, each SU chooses a  subset of channels over $n$ according
to some sensing/probing rules, where the multi-channel access strategy is denoted by $i$ and we have $f \in i \subset S$.
the transmission power for each channel is $\textsf{P}_t(i_f), 1 \le i_f \le k, 1 \le i \le N$, and the total
transmission power over a strategy $i$ is $\textsf{P}_t(i)$, and $\textsf{P}_t(i) = \sum\nolimits_{f = 1}^k {{\textsf{P}_t}(i_f)}$. Then, the instant data transmission rate $r_t(f)$
 for the SU at each selected channel $f$ is given as
 \begin{IEEEeqnarray*}{l}
\begin{array}{l}
\!\!{r_t}(f)\! =\! W{\log _2}(1 \!+ \!\vartheta {\textsf{SINR}}({\textsf{P}_t})) = W \cdot {\mathds{1}_t}(f) \cdot  \\
\!\!\!{\log}\left(\!\!1 \!+\! \frac{{\vartheta {\textsf{P}_t}(f){g_{ff}}(s_t^f)}}{{\underbrace
 {\sum\limits_{j \ne f}^{f \in K} {{\textsf{P}_t}(j){g_{fj}}(s_t^j)} }_{other \ CRs}\! +
 \! \underbrace {\sum\limits_{l = 1}^M {b_t^{fl}\psi _t^{_{fl}}} }_{PUs}\! + \!
  \underbrace {{a_t^{fJ}\varpi _t^{_{fJ}}} }_{Jamming}\! + \! \underbrace {(\sigma^{f}_t)^2}_{noise}}}\!\!\right),
\end{array}
 \end{IEEEeqnarray*}
where ${g_{ff}}(s_t^f)$ and ${g_{fj}}(s_t^j)$ are the respective
channel gains from itself and other SUs with instant channel states$ s_t^f$ and $s_t^j$, $b_t^f$ and
$\psi _t^{_{fl}}$ are the respective interfering power and channel gain from the PU $l$,  $a_t^f$ and $\varpi _t^{_{fa}}$ are the respective interfering power and channel gain from the jammer $J$, and $(\sigma^{f}_t)^2$ is the background noise power. The unit of $r_t(f)$ is
$nats/s$.

In traditional wireless communications,  the \textup{EE} of the multi-channel or OFDM (e.g. \cite{TVT2014}\cite{Merti15_1}) wireless systems with
 the number of subchannel (subcarrier) $k$ at timeslot $t$ is defined as
 \begin{IEEEeqnarray*}{l}
\quad \quad  \textup{EE}_t = \frac{{\sum\nolimits_{f = 1}^k {{r_t}(i_f)} }}{{\textsf{P}_c^t(i) + \sum\nolimits_{f = 1}^k {{\textsf{P}_t}(i_f)} }} \quad nats/J,
\IEEEyesnumber \label{eq:EECC}
 \end{IEEEeqnarray*}
 where $P_c^t$ is the processing and circuit power consumption at time $t$ while ${{\textsf{P}_t}(i_f)}$ is the
 transmission power for each sub-channel (sub-carrier) $f$.

 By contrast, the definition of EE in EECC is slightly different. Because the multi-channel
 CC is not restricted to a pre-defined fixed set of OFDM (OFDMA) channel sets, where multi-radio based spectrum sensing and
 channel probing are necessary to scan from a large spectrum pool \emph{separately} for a group of (potentially) \emph{nonconsecutive} and distributed channels with the
 best channel sensing/probing qualities for general CC systems \cite{MCSens2010}. As such, the measurement of \textup{EE} for EECC
 is from the view of each sensed/probed transmitting channel within the SPA scheme, i.e.,
  \begin{IEEEeqnarray*}{l}
\quad \quad  \textup{EE}_{t,f} = \frac{{{r_t}(i_f)}}{{\textsf{P}_c^t(i_f) +  {{\textsf{P}_t}(i_f)} }} \quad nats/J.
\IEEEyesnumber \label{eq:EECC2}
 \end{IEEEeqnarray*}
 Then, the overall average \textup{EE} for the each SPA strategy $i$ is given as
   \begin{IEEEeqnarray*}{l}
\quad \quad  \textup{EE}_t^{CC} =  \frac{1}{k}\sum\nolimits_{f = 1}^k  \textup{EE}_{t,f} \quad nats/J, \forall f \in i.
\IEEEyesnumber \label{eq:EECC3}
 \end{IEEEeqnarray*}
  Note that the
 sensing and probing energy consumption are also categorized into $P_c^t(i_f)$ with $\sum\nolimits_{f = 1}^k P_c^t(i_f) = \textsf{P}_c^t(i)$\footnote{
 More precisely, we could divide the circuit and processing power among the $k$ channels according to the
 bandwidth of each channel and calculate the sensing and probing energy cost based on the monitoring of each channel. Roughly speaking,
  we can simply do an energy-cost division among all channels $k$.} and $\sum\nolimits_{f = 1}^k {{\textsf{P}_t}(i_f)} = {{\textsf{P}_t}(i)}$. A simple fact about the relation of (\ref{eq:EECC}) and (\ref{eq:EECC3}) is that
 $\max_f\textup{EE}_t^{CC}(f) \ge \textup{EE}_t^{CC} \ge \textup{EE}_t \ge \min_f\textup{EE}_t^{CC}(f)$. When $\forall f=f'$, $\textup{EE}_{t,f}
 = \textup{EE}_{t,f'}$, $\textup{EE}_t^{CC} = \textup{EE}_t$.  Thus, maximize  $\textup{EE}_t^{CC}$ will push the
 fairness  of \textup{EE} among different channels. Incorporated with sensing and
probing, and after determined the channel access strategy $i$,  the EECC can be formulated as the following nonlinear program.
 \begin{IEEEeqnarray*}{l}
\begin{array}{l}
\quad \max \quad\quad\quad \ \textup{EE}_t^{CC}  \\
\emph{subject to} \quad    \textsf{P}_t(i) \le {{ \textsf{P}}_o},
\end{array}\IEEEyesnumber \label{eq:EECCOPT}
 \end{IEEEeqnarray*}
where  each SU has a power budget $\textsf{ P}_o$. By similar approaches in \cite{TVT2014} that the  problem  (\ref{eq:EECCOPT}) is also quasi-concave with respect to $\textsf{P}_t(i_f)$, where water-filling method can be used to resolve the problem. Moreover, the definition of $\textup{EE}_t^{CC}$ enables the  information sharing of $\textup{EE}_{t,f}$ for
each channel among different strategies, which is specially suitable for EECC design over large spectrum pools.

% When the channel qualities varies largely
% across different nonconsecutive channels, the model (\ref{eq:EECC3}) is a more accurate model than \label{eq:EECC}

% However, the introduction of $P_c^t$ will cause the existing
%iterative algorithm has high complexities (See \cite{TVT2014} for details). However, in our online
%learning problem, we do not face this difficulties.

%  exists and the receiver experiences very low data rate, the receiver needs to send feedback data rate information
% to the transmitter. Then, the transmitter can increase its transmission  power to combat the jamming attack.}

%In the receiver side, we assume that after transmitting  data packets, there are efficient message verifications and authentications
%as in \cite{Hoc09} \cite{QianJSAC12}.
%
%Regarding the privacy issue, we can use our proposed protocols to transmit messages
%of a key establishment  protocol to generate a secret key.

% We name ours as the Adaptive Online EECC (AOEECC) protocol due to its flexibility to achieve optimal performance in various scenarios.

\subsection{The Adaptive Online Learning for EECC: A Constrained Regret Minimization Formulation}
In reality, since no  secret is shared and no adversarial event is informed to the transceiver pair,
the multi-channel EECC in unknown environments are necessary to sensing/probing and hoping among different channels to dynamically access a subset to maximize its accumulated EE over time. Namely,
this sequential channel sensing/probing/accessing (SPA) problem is to determine when to conduct the channel hopping (multi-channel access) and
 power allocation repeated game with environments, without knowing instant channel states, for a pair of CR user transceiver so as to improve the EE of CC.  The difference of our
SPA problem (based on MAB) with the classic MAB problem is that,  at every timeslot $t$, the classic MAB receives a reward and repeat this
for $T$ timeslots; while for the SPA problem, at a timeslot $t$, we will not have any gain if the the CR users donot happen to
use the channel for data transmission after the sense/probe of the channel.

To address this issue, we only need to count the timeslots
 spent for sensing/probing a chosen channel a  \emph{round} (Or still say ``timeslot", if no confusion). The immediate
 following timeslot spent for data transmission over a chosen multiple channel set are not counted as a  round. However, we
 will calculate and treat the averaged \textup{EE} (\ref{eq:EECC3}) from  (\ref{eq:EECC2}) based on the previous transmitted
 data and the chosen transmission power and known circuit and processing power $\textsf{P}_c^t(i_f)$ for  each
  sensed/probed channel $f$, where its gain is  ${g_t}(f) = {\mathds{1}_t}(f) \cdot \frac{1}{{{Z_t}}}\int_0^{{Z_t}} {\textup{EE}_{t,f}} \cdot (1 - {\Pr _t}(PU,J,SU))$,
 where $Z_t$ denotes the time of the actual transmission and ${\Pr _t}(PU,J,SU)$ are the probability that the transmission will be
 destroyed by the return of PU, jammer or some other SUs within the $Z_t$ time duration. We set that $t_a=Z_t$.
Let $n$ be the number of sensing/probing timeslots executed during the whole run of the system evolution duration $T$, which should
satisfy the condition $n \cdot {t_{sp}} + \sum\nolimits_{t = 1}^n {{\mathds{1}_t}(i)}  \cdot {Z_t} \le T.$ The first part
is the time spent for sensing/probing and the second part is the time spent for multi-channel EECC.

For the multi-channel accessing part, let us denote $\{0,1\}^K$
 as the vector space of all $K$ channels. The strategy space for the transmitter
 is  denoted as $S \subseteq \{0,1\}^K$ of size $N=\binom{K}{k}$. If the $f^{th}$-channel is selected for
 transmitting  data, the value of the $f$-th entry of a vector (channel access strategy) is
 $1$, and 0 otherwise. In the case of the existence of jamming attack on a subset of $k_j$ channels, the
  strategy space for the jammer is denoted as $S_j \subseteq \{0,1\}^K$ of size $\binom{K}{k_j}$. For convenience,
  we say that the $f$-th channel is \emph{jammed} if the value of $f$-th entry is 0 and otherwise is 1.

%  At each timeslot, after choosing a strategy $s_r$, the value of the data rate (or called ``reward") $g_n(f)$ is revealed
%to the receiver if and only if $f$ is chosen as a transmitting  channel.

Formally, our MAB-based SPA problem is described
as follows: at each timeslot $n=1,2,3,...$, the transmitter  (as a decision maker) selects a strategy $I_n$ from $S_r$ with a power strategy $\textsf{P}_t(I_f),
\forall f \in I_n$. The cardinality of
$S_r$ is $|S_r|=N$. The reward  $g_n(f)$ is assigned to
each channel $f \in \{1,...,K\}$ and the SU only gets rewards in strategy $i \in S_r$.   The total reward of a strategy
$i$ in timeslot $n$ is ${g_n(i)} = \sum\nolimits_{f \in i} {{g_n(f)}}$. Then, on the one hand, the cumulative reward (or  EE) up to timeslot $n$ of the strategy
$i$ is $
{G_{n,i}} =\textup{EE}_n^{CC}= \sum\nolimits_{t = 1}^n {{g_t(i)}}  = \sum\nolimits_{f \in i} {\sum\nolimits_{t = 1}^n {{g_t(f)}} } .$
On the other hand, the total reward over all the chosen strategies by the receiver up to timeslot $n$ is
${\hat G_n} =\hat{\textup{ EE}}_n^{CC}= \sum\nolimits_{t = 1}^n {{g_{t}(I_t)}}  = \sum\nolimits_{t = 1}^n {\sum\nolimits_{f \in {I_t}} {{g_t(f)}} }$,
where the strategy $I_t$ is chosen randomly according to some distribution over $S_r$. The performance of this algorithm
is qualified by \emph{regret} $R(n)$, defined as the difference between the expected number of successfully received data packets
using our proposed algorithm and the expected rewards that use the best fixed solution up to $n$, i.e.,
 \begin{IEEEeqnarray*}{l}
R(n) = \mathop {\max }\limits_{i \in {S_r}} \mathbb{E}_{i \sim {\textbf{p}_n}}\{ {{G_{n,i}}} \} - \mathbb{E}[ {{{\hat G}_n}} ],
\IEEEyesnumber
\end{IEEEeqnarray*}
where $\textbf{p}_n$ is the decision probability vector over all strategies and the maximum is taken over all available strategies. However, if we
 use the \emph{gain} (reward) model,  we will face technical difficulties as presented in \cite{Bubeck12} (pages 25-28). Thus, we can introduce the \emph{loss} model by the simple trick of $  \ell_n(f)= 1- g_n(f)$ for each channel $f$ and $\ell_n(i)= k -g_n(i)$ for
 each strategy to avoid this issue. Then, we have $ L_n(i) = nk - G_{n,i}$ where ${L_n}(i) = \sum\nolimits_{t = 1}^n \ell_n(i)=  {\sum\nolimits_{t = 1}^n {
 \sum\nolimits_{f \in i} {{\ell_t(f)}} } }$, and similarly, we have $ \hat{L}_n = nk - \hat{G}_n$. Use $\mathbb{E}_n [\cdot]$ to denote expectations on
 realization of all strategies as random variables up to round $n$, the expected regret $R(n)$ can be rewritten as
  \begin{IEEEeqnarray*}{l}
  \begin{array}{l}
\!\! R(n) = \mathbb{E}\left[ {\sum\limits_{t = 1}^n {{\textbf{p}_t^T}{\ell_t}(i) - {\textbf{p}_t^T}{\ell_t}({I_t})} } \right] \\
\quad\quad =\mathbb{E}\left[ {{{\hat L}_n}} \right] - \mathop {\min}\limits_{i \in {S_r}} \mathbb{E}_{i \sim {\textbf{p}_n}} \left\{ {{L_n(i)}} \right\}\\
%\quad\quad = \mathbb{E} \sum\limits_{t = 1}^n {{\ell_n}({I_n})}  - \mathop {\min }\limits_{i \in {S_r}} \mathbb{E} \sum\limits_{t = 1}^n {{\ell_n}(i)} \\
  \quad\quad =\mathbb{E} [ \sum\limits_{t = 1}^n {\mathbb{E}_t [ {\sum\limits_{f \in {I_t}} {{\ell_t(f)}} } ]}]  - \mathop {\min }\limits_{i \in {S_r}} ( { \mathbb{E}   [  {\sum\limits_{t = 1}^n
 \mathbb{E}_t [ {\sum\limits_{f \in i} {{\ell_t(f)}}  }] }\!\! ]} ).
 \end{array}\IEEEyesnumber \label{eq:Regrets}
\end{IEEEeqnarray*}
The goal of the algorithm is to minimize the \emph{weak} regret \cite{Bubeck12}, or simply called \emph{regret}. For AOEECC, in addition to
rewards, there are power budget constraints on the decision of transmission power $\textsf{P}_t(f)$ that need to be satisfied. Particularly, for the  decision $\textbf{p}_n$ made by the learner for each channel access strategy, the
power budget constraint can be written as
  \begin{IEEEeqnarray*}{l}
\textbf{p}_n^\textup{T}\textsf{\textbf{P}}_n  \le  \textsf{P}_o\IEEEyesnumber \label{eq:Pbdgt2}.
\end{IEEEeqnarray*}
Note that the SUs of the CC need to make decisions  $\textbf{p}_n$ that attains maximal cummulative reward while satisfying
the additional constraints (\ref{eq:Pbdgt2}).

\begin{figure}
%\vspace{-.3cm}
\centering%.6
\includegraphics[scale=.42]{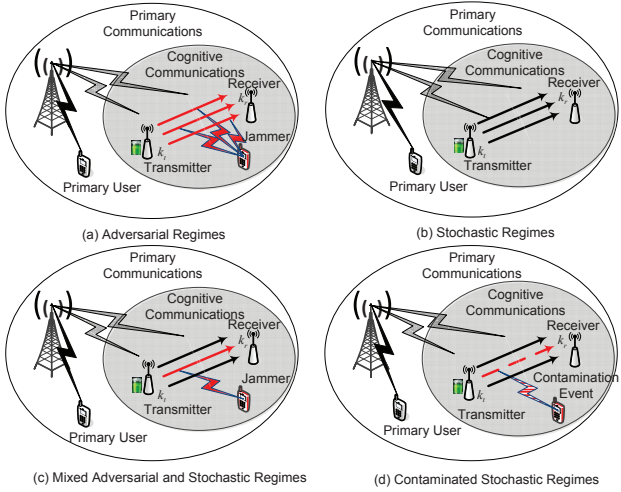}
\caption{Multi-channel EECC in Different Regimes}
\label{fig:digraph}
\vspace{-.5cm}
\end{figure}

%Within our setting, we consider repeated games with \emph{general} rewards model (That includes all the four typical
%regimes in the next subsections, i.e, adversarial, stochastic, mixed adversarial and stochastic and
%contaminated stochastic models) and \emph{stochastic} constraints model. That means the selections of the power allocation
%strategy is according to some random stochastic process in the set of the power strategy.

Within our setting, we refer this problem as the \emph{constrained regret minimization} problem. More precisely, let $\textbf{\textsf{P}}=\{
\textsf{P}(1),..., \textsf{P}(N)\}$ be the constraint vector defined over power allocation actions. In stochastic setting,
the vector $\mathcal{P}$ is not predetermined and is unknown to the learner. In each timeslot $t$, beyond the reward feedback, the SU
receives a random realization $\textbf{\textsf{P}}_t=\{\textsf{P}_t(1),...,\textsf{P}_t(i),..., \textsf{P}_t(N)\}$ of $\mathcal{P}$, where $\mathbb{E}[{\textsf{P}_t(i)}]=\textsf{P}(i)$. W.l.o.g., we assume $\textsf{P}_t \in [0,1]^N$ and $\textsf{P}_o \in [0,1]$. Formally,
the goal is to attain a gradually vanishing \emph{constrained regret} as
  \begin{IEEEeqnarray*}{l}
 \textup{Regret}_n={R(n)}_{\textbf{p}_n^\textup{T}\textsf{P}}  \le O(n^{1-\beta_1})\IEEEyesnumber \label{eq:Pbdgt}.
\end{IEEEeqnarray*}
Furthermore, the decision $\textbf{p}_n$ made by the learner are required to attain sublinear bound on the violation of the
constraint in \emph{long run}, i.e.,
  \begin{IEEEeqnarray*}{l}
 \textup{Violation}_n=  \left[\sum\limits_{t = 1}^n  (\textbf{p}_t^\textup{T}\textsf{P} - \textsf{P}_o)\right]_+  \le O(n^{1-\beta_2})
 \IEEEyesnumber \label{eq:longrun}.
\end{IEEEeqnarray*}
In contrast to the short-term constraints that the constraint  (\ref{eq:Pbdgt}) is required
to be satisfied at every timeslot, SUs are allowed to violate the constraints for some rounds in a controlled
way; but the constraints must hold on average for all rounds, i.e., $(\sum\nolimits_{t = 1}^n \textbf{p}_t^\textup{T}\textsf{P})/n \le \textsf{P}_o$.

\subsection{The Four Regimes of Wireless Environments}
Since our algorithm does not need to know the nature of the environments, there exist  different features of  the environments
that will affect the its performance. We categorize them into the four typical regimes as shown in Fig. 1.

\subsubsection{Adversarial Regime}
In this regime, there is a jammer sending interfering power or injecting garbage data packets
over all $K$ channels such that the transceiver's channel rewards are completely suffered by an unrestricted jammer (See Fig.1 (a)).
Usually, the EE will be significantly reduced in the adversarial regime. Note that, as a classic  model of the well known non-stochastic MAB problem \cite{non_MAB02}, the adversarial regime implies that the jammer often launches attack in every timeslot. It is the most general setting and other three regimes can be regarded as  special cases of the
  adversarial regime.

   %Obviously, a  strategy $i \in \mathop {\arg \min }\nolimits_{i' \in {S_r}} \{ {\mathbb{E}[ {\sum\nolimits_{t = 1}^n {{L_n}(i')} } ]} \}$ is known as a \emph{ best strategy in hindsight} for the first $n$ round.

%We focus on the following
%two type of jammers in the adversarial regime:¡¡%  a) \emph{oblivious jammer:}
%\begin{itemize}
\textbf{Attack Model:}
Different attack philosophies will lead to different level of effectiveness. We focus on the following
two type of jammers in the adversarial regime:¡¡
%\begin{itemize}

  a) \emph{Oblivious attacker:} an oblivious attacker attacks different channels with different attacking strength as a result of different
 EE reductions, which is independent of the past communication records it might have observed.

%As described in \cite{QianJSAC12}, oblivious attacker
%can use \emph{static} and \emph{random} strategies to attack the wireless channels.

 b) \emph{Adaptive attacker:} an adaptive attacker selects its attacking strength on the targeted (sub)set of channels by utilizing its
 past experience and observation of the previous communication records. It is  very powerful and can infer the
  SPR protocol and attack with different level of strength over a subset of channels during a
  single timeslot based on the historical monitoring records.  As shown in a recent work, no bandit algorithm can guarantee a sublinear
regret $o(t)$ against an adaptive adversary with unbounded memory, because the adaptive adversary can mimic the
behavior of SPR protocol to attack, which leads to a linear regret (the attack can not be defended). Therefore, we consider a more practical \emph{$\theta$-memory-bounded adaptive adversary} \cite{Arora12} model. It is an adversary constrained
to loss functions that depends only on the $\theta+1$ most recent strategies.

\subsubsection{Stochastic Regime}
In this regime, the SU's transceiver communicating  over $K$ stochastic channels within PC is  shown in Fig.1 (b). The channel loss $\ell_n(f), \forall f \in 1,...,K$ (Obtained
by transferring  the reward to loss $\ell_n(f)=1- g_n(f)$) of each channel $f$ are sampled independently from an unknown distribution that depends on $f$, but not on $n$. We use $\mu_f = \mathbb{E}
\left[ {{\ell_n(f)}} \right]$ to denote the expected loss of channel $f$. We define channel $f$ as the \emph{best channel}
if $\mu (f) = {\min _{f'}}\{ {\mu (f')} \}$ and \emph{suboptimal channel} otherwise; let $f^*$ denote some best channel. Similarly,  for each strategy
$i \in S_r$, we have  the \emph{best strategy} $\mu (i) = {\min _{i'}}\{ {\sum\nolimits_{f \in i'} {\mu (f)} } \}$ and \emph{suboptimal
strategy} otherwise; let $i^*$ denote some best strategy. For each channel $f$, we define the gap $\Delta (f) = \mu (f) - \mu ({f^*})$;
let $\Delta_f = {\min _{f:\Delta (f) > 0}}\left\{ {\Delta (f)} \right\}$ denote the minimal gap of channels.   Let $N_n(f)$   be the   number of times channel $f$   was played up to time $n$, the regret can be rewritten as
  \begin{IEEEeqnarray*}{l}
R(n) = \sum\nolimits_f {\mathbb{E}\left[ {{N_n}(f)} \right]} \Delta (f). % = \sum\limits_i {\mathbb{E}\left[ {{N_n}(i)} \right]} \Delta (i).
\IEEEyesnumber \label{eq:StocR}
\end{IEEEeqnarray*}
Note that we can calculate the regret either from the perspective of channels $f \in 1,...,K$ or from the perspective of strategies
$i \in S_r$. However, because of the set of strategies is of the size  $\binom{K}{k}$ that grows exponentially with respect to $K$ and
it does not exploit the channel dependency among different strategies, we thus calculate the regret from channels, where tight regret
bounds are achievable.

\subsubsection{Mixed Adversarial and Stochastic Regime}
This regime assumes that the jammer only attacks $k_j$ out of $k$ currently chosen channels at each timeslot shown in Fig.1 (c). There is always a
$k_j/k$ portion of channels under adversarial attack while the other $(k-k_j)/k$ portion is stochastically distributed.

% We call this regime  the mixed adversarial and stochastic regime.

\textbf{Attack Model:} We consider the same attack model as in the adversarial regime. The difference here is that the jammer
only attacks a subset of size $k_j$ over the total  $k$ channels.

%Detailed description of the
%attack models for the mixed adversarial and stochastic regime can be referred in \cite{QianJSAC12}, which can be regarded as a special case of our problem.

\subsubsection{Contaminated Stochastic Regime}
The definition of this regime comes from many practical observations that only a few channels
 and timeslots are exposed to the jammer or other disturbing events in CC. In this regime, for the oblivious jammer, it selects some slot-channel pairs $(t,f)$ as ``locations" to attack, while the remaining channel weights are generated the same as in the stochastic regime. We    define
the \emph{attacking strength} parameter $\zeta \in [0, 1/2)$. After certain $\tau$ timslots, for all $t> \tau$ the total number of contaminated
locations of each suboptimal channel up to time $t$ is $t\Delta(f)\zeta$ and the number of contaminated locations of each best
channel is $t\Delta_f \zeta$. We call a contaminated stochastic regime \emph{moderately contaminated}, if $\zeta$ is at most $1/4$, we can prove
that for all $t> \tau$ on the average over the stochasticity of the
loss sequence the attacker can reduce the gap of every channel by at most one half.

\begin{algorithm}
\caption{AOEECC-EXP3++: An $\epsilon$-SPA Scheme for multi-channel EECC}
\begin{algorithmic}
\STATE \textbf{Input}: $K, k, n$.  See text for definition of $\eta_n$, $\xi_n(f)$, and $\delta_n$.
\STATE \textbf{Initialization}: Set initial channel and strategy losses $\forall i \in [N], \tilde{L}_0(i)= 0$ and $\forall f \in [K], \tilde{\ell}_0(f)= 0$, respectively; Then the initial channel and strategy weights $\forall i \in [N],
W_0(i)= k$ and $\forall f \in [K], w_0(f)= 1$, respectively. The initial total strategy weight $W_0=N=\binom{K}{k}$.
\!\STATE \textbf{Set}: \!\! $\beta_n \!\! =\!\! \frac{1}{2}\sqrt {\frac{{\ln K}}{{nK}}}$; ${\varepsilon _n}\left( f \right) \!= \! \min \left\{ {\frac{1}{{2K}},{\beta _n},{\xi _n}\left( f \right)} \right\},\forall f \in \left[ K \right]$ and $\gamma_n= \sum\nolimits_{f=1}^{K} {{\varepsilon _n}(f)}$.
\FOR { timeslot $n=1,2,...$}
\STATE 1: Based on sensing and probing results,  randomly selects a channel access strategy $I_n$ according to the strategy't probability $p_n(i),\forall f \in \left[ K \right]$, with $p_n(i)$ computed as follows:
\begin{IEEEeqnarray*}{l}{p_n}(i) = \left\{ \begin{array}{l}
\!\!\!  ( 1 - \gamma_n )\frac{{{w_{n - 1}}\left( i \right)}}{{{W_{n - 1}}}} + \! \sum\nolimits_{f \in i} {{\varepsilon _n}(f)} \ \emph{if} \ i \in \mathcal{C} \\
\!\!\!  ( 1 - \gamma_n )\frac{{{w_{n - 1}}\left( i \right)}}{{{W_{n - 1}}}} \  \quad   \quad \quad \quad \
\text{\emph{if}} \ i \notin \mathcal{C}
\end{array} \right.\IEEEyesnumber \label{eq:Aleq1}
\end{IEEEeqnarray*}
%The computation is taken for the probability distributions over all strategies $p_n=(p_n(1),p_n(2),...,p_n(N))$.

\STATE 2: Computes the probability $q_n(f),\forall f \in \left[ K \right]$,
\begin{IEEEeqnarray*}{l}
\begin{array}{l}
{\rho_n}(f) = \sum\nolimits_{i:f \in i} {{p_n}(i)}
 = ( {1 - \gamma_n } )\frac{{\sum\nolimits_{i:f \in i} {{w_{n - 1}}\left( i \right)} }}{{{W_{n - 1}}}} \\
\quad \quad\quad \quad \quad \quad \quad \quad \ \ + \sum\nolimits_{f \in i} {{\varepsilon _n}(f)}
 \left| {\left\{ {i \in \mathcal{C}:f \in i} \right\}} \right|.
\end{array}\IEEEyesnumber \label{eq:Aleq2}
\end{IEEEeqnarray*}
%Then, the probability distributions over all channels are $q_n=(q_n(1),q_n(2),...,q_n(K))$.

\STATE 3: Sense and probe channels for $f \in {I_n}$.  Receive the scaled (i.e., in the range $[0, 1]$)
loss model (converted from reward model) of reward and power , and then calculate \textup{EE} for channel $f$, $\ell_{n-1}(f)$,  and
the  realization of power budget ${\textsf{P}}_n(f)$,  $\forall f \in I_n$. Update the estimated loss with
 augmented power allocation constraint $\tilde{\psi}_{n}(f), \forall f \in [K]$ as follows:
\begin{IEEEeqnarray*}{l}
{\tilde{\psi}_n}(f) = \left\{ \begin{array}{l}
{\mathbf{\tilde{\ell}}}_{n-1}(f) + \lambda_{n-1} \tilde{\textsf{P}}_{n-1}(f), \quad  \text{\emph{if}} \ f \in {I_n} \\
0         \ \ \quad\quad\quad\quad\quad\quad\quad\quad\quad\quad\ \emph{otherwise}.
\end{array} \right.\IEEEyesnumber \label{eq:Aleq3}
\end{IEEEeqnarray*}

\STATE 4: Update the Lagrangian Multiplier by the following equation:
 \begin{IEEEeqnarray*}{l}
 {\lambda _{n}} = [(1 - \delta_{n-1} \eta_{n-1} \sqrt{\gamma_{n-1}}){\lambda_{n-1}} \\
 \quad \quad \quad \quad \quad - \eta_{n-1} \sqrt{\gamma_{n-1}} ( \textsf{P}_o- \textbf{p}_{n-1}^\textup{T}\tilde{\textsf{P}}_{n-1})]_ + .
 \end{IEEEeqnarray*}

\STATE 5: The receiver updates all the weights as
 \begin{IEEEeqnarray*}{l}
{w_n}\left( f \right) = {w_{n - 1}}\left( f \right){e^{ - \eta_n \tilde{\psi}_{n}(f)}} = {e^{ - \eta_n \tilde{\Psi}_{n}(f)}}, \\
{{\bar w}_n}\left( i \right) = \prod\nolimits_{f \in i} {{w_n}(f)}  = {{\bar w}_{n - 1}}\left( i \right){e^{ - \eta_n {{\tilde \psi}_n}(i)}},
\end{IEEEeqnarray*}
The sum   weights  of all strategies is
${W_n} = \sum\nolimits_{i \in S_r} {{{\bar w}_n}\left( i \right)}$.

\STATE 6: Access each of the channel  $f \in {I_n}$ with probability $\epsilon$, i.e., set ${\mathds{1}_n}(f)=1$ with probability  $\epsilon$.
\ENDFOR
\end{algorithmic}
\end{algorithm}
\vspace{-.2cm}

\section{The AOEECC Algorithm}
In this section, we focus on developing an AOEECC algorithm for the SU. The design philosophy is that the transmitter  collects and learns the rewards of the
previously chosen channels, based on which it can decide the next timeslot channel access strategy, i.e., the SU
will decide whether to transmit data over the current channel set (called \emph{exploitation}) or to continue
sensing/probing some other channels for accessing (called \emph{exploration}).

We describe the Algorithm 1, namely AOEECC-EXP3++,  is a combinatorial variant based on
EXP3 algorithm. Before we present the algorithm, let us introduce the following vectors: $ {\bm {\tilde{\ell}}}_t$ is
all zero vector except in the $I_t$th  channel access strategy and so does the channel loss $\tilde{\mathbf{\ell}}_t(f)$ within
the $I_t$, $\forall f \in I_t$, we have $\tilde{\mathbf{\ell}}_t(f)= {\mathbf{\ell}}_t(f)/\rho_n(f)$. Similarly ${
 \bm \tilde {\textsf{P}}}_t$ is all zero vector except in $I_t$th  channel access strategy and so does the power $\tilde{\textsf{P}} _t(f)$ within
the $I_t$, $\forall f \in I_n$, where we have $\tilde{\textsf{P}}_t(f)= {\textsf{P}}_t(f)/\rho_n(f)$. It is easy to verify $\mathbb{E}
_{i_t}[{\mathbf{\tilde{\ell}}}_t(f)]= {\mathbf{{\ell}}}_t(f)$  and $ \mathbb{E} _{i_t}[\tilde{\textsf{P}}_t(f)]= {\textsf{P}}_t(f)$, where
$\bm {\rho}_{n,f} = (\rho_n(1),...,\rho_n(f),...,\rho_n(K))$ and $\textbf{p}_n = (p_n(1),...,p_n(i),...,p_n(N))$. In addition, we have
the following equalities  at step $5$ of Algorithm 1.
\begin{IEEEeqnarray*}{l}
\begin{array}{l}
\!\!\!\! {{\tilde \Psi}_n}(f) = {{\tilde \Psi}_{n - 1}}(f) + {{\tilde \psi}_{n - 1}}(f), {{\tilde \Psi}_n}(i) = {{\tilde \Psi}_{n - 1}}(i) + {{\tilde \psi}_{n - 1}}(i)\\
\!\!\!\!{{\tilde \Psi}_n}(f) =  {{\tilde{L}}}_{n-1}(f) + \lambda_{n-1} {{\tilde{\Gamma}}_{n-1}}(f), \\
 \!\!\!\! {{\tilde \Psi}_n}(i) =  {{\tilde{L}}}_{n-1}(i) + \lambda_{n-1} {{\tilde{\Gamma}}_{n-1}} (i), \\
\!\!\!\!{{\tilde L}_n}(f) = {{\tilde L}_{n - 1}}(f) + {{\tilde \ell}_{n - 1}}(f),{{\tilde L}_n}(i) = {{\tilde L}_{n - 1}}(i) + {{\tilde \ell}_{n - 1}}(i) \\
\!\!\!\! {{\tilde{\Gamma}}_n}(f) = {{\tilde{{\Gamma}}}_{n - 1}}(f) + {{\tilde{\textsf{P}}}_{n - 1}}(f), {{\tilde{\Gamma}}_n}(i) = {{\tilde{{\Gamma}}}_{n - 1}}(i) + {{\tilde{\textsf{P}}}_{n - 1}}(i) \\
\!\!\!\! {{\tilde \ell}_{n - 1}}(i) = \sum\nolimits_{f \in i} {{{\tilde \ell}_{n - 1}}(f)},  {{\tilde{\textsf{P}}}_{n - 1}}(i) = \sum\nolimits_{f \in i} {{{\tilde{\textsf{P}}}_{n - 1}}(f)},
\end{array}\IEEEyesnumber \label{eq:Aleq2d}
\end{IEEEeqnarray*}
where ${{\tilde L}_n}(f)$ and ${{\tilde{\Gamma}}_n}(f)$ are the respective accumulated estimated loss and allocated power on channel $f$ up to round $n$, and
${{\tilde L}_n}(i)$ and ${{\tilde{\Gamma}}_n}(i)$ are the respective accumulated estimated loss and allocated power on strategy $i$ up to round $n$. Moreover,
we have the exploration probability been decomposed for each channel, where we have $\gamma_n= \sum\nolimits_{f=1}^{K} {{\varepsilon _n}(f)} $.

 Our new algorithm uses the fact that when losses  (converted from rewards) and power of channels in the chosen strategy are revealed, it
also shares this information with the common channels of the other chosen strategies. During each timeslot, we assign
a channel weight that is dynamically adjusted based on the channel losses revealed. The weight of a strategy is determined
by the product of weights of all channels. Our algorithm has two control levers: the \emph{learning rate} $\eta_n$ and the exploration
parameters $\xi_n(f)$ for each channel $f$. To facilitate the  adaptive
channel access to optimal solutions without the knowledge about the nature of the environments, the crucial innovation is the introduction
of exploration parameters $\xi_n(f)$, which are tuned individually for each arm depending on the past observations.

A set of \emph{covering strategy} is defined to ensure that
each channel is sampled sufficiently often. It has the property that,  for each channel $f$, there is a strategy $i \in \mathcal{C}$ such that
$f \in i$. Since there are only $K$ channels and each strategy includes $k$ channels, we have $|\mathcal{C}| = \lceil {\frac{K}{{{k}}}} \rceil$.
 The value $ \sum\nolimits_{f \in i} {{\varepsilon _n}(f)}$ means the randomized
 exploration probability for each strategy $i \in \mathcal{C}$, which is the summation of each channel $f$'s exploration probability ${\varepsilon _n}\left( f \right)$ that belongs to the strategy $i$. The introduction of $\sum\nolimits_{f \in i} {{\varepsilon _n}\left( f \right)}$ ensures that $p_n(i) \ge  \sum\nolimits_{f \in i} {{\varepsilon _n}(f)}$ so that it is a mixture of exponentially
 weighted average distribution and uniform distribution  \cite{Bubeck12} over each strategy.

In the following discussion, to facilitate the AOEECC-EXP3++ algorithm without knowing about the nature of environments, we can apply the
 two control parameters simultaneously by setting $\eta_n = \beta_n$, $2kK\eta \le \delta = O(\eta_n)$ and use the control parameter $\xi_n(f)$  such that it can achieve
 the optimal ``root-n" regret in the adversarial regime and almost optimal ``logarithmic-n" regret in the stochastic regime.

 % In addition, we show that the new control lever $\xi_n(f)$ is even more powerful and allows to
% detect and exploit the gap in even more challenging situations of the contaminated stochastic regimes.

\section{Performance Results of EECC under $\epsilon$-SPA}
This section analyzes the regret and power budget violation performance of our proposed AOEECC-EXP3++ algorithm in different regimes.

\subsection{Adversarial Regime}
We first show that tuning $\eta_n$ and $\xi_n(f)$ together, we can get the optimal regret (of reward and violation) of AOEECC-EXP3++ in the adversarial regime, which is a
general result that holds for all other regimes. Define ${{{\hat G}_n}}(\epsilon)$ as the expected average EEs that can be achieved by the $\epsilon$-SPA scheme over $n$ rounds. The Theorem 1,
Theorem 3, Theorem 5, Theorem 7, Theorem 9 and Theorem 11 bound
the regret of EE, $\mathop {\max }\nolimits_{i \in {S_r}} \mathbb{E}_{i \sim {\textbf{p}_n}}\{ {{G_{n,i}}} \} - \mathbb{E}[ {{{\hat G}_n}(1)} ]$ when set $\epsilon =1$.

\textbf{Theorem 1.}  Under the {oblivious} jamming attack, no matter how the status of the channels change (potentially in an adversarial manner), for $\eta_n = \beta_n$, $\delta_n  = 2k\sqrt {\frac{{K\ln K}}{n}} $ and any
$\xi_n(f) = \tilde{O}(1/n)$, the regret of the AOEECC-EXP3++ algorithm for any $n$ satisfies:
\begin{IEEEeqnarray*}{l}
\begin{array}{l}
 \textup{Regret}_n={R(n)}_{\textbf{p}_n^\textup{T}\textsf{P}} \le 4{k}\sqrt {nK\ln K} = O(n^{1/2}). \\
  \textup{Violation}_n=   \mathbb{E}{[ {\sum\nolimits_{t = 1}^n {\textbf{p}_{t}^\textup{T}\tilde{\textsf{P}}_{t}-
 \textsf{P}_o} } ]_ + } \le O(n^{\frac{3}{4}}).
 \end{array}
\end{IEEEeqnarray*}

From Theorem 1, we can find that the regret is
order and leading factor optimal when compared to the results in the anti-jamming wireless communications \cite{QianJSAC12}. For the
power budget violation, we have a regret of  sublinear $O(n^{\frac{3}{4}})$. From the proof of the Theorem, this upper bound may be
very loose.

According to the $\epsilon$-SPA scheme, CR user will transmit $\epsilon n$ times in expectation during $n$ rounds. It is easy to
show $\mathbb{E}[{{\hat G}_n}(\epsilon )] = \epsilon \mathbb{E}[{{\hat G}_n}(1)]$, which implies $\mathbb{E}[{{\hat G}_n}(\epsilon )]
\ge \epsilon \mathop {\max }\nolimits_{i
\in {S_r}} \mathbb{E}[{{\hat G}_{n,i}}] - 4\epsilon k\sqrt {nK\ln K} $.
Let $G_{\max}$ be the large expected data rate of channel access strategies among all the strategies. We have $\mathop {\max }\nolimits_{i
 \in {S_r}} \mathbb{E}_{i \sim {\textbf{p}_n}}\{ {{G_{n,i}}} \} = n \cdot G_{\max}$. Assume $t_a = \alpha t_{sp}$ where constant $\alpha
 \gg 1$. Then we have

\textbf{Theorem 2.} The expected EE of $\epsilon$-SPA scheme of AOEECC-EXP3++ in the adversarial regime
 under the oblivious jammer is at least
 \begin{IEEEeqnarray*}{l}
 \frac{{{{\hat G}_n}(\epsilon ){t_a}}}{T} \! \ge \! \frac{{{G_{\max }}\! -\! 4k\sqrt {\frac{{K\ln K}}{n}} }}{{\frac{1}{{\alpha \epsilon }}\! +\! 1}}\! = \!
 \frac{{{G_{\max }} \!-\! 4k\sqrt {\frac{{(1 \! + \! \alpha \epsilon ){t_{sp}}K\ln K}}{T}} }}{{\frac{1}{{\alpha \epsilon }} \!+ \!1}},
\end{IEEEeqnarray*}
where $T = n{t_{sp}} + \epsilon n{t_a}$.

We find that when  $T$ is sufficiently large, the achievable expected EE is at least $\frac{{{G_{\max }}}}{{\frac{1}{{\alpha \epsilon }} + 1}}$, which
is maximized when $\epsilon=1$. Obviously, the expected EE that can be achieved is no more than $\frac{{{G_{\max }}{t_a}}}{{{t_{sp}} + {t_a}}} = \frac{{{G_{\max }}}}
{{\frac{1}{\alpha } + 1}}$, because each transmission takes at least $t_{sp} + t_a$ time while the
expected EE is no more than $G_{\max }$. Thus, when $T$ is sufficiently large, the $\epsilon$-SPA scheme of AOEECC-EXP3++ is almost optimal. Similar
conclusions holds also for the following Theorem 4, Theorem 6, Theorem 8,  Theorem 10, Theorem 12 and Theorem 14.

\textbf{Theorem 3.}  Under the \emph{$\theta$-memory-bounded adaptive} jamming attack, for $\eta_n = \beta_n$, $\delta_n  = 2k\sqrt {\frac{{K\ln K}}{n}} $  and any
$\xi_n(f) \ge 0$, the regret of the AOEECC-EXP3++ algorithm for any $n$ is upper bounded by:
\begin{IEEEeqnarray*}{l}
\begin{array}{l}
 \textup{Regret}_n={R(n)}_{\textbf{p}_n^\textup{T}\textsf{P}} \le  O((\theta + 1){(4{k}\sqrt {K\ln K} )^{\frac{2}{3}}}{n^{\frac{2}{3}}}). \\
  \textup{Violation}_n=   \mathbb{E}{[ {\sum\nolimits_{t = 1}^n {\textbf{p}_{t}^\textup{T}\tilde{\textsf{P}}_{t}-
 \textsf{P}_o} } ]_ + } \le O(n^{\frac{3}{4}}).
 \end{array}
\end{IEEEeqnarray*}

\textbf{Theorem 4.} The expected EE of $\epsilon$-SPA scheme of AOEECC-EXP3++ in the adversarial regime
 under the \emph{$\theta$-memory-bounded adaptive}  jammer is at least
 \begin{IEEEeqnarray*}{l}
\frac{{{{\hat G}_n}(\epsilon ){t_a}}}{T}\! \ge \! \frac{{{G_{\max }}\! - \! (\theta+1){{( {4k \sqrt{K\ln KT} }
 )}^{\frac{2}{3}}}{{( {(1 + \alpha \epsilon ){t_{sp}}} )}^{ \frac{1}{3}}}}}{{\frac{1}{{\alpha \epsilon }} + 1}},
\end{IEEEeqnarray*}
where $T = n{t_{sp}} + \epsilon n{t_a}$. With sufficiently large $T$, our $\epsilon$-SPA scheme of AOEECC-EXP3++
is almost optimal.

\subsection{Stochastic Regime}
%Now we show that for any $\eta_n \ge \beta_n$, tuning the exploration parameters $\xi_n(f)$ is sufficient to control the regret of the algorithm
%in the stochastic regime.
 We consider a different number of ways of tuning the exploration parameters $\xi_n(f)$ for different practical
 implementation considerations, which will lead to different regret performance of AOEECC-EXP3++. We begin with an idealistic assumption that the gaps $\Delta(f), \forall f \in K$ is known in Theorem 5, just to
give an idea of what is the best result we can have and our general idea for all our proofs.

%\subsubsection{A Idealistic Case of knowing the gap}

% and $n^*_1= \max_{\{f \in K\}} n^* {{(f)}}$
\textbf{Theorem 5}. Assume that the gaps $\Delta(f), \forall f \in K,$  are known. Let $n^*$ be the
minimal integer that satisfy ${n^*(f)} \ge \frac{{4{c^2}K\ln {{({n^*(f)}\Delta {{(f)}^2})}^2}}}{{\Delta {{(f)}^4}\ln (K)}}$.
For any choice of $\eta_n = {\beta _n}$
 and any $c \ge 18$, the regret of the AOEECC-EXP3++ algorithm with $\xi_n(a)=\frac{{c\ln (n\Delta {{(f)}^2})}}{{n\Delta {{(f)}^2}}}$ and
 $\delta_n  = 2k\sqrt {\frac{{K\ln K}}{n}} $ in the
 stochastic regime satisfies:
 \begin{IEEEeqnarray*}{l}
 \hspace{-2.43em}\begin{array}{l}
    \textup{Regret}_n={R(n)}_{\textbf{p}_n^\textup{T}\textsf{P}}  \mathop  \le O(kK\frac{{c\ln {{(n)}^2}}}{{\Delta_f}}) \\
  \textup{Violation}_n=   \mathbb{E}{[ {\sum\nolimits_{t = 1}^n {\textbf{p}_{t}^\textup{T}\tilde{\textsf{P}}_{t}-
 \textsf{P}_o} } ]_ + }  \le O(n^{\frac{3}{4}}).
\end{array}\IEEEyesnumber \label{eq:CstReg3}
\end{IEEEeqnarray*}
  From the upper bound results, we note that the leading constants $k$  and $K$ are optimal and tight as indicated in CombUCB1 \cite{Branislav2015} algorithm. However, we have a factor of $\ln(n)$ worse of the regret performance than
the optimal ``logarithmic" regret as in \cite{Robbins1985}\cite{Branislav2015}, where the
performance gap is
trivially negligible (See numerical results in Section IX).

\textbf{Theorem 6.} The expected EE of $\epsilon$-SPA scheme of AOEECC-EXP3++ in the stochastic regime  is at least
 \begin{IEEEeqnarray*}{l}
\frac{{{{\hat G}_n}(\epsilon ){t_a}}}{T} \ge \frac{{{G_{\max }} - \frac{{2ckK\ln {{({{(1 + \alpha \epsilon ){t_{sp}}} \mathord{\left/
 {\vphantom {{(1 + \alpha \epsilon ){t_{sp}}} T}} \right.
 \kern-\nulldelimiterspace} T})}^2}}}{{{{\Delta _f(1 + \alpha \epsilon ){t_{sp}}} \mathord{\left/
 {\vphantom {{(1 + \alpha \epsilon ){t_{sp}}} T}} \right.
 \kern-\nulldelimiterspace} T}}}}}{{\frac{1}{{\alpha \epsilon }} + 1}},
\end{IEEEeqnarray*}
where $T = n{t_{sp}} + \epsilon n{t_a}$. With sufficiently large $T$, our $\epsilon$-SPA scheme of AOEECC-EXP3++
is almost optimal.

%, which is an open question. We believe that this gap
%can finally be closed by more advanced techniques.

\subsubsection{A Practical Implementation by estimating the gap}
Because of the gaps $\Delta(f), \forall f \in K$ can not be known in advance before running the algorithm.
In the next, we show a more practical result that using the empirical gap as an estimate
of the true gap. The estimation process can be performed in background for each channel $f$ that
starts from the running of the algorithm, i.e.,
 \begin{displaymath}
{{\hat \Delta }_n}(f) = \min \{ {1,\frac{1}{n}({{\tilde L}_n}(f) -\mathop {\min }\limits_{f'} ({{\tilde L}_n}(f')) )} \}.
\IEEEyesnumber \label{eq:EstD}
 \end{displaymath}
 This is a first algorithm that can be used in many real-world applications.

\textbf{Theorem 7.} Let $c \ge 18$ and $\eta_n = \beta_n$. Let $n^*$ be the minimal integer that satisfies $n^*
 \ge \frac{{4{c^2}\ln {{(n^*)}^4} K}}{{\ln (K)}}$, and let ${n^*}(f) = \max \{ {{n^*},\lceil {{e^{1/\Delta {{(f)}^2}}}} \rceil } \}$ and
$n^*=max_{\{f \in K\}} n^* {{(f)}}$. The regret of the AOEECC-EXP3++ algorithm
with ${\xi _n}(f) = \frac{{c{{( {\ln n} )}^2}}}{{n{{\hat \Delta }_{n-1}}{{(f)}^2}}}$ and
 $\delta_n  = 2k\sqrt {\frac{{K\ln K}}{n}} $, termed as
AOEECC-EXP3++$^\emph{AVG}$, in the stochastic regime satisfies:
 \begin{IEEEeqnarray*}{l}
 \hspace{-2.43em}\begin{array}{l}
    \textup{Regret}_n={R(n)}_{\textbf{p}_n^\textup{T}\textsf{P}}  \mathop  \le O(kK\frac{{c\ln {{(n)}^3}}}{{\Delta_f}}) \\
  \textup{Violation}_n=   \mathbb{E}{[ {\sum\nolimits_{t = 1}^n {\textbf{p}_{t}^\textup{T}\tilde{\textsf{P}}_{t}-
 \textsf{P}_o} } ]_ + }  \le O(n^{\frac{3}{4}}).
\end{array}\IEEEyesnumber \label{eq:CstReg5}
\end{IEEEeqnarray*}
From the theorem, we see in this more practical case,  another factor of $ln(n)$ worse of the regret performance when compared to the idealistic
case for EECC.

%Also, the additive constants $n^*$ in this theorem can be  very large. However, our experimental results show that
%a minor modification of this algorithm performs comparably to ComUCB1 \cite{Branislav2015} in the stochastic regime.

\textbf{Theorem 8.} The expected EE of $\epsilon$-SPA scheme of AOEECC-EXP3++$^\emph{AVG}$ in the stochastic regime
 under the oblivious jammer is at least
 \begin{IEEEeqnarray*}{l}
\frac{{{{\hat G}_n}(\epsilon ){t_a}}}{T} \ge \frac{{{G_{\max }} - \frac{{2ckK\ln {{({{(1 + \alpha \epsilon ){t_{sp}}} \mathord{\left/
 {\vphantom {{(1 + \alpha \epsilon ){t_{sp}}} T}} \right.
 \kern-\nulldelimiterspace} T})}^3}}}{{{\Delta _f {(1 + \alpha \epsilon ){t_{sp}}} \mathord{\left/
 {\vphantom {{(1 + \alpha \epsilon ){t_{sp}}} T}} \right.
 \kern-\nulldelimiterspace} T}}}}}{{\frac{1}{{\alpha \epsilon }} + 1}} ,
\end{IEEEeqnarray*}
where $T = n{t_{sp}} + \epsilon n{t_a}$. With sufficiently large $T$, our $\epsilon$-SPA scheme of AOEECC-EXP3++ is almost optimal.

\subsection{Mixed Adversarial and  Stochastic Regime}
The mixed adversarial and stochastic regime can be regarded as a special case of mixing
adversarial and stochastic regimes. Since there is  always a jammer randomly attacking $k_j$ transmitting channels
constantly over time,  we will have the following theorem for the AOEECC-EXP3++$^\emph{AVG}$ algorithm, which is
a much more refined regret performance bound than the general regret bound in the adversarial regime.

\textbf{Theorem 9.} Let $c \ge 18$ and $\eta_n = \beta_n$. Let $n^*$ be the minimal integer that satisfies $n^*
 \ge \frac{{4{c^2}\ln {{(n^*)}^4} K}}{{\ln (K)}}$, and Let ${n^*}(f) = \max \{ {{n^*}, \lceil {{e^{1/\Delta {{(f)}^2}}}} \rceil } \}$ and
$n^*=max_{\{f \in K\}} n^* {{(f)}}$. The regret of the AOEECC-EXP3++ algorithm
with ${\xi _n}(f) = \frac{{c{{( {\ln n} )}^2}}}{{n{{\hat \Delta }_{n-1}}{{(f)}^2}}}$ and
 $\delta_n  = 2k\sqrt {\frac{{K\ln K}}{n}} $, termed as
AOEECC-EXP3++$^\emph{AVG}$ under \emph{oblivious jamming} attack, in the mixed stochastic and adversarial regime satisfies:
 \begin{IEEEeqnarray*}{l}
  \begin{array}{l}
    \textup{Regret}_n={R(n)}_{\textbf{p}_n^\textup{T}\textsf{P}}  \mathop  \le O(K(k-k_j)\frac{{c\ln {{(n)}^3}}}{{\Delta_f}})
     \\
     \hspace{3.2cm}+ O(4k_j\sqrt {tK\ln K}) \\
  \textup{Violation}_n=   \mathbb{E}{[ {\sum\nolimits_{t = 1}^n {\textbf{p}_{t}^\textup{T}\tilde{\textsf{P}}_{t}-
 \textsf{P}_o} } ]_ + }  \le O(n^{\frac{3}{4}}).
\end{array}\IEEEyesnumber \label{eq:CstReg4}
\end{IEEEeqnarray*}
Note that the results in Theorem 9 has better regret performance than the classic results obtained in adversarial regime shown
 in Theorem 1 and the anti-jamming algorithm in \cite{QianJSAC12}.

\textbf{Theorem 10.} The expected EE of $\epsilon$-SPA scheme of AOEECC-EXP3++$^\emph{AVG}$, in the mixed stochastic and adversarial regime
 under \emph{oblivious jamming} attack  is at least
 \begin{IEEEeqnarray*}{l}
\begin{array}{l}
\frac{{{{\hat G}_n}(\epsilon ){t_a}}}{T} \ge \frac{{{G_{\max }} - \frac{{2cK\left( {k - {k_j}} \right)\ln {{({{(1 + \alpha \epsilon ){t_{sp}}} \mathord{\left/
 {\vphantom {{(1 + \alpha \epsilon ){t_{sp}}} T}} \right.
 \kern-\nulldelimiterspace} T})}^3}}}{{{\Delta _f{(1 + \alpha \epsilon ){t_{sp}}} \mathord{\left/
 {\vphantom {{(1 + \alpha \epsilon ){t_{sp}}} T}} \right.
 \kern-\nulldelimiterspace} T}}}}}{{\frac{1}{{\alpha \epsilon }} + 1}}\\
\hspace{1.5cm} - \frac{{4{{{k_j}}}\sqrt {\frac{{(1 + \alpha \epsilon ){t_{sp}}K\ln K}}{T}} }}{{\frac{1}{{\alpha \epsilon }} + 1}},
\end{array}
\end{IEEEeqnarray*}
where $T = n{t_{sp}} + \epsilon n{t_a}$. With sufficiently large $T$, our $\epsilon$-SPA scheme of AOEECC-EXP3++
is almost optimal.

\textbf{Theorem 11.}  Let $c \ge 18$ and $\eta_n \ge \beta_n$. Let $n^*$ be the minimal integer that satisfies $n^*
 \ge \frac{{4{c^2}\ln {{(n^*)}^4} K}}{{\ln (K)}}$, and Let ${n^*}(f) = \max \{ {{n^*},\lceil {{e^{1/\Delta {{(f)}^2}}}} \rceil } \}$ and
$n^*=max_{\{f \in K\}} n^* {{(f)}}$. The regret of the AOEECC-EXP3++ algorithm
with ${\xi _n}(f) = \frac{{c{{( {\ln n} )}^2}}}{{n{{\hat \Delta }_{n-1}}{{(f)}^2}}}$ and
 $\delta_n  = 2k\sqrt {\frac{{K\ln K}}{n}} $, termed as
AOEECC-EXP3++$^\emph{AVG}$ \emph{$\theta$-memory-bounded adaptive} jamming attack, in the mixed stochastic and adversarial regime satisfies:
 \begin{IEEEeqnarray*}{l}
  \begin{array}{l}
    \textup{Regret}_n={R(n)}_{\textbf{p}_n^\textup{T}\textsf{P}}  \mathop  \le O(K(k-k_j)\frac{{c\ln {{(n)}^3}}}{{\Delta_f }}) \\
       \hspace{2.2cm}+ O((\theta + 1){(4{k_j}\sqrt {K\ln K} )^{\frac{2}{3}}}{n^{\frac{2}{3}}}) \\
  \textup{Violation}_n=   \mathbb{E}{[ {\sum\nolimits_{t = 1}^n {\textbf{p}_{t}^\textup{T}\tilde{\textsf{P}}_{t}-
 \textsf{P}_o} } ]_ + }  \le O(n^{\frac{3}{4}}).
\end{array}\IEEEyesnumber \label{eq:CstReg4}
\end{IEEEeqnarray*}
%The results shown in Theorem 5 provides the first quantitative regret performance under
%adaptive jamming attack, while the related work \cite{QianJSAC12} with the similar adversary model and the
%same communication scenario in this case only provided
%simulation results demonstrations.

\textbf{Theorem 12.} The expected EE of $\epsilon$-SPA scheme of AOEECC-EXP3++$^\emph{AVG}$, in the mixed stochastic and adversarial regime
 under \emph{$\theta$-memory-bounded adaptive}
jamming attack is at least
 \begin{IEEEeqnarray*}{l}
\begin{array}{l}
\frac{{{{\hat G}_n}(\epsilon ){t_a}}}{T} \ge \frac{{{G_{\max }} - \frac{{2cK\left( {k - {k_j}} \right)\ln {{({{(1 + \alpha
\epsilon ){t_{sp}}} \mathord{\left/
 {\vphantom {{(1 + \alpha \varepsilon ){t_{sp}}} T}} \right.
 \kern-\nulldelimiterspace} T})}^3}}}{{{\Delta _f {(1 + \alpha \epsilon ){t_{sp}}} \mathord{\left/
 {\vphantom {{(1 + \alpha \epsilon ){t_{sp}}} T}} \right.
 \kern-\nulldelimiterspace} T}}}}}{{\frac{1}{{\alpha \epsilon }} + 1}}\\
\hspace{1.5cm} - \frac{{{{(\theta  + 1){}}}{{\left( {4k_j\sqrt {K\ln K T} } \right)}^{2/3}}{{\left( {(1 +
\alpha \epsilon ){t_{sp}}} \right)}^{ \frac{1}{3}}}}}{{\frac{1}{{\alpha \epsilon }} + 1}},
\end{array}
\end{IEEEeqnarray*}
where $T = n{t_{sp}} + \epsilon n{t_a}$. With sufficiently large $T$, our $\epsilon$-SPA scheme of AOEECC-EXP3++
is almost optimal.

\subsection{Contaminated Stochastic Regime}
We  show that the algorithm AOEECC-EXP3++$^\emph{AVG}$ can still retain ``polylogarithmic-n" regret in the  contaminated stochastic
regime with a potentially large leading constant in the performance. The following is the result for the \emph{moderately contaminated stochastic regime}.

%The \emph{moderately contaminated stochastic
%regime} without a significant deterioration in performance. We also define the \emph{severely contaminated stochastic
%regime}. The general result for the contaminated stochastic regime is summarized in the following theorem.

\textbf{Theorem 13.} Under the setting of all parameters given in Theorem 2, for ${n^*}(f) = \max \{ {{n^*},\lceil {{e^{4
/\Delta {{(f)}^2}}}} \rceil } \}$, where $n^*$ is defined as before  and
$n_3^*=max_{\{f \in K\}} n^* {{(f)}}$, and the attacking strength parameter $\zeta \in [0,1/4)$ the regret of the \rm{AOEECC-EXP3++} algorithm in the
contaminated stochastic regime that is contaminated after $\tau$ steps satisfies:
 \begin{IEEEeqnarray*}{l}
 \hspace{-2.43em}\begin{array}{l}
    \textup{Regret}_n={R(n)}_{\textbf{p}_n^\textup{T}\textsf{P}}  \mathop  \le {O( {\frac{{K k\ln {{(n)}^3}}}
 {{(1-2\zeta)\Delta_{f}}}} )} + K n_3^* \\
  \textup{Violation}_n=   \mathbb{E}{[ {\sum\nolimits_{t = 1}^n {\textbf{p}_{t}^\textup{T}\tilde{\textsf{P}}_{t}-
 \textsf{P}_o} } ]_ + }  \le O(n^{\frac{3}{4}}).
\end{array}\IEEEyesnumber \label{eq:CstReg5}
\end{IEEEeqnarray*}

Note that $\zeta$ can be within the interval  $[0, 1/2
)$. If $\zeta \in (1/4, 1/2)$, the leading factor $1/(1-2\zeta)$ will be very large, which is \emph{severely} contaminated. Now, the obtained regret bound is not quite meaningful, which could be much worse than the regret performance in the adversarial regime for
both oblivious and adaptive adversary.

\textbf{Theorem 14.} The expected EE of $\epsilon$-SPA scheme of AOEECC-EXP3++ in the
contaminated stochastic regime that is contaminated is at least
 \begin{IEEEeqnarray*}{l}
\frac{{{{\hat G}_n}(\varepsilon ){t_a}}}{T} \ge \frac{{{G_{\max }} - \frac{{2ckK\ln {{({{(1 + \alpha \varepsilon ){t_{sp}}} \mathord{\left/
 {\vphantom {{(1 + \alpha \varepsilon ){t_{sp}}} T}} \right.
 \kern-\nulldelimiterspace} T})}^3}}}{{{{(1 - 2){\Delta _f}(1 + \alpha \varepsilon ){t_{sp}}} \mathord{\left/
 {\vphantom {{(1 - 2){\Delta _f}(1 + \alpha \varepsilon ){t_{sp}}} T}} \right.
 \kern-\nulldelimiterspace} T}}}}}{{\frac{1}{{\alpha \varepsilon }} + 1}},
\end{IEEEeqnarray*}
where $T = n{t_{sp}} + \epsilon n{t_a}$. With sufficiently large $T$, our $\epsilon$-SPA scheme of AOEECC-EXP3++
is almost optimal.

 %  \begin{displaymath}
%         \begin{array}{l}
% R(n) \le \sum\limits_{f = 1}^K {O\left( {\frac{{k\ln {{(n)}^3}}}
% {{(1-2\zeta)\Delta {{(f)}}}}} \right)}  + \sum\limits_{f = 1}^K \Delta {{(f)}} \max\{n^* {{(f)}}, \tau\}. \\
%  \hspace{.7cm}= {O\left( {\frac{{K k\ln {{(n)}^3}}}
% {{(1-2\zeta)\Delta_{f}}}} \right)} + K n_3^*.
%       \end{array}
%   \end{displaymath}

%In this theorem, we can view the parameter $\zeta$ as the attacking strength parameter.

%For the \emph{moderately} contaminated stochastic regime, we set the attacking strength parameter $\zeta$ to be at most $1/4$.
%The price that is paid for moderate contamination after $\tau$ steps is the scaling of $\Delta(f)$ by  a
%factor of most $1/2$ ($\zeta = 1/4$) and the introduction of a new factor of $\tau$.

\subsection{Further Discussions on the SPA Scheme}
Besides the study of performance bounds in different regimes, we further discuss the following important issues on the sensing and probing phases.
\subsubsection{Impact of Sensing Time} Usually, the false alarm probability of sensing affects the
performance of SPA scheme, which we do not analyze it yet. Since we consider energy detector for channel sensing,
the false alarm probability is calculated by \cite{CRSes11}
$
{P_{fa}}({t_s}) = Q( {(\frac{{{\epsilon _0}}}{{\sigma _u^2}} - 1)\sqrt {{t_s}{f_s}} } )
$, where $\frac{{{\epsilon _0}}}{{\sigma _u^2}} $ is the decision threshold for sensing, $f_s$ is  the channel bandwidth,
and $Q()$ is the $Q$-function for the tail probability of the standard normal distribution. Consider the
false alarm probability, we have

\textbf{Corollary 15.} The expected EE of $\epsilon$-SPA scheme of the AOEECC-EXP3++ Algorithm is at least
 \begin{IEEEeqnarray*}{l}
 \frac{{{{\hat G}_n}(\epsilon ){t_a}}}{T} \! \ge \!
 \frac{{{G_{\max }} \!-\! 4k\sqrt {\frac{{(1 + \alpha {\epsilon _0}){t_{sp}}K\ln K}}{{(1 - {P_{fa}})T}}} }}{{\frac{1}{{\alpha \epsilon }} \!+ \!1}}.
 \end{IEEEeqnarray*}

 The proof is similar to that of Theorem 2. For each round, each channel is sensed and probed successfully with probability $1-P_{fa}$ in
 expectation. Replacing $T$ with $(1 - {P_{fa}})T$ we get the above result. Here $P_{fa}$ is a function of $t_s$ and $\alpha = \frac{t_a}{
 t_s + t_p}$. Treating $t_s$ as a variable, we can compute the optimal $t_s$ which maximizes the expected throughput by numerical analysis. By similar
 argument, we can have similar counterpart corollaries related to Theorem 4, Theorem 6, Theorem 8,  Theorem 10, Theorem 12 and Theorem 14. We omit
 here for brevity.

\subsubsection{Impact of Probing Time and Others} As pointed out, the step of probing is not necessary in our problem. The reason
why we need it is  that we want to make sure the EE to be high enough, which is important when the channel qualities
are very bad. On the contrary, when channel qualities are good enough, a sensing/probing scheme may achieve better EE since
there is no probing overhead. Our $\epsilon$-SPA scheme also can be extended to a simplified $\epsilon$-SA scheme without
probing steps, in which we only get the observation on the transmission rate after each successful transmission to calculate the
EE. Thus, in $n$ round, $\epsilon n$ expected data rates will be observed by $\epsilon$-SA scheme. Hence, we can show
that the expected throughput of  $\epsilon$-SA scheme is $\frac{{{G_{\max }} \!-\! 4k\sqrt {\frac{{(1 + \alpha' {\epsilon _0}){t_{s}}K\ln K}}{{
\epsilon T}}} }}{{\frac{1}{{\alpha' \epsilon }} \!+ \!1}}$ where $\alpha' = \frac{t_a}{t_s}$. Let $t_p^*$ be the probing time
which satisfies $\frac{{{G_{\max }} \!-\! 4k\sqrt {\frac{{(1 + \alpha' {\epsilon _0}){t_{s}}K\ln K}}{{
\epsilon T}}} }}{{\frac{1}{{\alpha' \epsilon }} \!+ \!1}} = \frac{{{G_{\max }} \!-\! 4k\sqrt {\frac{{(1 + \alpha {\epsilon _0}){(t_{s}+t_{p}^*)}K\ln K}}{{
\epsilon T}}} }}{{\frac{1}{{\alpha \epsilon }} \!+ \!1}} $. When $t_p \le t_p^*$, we will use $\epsilon$-SPA, otherwise, we use $\epsilon$-SA.

In addition, the knowledge of $p_I$ can be used to optimize $t_a$ to maximum the expected EE, if SUs possess this statistical
information.
%\section{Performance Results in Different Regimes}

\section{Cooperative Learning among Multiple SUs}
%We have studied the basic routing problem for a single source-destination pair with single-path routing. In this section, we
%focus on more general scenarios that concerns with multi-strategies probing, cooperative learning between multiple source-destination
%pairs.  We first focus on the multi-strategies routing case, where probing the potential path and its selection can be decoupled.

\begin{figure}
%\vspace{-.3cm}
\centering%.6
\includegraphics[scale=.47]{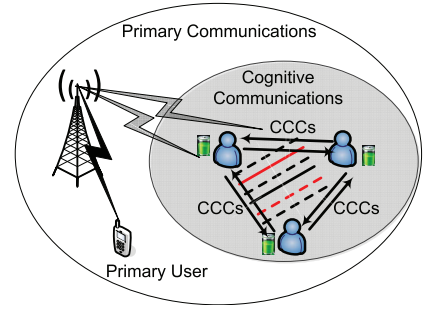}
\caption{Cooperative Bandit Learning among Multiple SUs}
\label{fig:digraph}
\vspace{-.5cm}
\end{figure}

The focus on the previous sections are from a single SU's perspective, where the proposed AOEECC-EXP3++ is an
 uncoordinated algorithm without cooperation with other SUs. It is well known that
exploiting the cooperative behaviors among multiple SUs, such as cooperative spectrum sensing \cite{CRSes11} and spectrum sharing \cite{CRAHN09},
are  effective approaches to improve the communication performance and EE of SUs. This section focuses on the accelerated learning by cooperative learning among multiple SUs.  As we have noticed,  considering information sharing among multi-users in  MAB setting is recently a novel research
direction, and we have seen initial results for stochastic MAB in  \cite{DistInfo15}. Thus, our work can be regarded as the first one
 for both adversarial and stochastic MABs. %All important proofs are put in Section VIII.

The cooperative learning can use Common Control Channels (CCCs) \cite{CRAHN09} to sharing information as illustrated in Fig. 2.
Intuitively, when multiple SUs sening/probing multiple strategies simultaneously and exchange this information
 among them, this would offer more information for decision making,  which results in faster learning and smaller regret value. At each timeslot $n$,
 suppose there are $L_n$ SUs cooperatively perform the $\epsilon$-SPA who wish to explore a total of $L_n$ strategeis over the
 total of the $N$ ($1 \le {L_n} \le N$) and picks a subsect ${\mathcal{O}_n} \subseteq \left\{ {1,...,N} \right\}$ of $L_n$ strategies to probe and
observe the channel losses and power strategies to get the EE. Note that the channels losses that belong to the un-sensed and un-probed set of
strategies $\mathcal{P} \setminus \mathcal{O}_n$ are still unrevealed. Accordingly, we have the probed and
observed set of channels $\tilde{\mathcal{O}}_n$ with the simple property $ f \in \tilde{\mathcal{O}}_n,  \forall f \in i \in \mathcal{O}_n$.   The proposed algorithm 2 based on Algorithm 1  that considers fully information sharing among $L_n$ SUs such that for a single SU $u$.
The probability
$\bm \varrho_{n,u}= \bm  \varrho_n=(\varrho_n(\mathbf{1}),...,
\varrho_n(\mathbf{N}))$ of each observed strategy is
 \begin{IEEEeqnarray*}{l}
\!\!{\varrho _n}(i) =
{{ p_n (i)}} +  \left( {1 - {{p_n (i)}}} \right)\frac{{{M_n} - 1}}{{{N} - 1}},
  \ if \ i \ \in
 \mathcal{O}_n,
\IEEEyesnumber \label{eq:Lim1}
 \end{IEEEeqnarray*}
 where a mixture of  the new exploration probability $(L_n-1)/(N-1)$ is introduced and $p_n (i)$ is defined in (\ref{eq:Aleq1}).
  Similarly,
the channel probability $ { \tilde\varrho}_n=({ \tilde\varrho}_n({1})...,{ \tilde\varrho}_n({n}))$ is computed as
 \begin{IEEEeqnarray*}{l}
\!\!{\tilde \varrho _n}(f) =
{\rho _n}(f) +  \left( {1 - {\rho _n}(f)} \right)\frac{{{m_n} - 1}}{{{K} - 1}},
  \ if \ f \ \in
  \mathcal{\tilde O}_n.
\IEEEyesnumber \label{eq:Lim2}
 \end{IEEEeqnarray*}
Here, we have a channel-level new mixing exploration probability
$(m_n-1)/(n-1)$ and ${\rho _n}(f)$ is defined in (\ref{eq:Aleq2}). The probing rate $m_n$ denotes the
number of simultaneous sensed/probed/accessed non-overlapping channels among all SUs at timeslot $n$. Assume the  weights of channels measured by different probes of different  SUs within
the same time slot also satisfy the assumption in Section II-A.   The design of (\ref{eq:Lim1}) and (\ref{eq:Lim2}) is well  thought out and the proof of all results
in this section are non-trivial tasks in our unified framework.

% Journal Version
%In the weighted major type of algorithm, our problem
%can be regarded as the prediction with \emph{limited} expert advice, where the

%\subsection{Cooperative AOEECC-EXP3++  among Multiple SUs}
\begin{algorithm}
\caption{Cooperative AOEECC-EXP3++:  Multi-user Multi-channel $\epsilon$-SPA scheme }
\begin{algorithmic}
\STATE \textbf{Input}: $M_1, M_2,...,$,  such that  $M_n \in \mathcal{P}$. Set $\beta_n, {\varepsilon _n}\left( f \right), {\xi _n}\left( f \right)$
at in Alg. 1.
  $\forall i \in \mathcal{P}, \tilde{L}_0(i)= 0$ and $\forall f \in E, \tilde{\ell}_0(f)= 0$.

\FOR { timeslot $n=1,2,...$}
\STATE 1:  Choose one SPA strategy $H_n$ according to $\rho_n$ (\ref{eq:Aleq1}). Get advice $\pi_n^{H_n}$ as the selected strategy.
Sample $M_n-1$ additional strategies uniformly over $N$.  Denote the set of sampled strategies by $\mathcal{O}_n$, where $H_n \in \mathcal{O}_n$
 and $|\mathcal{O}_n| = M_n$. Let $\mathds{1}_n^h= \mathds{1}_{\{h \in \mathcal{O}_n\}}$.

 \STATE 2: Update the  probabilities $\varrho_n(i)$ according to (\ref{eq:Lim1}). The  loss of the observed strategy is
  \begin{IEEEeqnarray*}{l}
{\tilde{\psi}_n}(i)  = \frac{{{{ \ell}_n}(i)+ \lambda_{n-1} {\textsf{P}}_{n-1}(i)}}{\varrho_n(i) } \mathds{1}_n^h,
 \forall i \in {\mathcal{O}_n}.
\IEEEyesnumber
 \end{IEEEeqnarray*}

% {\tilde{\psi}_n}(f) = \left\{ \begin{array}{l}
%{\mathbf{\tilde{\ell}}}_{n-1}(f) + \lambda_{n-1} \tilde{\textsf{P}}_{n-1}(f), \quad  \text{\emph{if}} \ f \in {I_n}

\STATE 3: Compute the probability of choosing each channel ${{\tilde \rho}_n}(f)$ that belongs to $i$ according to (\ref{eq:Aleq2}).

\STATE 4: let $\mathds{1}(f)_n = \mathds{1}(f)_{f \in h \in \mathcal{O}_n}$. Update the channel  probabilities $\tilde \varrho_n(f)$
 according to (\ref{eq:Lim2}). The  loss of the observed  channel is
   \begin{IEEEeqnarray*}{l}
{\tilde{\psi}_n}(f) = \frac{{{{ \ell}_n}(f)+ \lambda_{n-1} {\textsf{P}}_{n-1}(f)}}{{\tilde \varrho _n}(f) } \mathds{1}(f)_n,
\forall {f} \in {\tilde{\mathcal{O}}_n}.
\IEEEyesnumber
 \end{IEEEeqnarray*}

\STATE 5:  Updates all weights ${w_n}\left( f \right), {{\bar w}_n}\left( i \right), W_n$, Lagrangian multiplier and
channel access probability $\epsilon$ as in Algorithm 1.
\ENDFOR
\end{algorithmic}
\end{algorithm}
%\vspace{-.2cm}

\subsection{The Performance Results of $\epsilon$-SPA with $\epsilon =1$}
If $m_n$ is a constant or lower
 bounded by $m$, we have the following results.
 Define ${{{\hat G}_n}}(\epsilon)$ as the expected average EEs that can be achieved by the $\epsilon$-SPA scheme over $n$ rounds. The
Theorem 16, Theorem 17, Theorem 18, Theorem 19, Theorem 20 and Theorem 21 bound
the regret of EE, $\mathop {\max }\nolimits_{i \in {S_r}} \mathbb{E}_{i \sim {\textbf{p}_n}}\{ {{G_{n,i}}} \} - \mathbb{E}[ {{{\hat G}_n}(1)} ]$
when set $\epsilon =1$. From these results, we see a rate of $m$ in accelerating of learning performance.
 %The result under accelerated learning is give as:
%We first show that tuning $\eta_n$ is sufficient to control the regret of  AOEECC-EXP3++ in the adversarial regime, which is a
%general result that holds for all other regimes.
%$\hspace{-0.852cm}\cdot  $

 \textbf{Theorem 16.}  Under the \emph{oblivious} attack with same setting of Theorem 1,
 the regret of the  AOEECC-EXP3++ algorithm in the cooperative learning with probing rate $m$ satisfies
\begin{IEEEeqnarray*}{l}
 \hspace{-2.43em}\begin{array}{l}
    \textup{Regret}_n={R(n)}_{\textbf{p}_n^\textup{T}\textsf{P}}  \mathop  \le O(4\sqrt {n\frac{K}{m}\ln K}) \\
  \textup{Violation}_n=   \mathbb{E}{[ {\sum\nolimits_{t = 1}^n {\textbf{p}_{t}^\textup{T}\tilde{\textsf{P}}_{t}-
 \textsf{P}_o} } ]_ + }  \le O(n^{\frac{3}{4}}).
\end{array}\IEEEyesnumber \label{eq:AL1}
\end{IEEEeqnarray*}

%\begin{IEEEproof}See Appendix A. \end{IEEEproof}

\textbf{Theorem 17.}  Under the \emph{$\theta$-memory-bounded adaptive} attack
with same setting of Theorem 3, the regret of the  AOEECC-EXP3++ algorithm  in the cooperative learning with probing rate $m$ satisfies
\begin{IEEEeqnarray*}{l}
\begin{array}{l}
 \textup{Regret}_n={R(n)}_{\textbf{p}_n^\textup{T}\textsf{P}} \le  O((\theta + 1){(4{k}\sqrt {\frac{K}{m}\ln K} )^{\frac{2}{3}}}{n^{\frac{2}{3}}}). \\
  \textup{Violation}_n=   \mathbb{E}{[ {\sum\nolimits_{t = 1}^n {\textbf{p}_{t}^\textup{T}\tilde{\textsf{P}}_{t}-
 \textsf{P}_o} } ]_ + } \le O(n^{\frac{3}{4}}).
 \end{array}
\end{IEEEeqnarray*}

%\label{eq:EstD}
%\subsubsection{Stochastic Regime}
Considering the practical implementation in the stochastic regime by estimating the gap as
 (\ref{eq:EstD}), then we have

\textbf{Theorem 18.} With all other parameters hold as in Theorem 4, the regret of the  AOEECC-EXP3++ algorithm
with ${\xi _n}(f) = \frac{{c{{\left( {\ln n} \right)}^2}}}{{mt{{\hat \Delta }_{n-1}}{{(f)}^2}}}$ and
 $\delta_n  = 2\frac{k}{m}\sqrt {\frac{{K\ln K}}{n}} $ in the cooperative learning with probing rate $m$,
 in the stochastic regime satisfies
    \begin{IEEEeqnarray*}{l}
 \hspace{-2.43em}\begin{array}{l}
    \textup{Regret}_n={R(n)}_{\textbf{p}_n^\textup{T}\textsf{P}}  \mathop  \le O(\frac{kK}{m}\frac{{c\ln {{(n)}^3}}}{{\Delta_f}}) \\
  \textup{Violation}_n=   \mathbb{E}{[ {\sum\nolimits_{t = 1}^n {\textbf{p}_{t}^\textup{T}\tilde{\textsf{P}}_{t}-
 \textsf{P}_o} } ]_ + }  \le O(n^{\frac{3}{4}}).
\end{array}\IEEEyesnumber \label{eq:Al3}
\end{IEEEeqnarray*}
%
%From the theorem, we see in this more practical case,  another factor of $ln(n)$ worse of the regret performance when compared to the idealistic
%case. Also, the additive constants $n^*$ in this theorem can be  very large. However, our experimental results show that
%a minor modification of this algorithm performs comparably to ComUCB1 \cite{Branislav2015} in the stochastic regime.

%\subsubsection{Mixed Adversarial and  Stochastic Regime}
%The mixed adversarial and stochastic regime can be regarded as a special case of mixing
%adversarial and stochastic regimes. Since there is  always a jammer randomly attacking $k_a$ channels
%constantly over time,  we will have the following theorem for the  AOEECC-EXP3++$^\emph{AVG}$ algorithm, which is
%a much more refined regret performance bound than the general regret bound in the adversarial regime.
 %$\\ $
\textbf{Theorem 19.} With all other parameters hold as in Theorem 9, the regret of the  AOEECC-EXP3++ algorithm
with ${\xi _n}(f) = \frac{{c{{\left( {\ln n} \right)}^2}}}{{mt{{\hat \Delta }_{n-1}}{{(f)}^2}}}$  and
 $\delta_n  = 2\frac{k}{m}\sqrt {\frac{{K\ln K}}{n}} $
 under \emph{oblivious jamming} attack in the cooperative learning with probing rate $m$, in the mixed stochastic and adversarial regime
 satisfies
 \begin{IEEEeqnarray*}{l}
  \begin{array}{l}
    \textup{Regret}_n={R(n)}_{\textbf{p}_n^\textup{T}\textsf{P}}  \mathop  \le O(\frac{K(k-k_j)}{m}\frac{{c\ln {{(n)}^3}}}{{\Delta_f}})
     \\
     \hspace{3.2cm}+ O(\frac{4k_j}{m}\sqrt {tK\ln K}) \\
  \textup{Violation}_n=   \mathbb{E}{[ {\sum\nolimits_{t = 1}^n {\textbf{p}_{t}^\textup{T}\tilde{\textsf{P}}_{t}-
 \textsf{P}_o} } ]_ + }  \le O(n^{\frac{3}{4}}).
\end{array}\IEEEyesnumber \label{eq:Al4}
\end{IEEEeqnarray*}
%Note that the results in Theorem 5 has better regret performance than the results obtained by adversarial MAB as shown
% in Theorem 1 and the anti-jamming algorithm in \cite{QianJSAC12}.

\textbf{Theorem 20.} With all other parameters hold as in Theorem 11, the regret of the  AOEECC-EXP3++ algorithm
with ${\xi _n}(f) = \frac{{c{{\left( {\ln n} \right)}^2}}}{{mt{{\hat \Delta }_{n-1}}{{(f)}^2}}}$  and
 $\delta_n  = 2\frac{k}{m}\sqrt {\frac{{K\ln K}}{n}} $ in the cooperative learning with probing rate $m$, in the mixed stochastic and adversarial regime satisfies
 \begin{IEEEeqnarray*}{l}
  \begin{array}{l}
    \textup{Regret}_n={R(n)}_{\textbf{p}_n^\textup{T}\textsf{P}}  \mathop  \le O(\frac{K(k-k_j)}{m}\frac{{c\ln {{(n)}^3}}}{{\Delta_f }}) \\
       \hspace{2.2cm}+ O((\theta + 1)\frac{k_j}{m}{(4{k}\sqrt {K\ln K} )^{\frac{2}{3}}}{n^{\frac{2}{3}}}) \\
  \textup{Violation}_n=   \mathbb{E}{[ {\sum\nolimits_{t = 1}^n {\textbf{p}_{t}^\textup{T}\tilde{\textsf{P}}_{t}-
 \textsf{P}_o} } ]_ + }  \le O(n^{\frac{3}{4}}).
\end{array}\IEEEyesnumber \label{eq:Al5}
\end{IEEEeqnarray*}

\textbf{Theorem 21.}  With all other parameters hold as in Theorem 13, the regret of the \rm{ AOEECC-EXP3++} algorithm in the cooperative learning with probing rate $m$ in the
contaminated stochastic regime satisfies
 \begin{IEEEeqnarray*}{l}
 \hspace{-2.43em}\begin{array}{l}
    \textup{Regret}_n={R(n)}_{\textbf{p}_n^\textup{T}\textsf{P}}  \mathop  \le {O( {\frac{{K k\ln {{(n)}^3}}}
 {{m(1-2\zeta)\Delta_{f}}}} )} + K n_3^* \\
  \textup{Violation}_n=   \mathbb{E}{[ {\sum\nolimits_{t = 1}^n {\textbf{p}_{t}^\textup{T}\tilde{\textsf{P}}_{t}-
 \textsf{P}_o} } ]_ + }  \le O(n^{\frac{3}{4}}).
\end{array}\IEEEyesnumber \label{eq:CstReg5}
\end{IEEEeqnarray*}

\subsection{$\epsilon$-SPA and Cooperative Sensing/Probing Issues}
Due to space limit, we do not plan to present the least EE performance guarantees in the
cooperative learning scenarios for $\epsilon$-SPA with general $\epsilon$. Obviously, it is simply a division of factor $m$ in the
regret bound parts as in the related Theorem 2, Theorem 4, Theorem 6, Theorem 8, Theorem 10, Theorem 12 and Theorem 14.

 In addition, cooperative sensing are necessary to be adopted, where the cooperative sensing gain will improve the sensing/probing performance. The formula
 by  considering  energy detector with cooperative sensing can be found in many existing works, such as in \cite{CRSes11}. Hence, the analysis of cooperative  $\epsilon$-SPA-scheme of the AOEECC-EXP3++ on the issues of  1) Impact of sensing time and 2) Impact of probing time and others can follow the same line as in Section V.E.
\subsection{Distributed Protocols with Multiple Users}
While the cooperative learning scheme offers the optimal learning performance, the design of decentralized protocol without
using a CCC is challenging issue \cite{DistInfo15}. As presented in Section III-Section V, the energy detector cannot differentiate
spectrum usage of PUs and SUs, the opinion of channel qualities of each SU is also affected by other SUs. Notice that
the $\epsilon$-SPA scheme of the AOEECC-EXP3++ developed previously are applicable for all SUs, where their observation of
the best multi-channel access strategies are the same, if each of these channels has the same mean across the players.

For fully distributed solutions, we already have proposed solution as shown in  Section III-Section V to let each SU run the $\epsilon$-SPA scheme
based on their own observation. However, this will come to the situation that all the users will access the same set of channels
that result in low efficiency. This could be resolved by applying an approach similar to the TDFS scheme \cite{ZhaoTSP10} by
introducing round-robin schemes among SUs. We leave the details on this point in future works.
%It is challenging to design optimal decentralized protocol without using a common control channel (CCC). Since the energy-detection cannot differentiate spectrum usage of PUs and SUs, the view of each SU is also affected by other SUs. The channels quality and availability are thus dynamic
%and nonstochastic. Notice that the protocol ?-SPA developed
%previously applies to a more general observation model as long
%as SUs have the common set of k best channels and each of
%these channels has the same mean across players.  This
%could be proved by applying an approach similar to the TDFS
%scheme [16].

\section{Proofs of Regrets in Different Regimes}
 We prove the theorems of the performance results from the previous section in the order they were presented.
%( \textsf{P}_o- \textbf{q}_n^\textup{T}\tilde{\textsf{P}}_{n-1})

\textbf{Lemma 1.} (Dual Inequality) Let ${f_t}(\lambda ) = \frac{\delta_{t} }{2}{\lambda ^2} + \lambda (
\textbf{p}_{t-1}^\textup{T}\tilde{\textsf{P}}_{t-1}-\textsf{P}_o), {\lambda _t} = {[({\lambda _{t - 1}} - \eta_{t-1}
 \sqrt{\gamma_{t-1}}  \nabla {f_{t - 1}}({\lambda _{t - 1}}))]_ +}$ and
$\lambda_1=0$. Assuming $\eta_n > 0$, we have
  \begin{IEEEeqnarray*}{l}
 \hspace{-.6cm}\sum\limits_{t = 1}^n \!{\left( {\lambda _t  \! - \!{\lambda  }} \right)} (\textsf{P}_o- \textbf{p}_{t}^\textup{T}\tilde{\textsf{P}}_{t})\! + \!\!\frac{\delta_n }{2}\sum\limits_{t = 1}^n\! {\left( {\lambda _t^2 \!-\! {\lambda ^2}} \right)}\!   \le \! \frac{1}{{{\eta \sqrt{\gamma_{n}}  }}}{{\lambda} ^2}\!  \\
 \hspace{3.16cm} + \sum\limits_{t = 1}^n  \sqrt{\gamma_{t}} {\eta_t} + \sum\limits_{t = 1}^n \frac{{{\eta _{t}}}}{2} {(\textbf{p}_{t }^\textup{T}\tilde{\textsf{P}}_{t })^2}\!. \IEEEyesnumber \label{eq:Dual1}
\end{IEEEeqnarray*}

  \begin{IEEEproof}
  First, we note that
  \begin{IEEEeqnarray*}{l}
\hspace{-.2cm}  \begin{array}{l}
{\lambda _t} = {[({\lambda _{t - 1}} - {\eta _{t - 1}}\sqrt{\gamma_{t-1}}\nabla {f_{t - 1}}({\lambda _{t - 1}}))]_ + }\\
 \quad \ = {[(1 - \delta_{t - 1} {\eta _{t - 1}\sqrt{\gamma_{t-1}}}){\lambda _{t - 1}} \! + \! {\eta _{t - 1}\sqrt{\gamma_{t-1}}}(\textsf{P}_o \!\!-  \!\! \textbf{p}_{t-1}^\textup{T}\tilde{\textsf{P}}_{t-1})]_ + }\\
 \quad  \ \le {[(1 - \delta_{t - 1} {\eta _{t - 1}\sqrt{\gamma_{t-1}}}){\lambda _{t - 1}} \! + \! {\eta _{t - 1}\sqrt{\gamma_{t-1}}}{\textsf{P}_0}]_ + }
\end{array}
\end{IEEEeqnarray*}
By induction on $\lambda_t$, we can obtain ${\lambda _t} \le \frac{{{\textsf{P}_0}}}{\delta_t }$. Applying the standard analysis of online gradient descent \cite{Zinkevich03} yields
  \begin{IEEEeqnarray*}{l}
\begin{array}{l}%\!\!\!\!\!\!\!\!\!
\hspace{-.2cm}{| {{\lambda _t} \!-\! \lambda } |^2}\!  \\
 ={| {{\Pi _ + }[({\lambda _{t - 1}}\! - \!{\eta _{t - 1}}\sqrt{\gamma_{t-1}}(\delta_{t - 1} {\lambda _{t - 1}} + \textsf{P}_o
\! - \! \textbf{p}_{t-1}^\textup{T}\tilde{\textsf{P}}_{t-1}))]\! -\! \lambda } |^2}\\
  \le {| {{\lambda _{t - 1}} - \lambda } |^2} + {\eta _{t - 1}}^2\gamma_{t-1}{| {\nabla {f_{t - 1}}({\lambda _{t - 1}})} |^2} \\
  \quad - 2({\lambda _{t - 1}} - \lambda )( {{\eta _{t - 1}}\sqrt{\gamma_{t-1}}\nabla {f_{t - 1}}({\lambda _{t - 1}})} )\\
   \le {| {{\lambda _{t - 1}} - \lambda } |^2} + {\eta _{t - 1}}^2\gamma_{t-1}{| {\nabla {f_{t - 1}}({\lambda _{t - 1}})} |^2} \\
   \quad + 2{\eta _{t - 1}}\sqrt{\gamma_{t-1}}( {  {f_{t - 1}}(\lambda ) -   {f_{t - 1}}({\lambda _{t - 1}})} )
\end{array}
\end{IEEEeqnarray*}
Then, rearrange terms, we get,
  \begin{IEEEeqnarray*}{l}
\begin{array}{l}
\!\!\!  {f_{t - 1}}({\lambda _{t - 1}})  \! - \!    {f_{t - 1}}(\lambda ) \le \frac{1}{{2{\eta _{t - 1}}\sqrt{\gamma_{t-1}}}}({\left| {\lambda \!   - \!   {\lambda _{t - 1}}} \right|^2} \!  -  \! {\left| {\lambda \!  -  \! {\lambda _{t - 1}}} \right|^2}) \\
\quad\quad\quad\quad\quad\quad  \quad\quad\quad + \frac{{\sqrt{\gamma_{t-1}}{\eta _{t - 1}}}}{2}{\left| {\nabla {f_{t - 1}}({\lambda _{t - 1}})} \right|^2}
\\
  \hspace{3cm}\le \frac{1}{{2\sqrt{\gamma_{t-1}}{\eta _{t - 1}}}}({\left| {\lambda  \!- \!{\lambda _{t - 1}}} \right|^2}\!
  - \!{\left| {\lambda  \!  - \!  {\lambda _{t - 1}}} \right|^2}) \\
\quad\quad\quad\quad\quad\quad  \quad\quad\quad + \frac{{{\sqrt{\gamma_{t-1}} \eta _{t - 1}}}}{2}{(\textbf{p}_{t-1}^\textup{T}\tilde{\textsf{P}}_{t-1})^2}  + {\eta _{t - 1}}.
\end{array}
\end{IEEEeqnarray*}
Note that $\eta _{t}$ varies with $t$. For the first term leading factor
in the r.h.s of the above inequality, use the trick by letting  $\eta_n=\eta _{t}$ as indicated in  \cite{Bubeck12} (page 25), e.g.
if $\eta _{t}= \frac{1}{{\sqrt t }}$, we have $\sum\nolimits_{t = 1}^n {\frac{1}{{\sqrt t }}}  \le \smallint _0^n\frac{1}{{\sqrt t }}dt = 2\sqrt n$,
a factor of $2$ gap between $\eta_n$ and $\eta _{t}$.  Thus,
substitute $\eta _{t}$ by $\eta_n/2$ and  expanding the terms on l.h.s and taking the sum over $t$, we obtain the inequality.
  \end{IEEEproof}

  \subsection{The Adversarial Regime}
 The proof of Theorem 1 borrows some of  the analysis of EXP3 of the loss model in  \cite{Bubeck12}. However, the introduction
 of the new mixing  exploration parameter and the truth of channel dependency as a special type of combinatorial MAB
 problem in the loss model  makes the proof a non-trivial task, and we prove it for the first time.

 \emph{Proof of Theorem 1.}
  \begin{IEEEproof}
Note first that the following equalities can be easily verified:
${\mathbb{E}_{i \sim {\textbf{p}_n}}}{\tilde{\ell}_n}(i) = {\ell_n}({I_n}),{\mathbb{E}_{{\tilde{\ell}_n} \sim {\textbf{p}_n}}}{\ell_n}(i) = {\ell_n}(i),{\mathbb{E}_{i \sim {\textbf{p}_n}}}{\tilde{\ell}_n}{(i)^2} = \frac{{{\ell_n}{{({I_n})}^2}}}{{{p_n}({I_n})}}$ and ${\mathbb{E}_{{I_n} \sim {\textbf{p}_n}}}\frac{1}{{{p_n}({I_n})}} = N$.

%= \mathbb{E}_n \left[ \sum\limits_{s = 1}^n {{\ell_s}({I_s})}  - \sum\limits_{s = 1}^n {{\ell_s}(i)} \right] \\
%\hspace{.7cm}
Let $   \tilde{ \Phi}_n(i) = {{\tilde { \ell}}_{n}}(i) +  \lambda_{n}{{  \tilde{ \textsf{P}}}_{n}}(i)$.
The regret with respect to $ \tilde{ \Phi}_n(i)$ , $R_{\tilde{ \Phi}_n(i)}(n)$, is
\begin{IEEEeqnarray*}{l}
R_{\tilde{ \Phi}_n(i)}(n)  =\mathbb{E}_n \left[\sum\limits_{t = 1}^n {{\mathbb{E}_{i \sim {\textbf{p}_t}}}\tilde{ \Phi}_n(i)}  - \sum\limits_{t = 1}^n {{\mathbb{E}_{{I_t} \sim {\textbf{p}_t}}}\tilde{ \Phi}_n(i)} \right].
\end{IEEEeqnarray*}

The key step here is to consider the expectation of the cumulative losses ${\tilde{\ell}_n}(i)$ in the sense of distribution
$i \sim {\textbf{p}_n}$. For all strategies, we have the distribution vector
$\textbf{q}_n = (q_n(1),...,q_n(N))$ with ${q _n}(i) = \frac{{{w_{n - 1}}\left( i \right)}}{{{W_{n - 1}}}}$
 and  for all the channels, we have vector $\textbf{q}_{n,f} = (q_{n,f} (1),...,q_{n,f} (K))$ with
${q_{n,f}}(f') = \frac{{\sum\nolimits_{i:f \in i} {{w_{n - 1}}\left( i \right)} }}{{{W_{n - 1}}}}$. Let ${\varepsilon_n}(i)=\sum\nolimits_{f \in i} {{\varepsilon _n}(f)}$.  However, because of the mixing terms of $p_n$, we need to introduce a few more notations. Let $\textbf{u }=
( {\underbrace {\sum\nolimits_{f \in 1}
{{\varepsilon _n}(f)},...,\sum\nolimits_{f \in i}
{{\varepsilon _n}(f)},...,\sum\nolimits_{f \in |\mathcal{C}|}
 {{\varepsilon _n}(f)}}_{i \in \mathcal{C}},\underbrace {0,...,0}_{i
\notin \mathcal{C}}} )$ be the distribution over all the strategies. Let ${\textbf{q}_n} = \frac{{{\textbf{p}_n} - \textbf{u}}}{{1 - \sum\nolimits_{f} {{\varepsilon _n}(f)} }}$
 be the distribution induced by AOEECC-EXP3++ at the time $n$ without mixing. Then we have:
\begin{IEEEeqnarray*}{l}
\!\!\begin{array}{l}
{\mathbb{E}_{i \sim {\textbf{p}_t}}}{\tilde{ \Phi}_t}(i) = ( {1 - \sum\nolimits_{f} {{\varepsilon _t}(f)} } ){\mathbb{E}_{i \sim {\textbf{q}_t}}}{\tilde{ \Phi}_t}(i) + {\varepsilon _t}(i){\mathbb{E}_{i \sim \textbf{u}}}{\tilde{ \Phi}_t}(i)\\
\quad\quad \quad \quad \ \  = ( {1 - \sum\nolimits_{f} {{\varepsilon _t}(f)} } )(\frac{1}{{{\eta_t}}}\ln {\mathbb{E}_{i \sim {\textbf{q}_t}}}\exp ( - {\eta_t}({\tilde{ \Phi}_t}(i) \\
\quad\quad \quad \quad \quad \ - {\mathbb{E}_{j \sim {\textbf{q}_t}}} {\tilde{ \Phi}_t}(j))))\\
 \quad\quad \quad \quad \quad \ - \frac{( {1 - \sum\nolimits_{f} {{\varepsilon _t}(f)} } )}{{{\eta_t}}}\ln {\mathbb{E}_{i \sim {\textbf{q}_t}}}\exp ( - {\eta_t}{\tilde{ \Phi}_t}(i))) \\
\quad\quad \quad \quad \quad \  + {\mathbb{E}_{i \sim \textbf{u}}}{\tilde{ \Phi}_t}(i).
\end{array}\IEEEyesnumber \label{eq:AppenA}
\end{IEEEeqnarray*}

In the second step, we use the inequalities $lnx \le x-1$ and $exp(-x) -1 +x \le x^2/2$, for all $x \ge 0$, to obtain:
\begin{IEEEeqnarray*}{l}
\begin{array}{l}
\ln {\mathbb{E}_{i \sim {\textbf{q}_t}}}\exp ( - {\eta_t}({\tilde{ \Phi}_t}(i) - {\mathbb{E}_{j \sim {\textbf{q}_t}}}{\tilde{ \Phi}_t}(j)))\\
\quad\quad \quad = \ln {\mathbb{E}_{i \sim {\textbf{q}_t}}}\exp ( - {\eta_t}{\tilde{ \Phi}_t}(i)) + {\eta_t}{\mathbb{E}_{j \sim {\textbf{q}_t}}}{\tilde{ \Phi}_t}(j)\\
\quad\quad \quad \le {\mathbb{E}_{i \sim {\textbf{q}_t}}}( {\exp ( - {\eta_t}{\tilde{ \Phi}_t}(i)) - 1 + {\eta_t}{\tilde{ \Phi}_t}(j)} )\\
\quad\quad \quad \mathop  \le \limits^{(a)}  {\mathbb{E}_{i \sim {\textbf{q}_t}}}\frac{{\eta_t^2{\tilde{ \Phi}_t}{{(i)}^2}}}{2} \mathop  \le \limits^{(b)}  {\eta_t^2 \mathbb{E}_{i \sim {\textbf{q}_t}}} {{{ \tilde  \ell_t}{{(i)}^2}}} + \lambda_t^2 \eta_t^2 {\mathbb{E}_{i \sim {\textbf{q}_t}}} {{ {{  \tilde{ \textsf{P}}}_{n}}(i) }^2} ,
\end{array}\IEEEyesnumber \label{eq:Apart1}
\end{IEEEeqnarray*}
where we used $\frac{{{q_n}(i)}}{{{p_n}(i)}} \le \frac{1}{{1 - \sum\nolimits_{f} {{\varepsilon _t}(f)} }}$ in the above inequality $(a)$ and
the fact ${(x + y)^2} \le 2{x^2} + 2{y^2}$ in the above inequality $(b)$. Moreover, take expectations over all random strategies of losses ${{\tilde \ell}_t}{(i)^2}$, we have
\begin{IEEEeqnarray*}{l}
\begin{array}{l}
%\!\!\!\!\!\!
{\mathbb{E}_n}\left[{\mathbb{E}_{i \sim {\textbf{q}_t}}}{{\tilde \ell}_t}{(i)^2} \right] =
{\mathbb{E}_n}\left[\sum\limits_{i = 1}^N {{{q}_t}(i){{\tilde \ell}_t}{{(i)}^2}} \right]\\
= {\mathbb{E}_n}\!\!\left[\sum\limits_{i = 1}^N {{{q}_t}(i){{(\sum\limits_{f \in i} {{{\tilde \ell}_t}(f)} )}^2}}  \right]
 \! = \! {\mathbb{E}_n}\!\!\left[ \sum\limits_{i = 1}^N {{{q}_t}(i){k}\!\!\sum\limits_{f \in i} {{{\tilde \ell}_t}{{(f)}^2}} }\right]
 \\= \!{\mathbb{E}_n}k \left[\! \sum\limits_{f = 1}^K {{{\tilde \ell}_t}{{(f)}^2}} \!\!\!\!\sum\limits_{i \in {S_r}:f \in i} \!\!\!{{{q}_t}(i)} \right]\! = \!
 {k}{\mathbb{E}_n}\!\! \left[\!\sum\limits_{f' = 1}^K \!{{{\tilde \ell}_t}{{(f')}^2}{q_{t,f}}(f')} \right]\\
 \end{array}
\end{IEEEeqnarray*}
\begin{IEEEeqnarray*}{l}
\begin{array}{l}
 = {k}{\mathbb{E}_n}\left[ {\sum\limits_{f' = 1}^K {{{\left( {\frac{{{\ell_n}(f')}}{{{\rho_n}(f')}}{\mathds{1}_n}(f')} \right)}^2}} {q _{t,f}}(f')} \right]\\
 \le {k}{\mathbb{E}_n}\left[ {\sum\limits_{f' = 1}^K {\frac{{{q _{t,f}}(f')}}{{{\rho_n}{{(f')}^2}}}{\mathds{1}_n}(f')} } \right] =
  {k}\sum\limits_{f' = 1}^K {\frac{{{q _{t,f}}(f')}}{{{\rho_n}(f')}}} \\
 = {k}\sum\limits_{f' = 1}^K {\frac{{{q _{t,f}}(f')}}{{\left( {1 - \sum\nolimits_{f} {{\varepsilon _n}(f)} } \right)
{q _{t,f}}(f')
 + {\sum\nolimits_{f \in i} {{\varepsilon _n}(f)} } \left| {\left\{ {i \in \mathcal{C}:f \in i} \right\}} \right|}}}
 \le 2k K,
\end{array}\IEEEyesnumber \label{eq:Apart1s}
\end{IEEEeqnarray*}
where the last inequality follows the fact that $( {1 - \sum\nolimits_{f} {{\varepsilon _n}(f)} }) \ge \frac{1}{2}$ by the definition of
${{\varepsilon _n}(f)}$. Similarly,
\begin{IEEEeqnarray*}{l}
\begin{array}{l}
%\!\!\!\!\!\!
{\mathbb{E}_n}\left[{\mathbb{E}_{i \sim {\textbf{q}_t}}} {{ {{  \tilde{ \textsf{P}}}_{n}}(i) }^2}  \right]  \le 2k K,
\end{array}\IEEEyesnumber \label{eq:ApartConst2}
\end{IEEEeqnarray*}

In the third step, note that ${{\tilde L}_0}(i) = 0$. Let ${\Upsilon _n}(\eta ) = \frac{1}{\eta }\ln \frac{1}{N}\sum\nolimits_{i = 1}^N {\exp ( - \eta {{\tilde L}_n}(i))}$ and ${\Upsilon _0}(\eta )=0 $. The second term in (\ref{eq:AppenA}) can be bounded by using the same technique in \cite{Bubeck12} (page 26-28).  Let us substitute  inequality (\ref{eq:Apart1s}) into (\ref{eq:Apart1}), and then substitute (\ref{eq:Apart1}) into equation  (\ref{eq:AppenA}) and sum over $n$.
Use the fact that the sum of expectation on $u$ and $\textbf{q}_t$ with respect to ${{{\tilde \Phi}_t}(j)}$ is less than
${\mathbb{E}_{{I_t} \sim {\textbf{p}_t}}} {{{\tilde \Phi}_t}(j)}$. Take expectation over all random strategies of losses up to time $n$, we obtain
\begin{IEEEeqnarray*}{l}
\begin{array}{l}%\!\! \sum\limits_{t = 1}^n  \!{\mathbb{E}_{i \sim u}}{{\tilde \Phi}_t}(i)
\hspace{-.1cm}{\mathbb{E}_n}\left[ \sum\limits_{t = 1}^n {\mathbb{E}_{i \sim {\textbf{p}_t}}}{{\tilde \Phi}_t}(i) \right]
 \le 2k K \!\sum\limits_{t = 1}^n \eta_t +  2 k K \!\sum\limits_{t = 1}^n \eta_t \lambda_t^2 +  \frac{{\ln N}}{\eta_n } \\
\hspace{.3cm} + {
\mathbb{E}_n}\left[\sum\limits_{t = 1}^n {\mathbb{E}_{{I_t} \sim {\textbf{p}_t}}} {{{\tilde \Phi}_t}(j)} \right] + {\mathbb{E}_n}\left[ \sum\limits_{t = 1}^{n - 1} ({{\Upsilon _t}({\eta _{t + 1}}) - {\Upsilon _t}({\eta _t}))})\right]\!.
\end{array}
\end{IEEEeqnarray*}
The last term in the r.h.s of the inequality is less than or equals to zero as indicated in \cite{Bubeck12}.  Then, we get
\begin{IEEEeqnarray*}{l}
R_{\tilde{ \Phi}_n(i)}(n)=  \mathbb{E}_n \left[\sum\limits_{t = 1}^n {{\mathbb{E}_{i \sim {\textbf{p}_t}}}\tilde{ \Phi}_n(i)}  - \sum\limits_{t = 1}^n {{\mathbb{E}_{{I_t} \sim {\textbf{p}_t}}}\tilde{ \Phi}_n(i)} \right] \\
\hspace{1.4cm} \le 2 k K \!\sum\limits_{t = 1}^n \eta_t + \frac{{\ln N}}{\eta_n } +  2 k K \!\sum\limits_{t = 1}^n \eta_t \lambda_t^2 \\
\end{IEEEeqnarray*}
\begin{IEEEeqnarray*}{l}
%\hspace{.7cm}\mathop  \le \limits^{(a)} k K \!\sum\limits_{t = 1}^n \eta_t + \frac{{\ln N}}{\eta_n }
%+ k \sum\limits_{t = 1}^n {\sum\limits_{f = 1}^K {{\varepsilon _t}(f)} } \\
%\hspace{.7cm}\mathop  \le \limits^{(b)} 2k K \!\sum\limits_{t = 1}^n \eta_t + \frac{{\ln N}}{\eta_n } \\
\hspace{.7cm}\mathop  \le \limits^{(a)} 2k K \!\sum\limits_{t = 1}^n \eta_t + k\frac{{\ln K}}{\eta_n } +   2 k K \!\sum\limits_{t = 1}^n \eta_t \lambda_t^2.
%\le k K \!\sum\limits_{t = 1}^n \eta_t + \frac{{\ln N}}{\eta_n }
\IEEEyesnumber \label{eq:CstReg1}
\end{IEEEeqnarray*}
Note that, the
 inequality $(a)$ is due to the fact that $N \le K^{k}$. %Setting $\eta_n=b_n$, we prove the theorem.

Combine  (\ref{eq:Dual1}) and  (\ref{eq:CstReg1}) gives that
\begin{IEEEeqnarray*}{l}
\begin{array}{l}
\mathbb{E}\left[ {\sum\limits_{t = 1}^n {{\textbf{p}^T_t}{\ell_t}(i) - {\textbf{p}^T_t}{\ell_t}({I_t})} } \right] + \\
\hspace{2.4cm} \mathbb{E}\left[ {\sum\limits_{t = 1}^n {\lambda (\textbf{p}_{t}^\textup{T}\tilde{\textsf{P}}_{t}- \textsf{P}_o)
-  \left( {\frac{{\delta_n n}}{2} + \frac{1}{\eta  \sqrt{\gamma_{t}} }} \right){\lambda ^2}} } \right]\\
\mathop  \le \limits^{(a)} 2kK\sum\limits_{t = 1}^n {{\eta _t}}
 + k\frac{{\ln K}}{{{\eta _n}}}+  \sum\limits_{t = 1}^n  {\eta_t \sqrt{\gamma_{t}}  } + \left( {kK\eta_n  - \frac{\delta_n }{2}} \right)\sum\limits_{t = 1}^n {\lambda _t^2}   \\
\hspace{3.3cm}+ \mathbb{E}\left[ {\sum\limits_{t = 1}^n {{\lambda _t}
(\textbf{p}_{t}^\textup{T}\tilde{\textsf{P}}_{t}- \textsf{P}_o)} } \right].
\end{array}
\end{IEEEeqnarray*}
For the above inequality $(a)$, we use the trick by letting $\eta_n = \eta_t$ indicated in [16] (page 25) again to extract the $\eta_t$ from the
sum of $\lambda_t^2$ over $n$ and the inequality in (\ref{eq:ApartConst2}).   Let $ {kK\eta_n \le \frac{\delta_n}{2}}$ by setting properly the values  such that  $\eta_n = O(\delta_n)$ (shown
in the next). Thus, the last two terms in the r.h.s of the above inequality is non-positive.  By taking maximization over $\lambda$,
we have
\begin{IEEEeqnarray*}{l}
\begin{array}{l}
\!\!\mathbb{E}\! \left[ {\mathop {\max }\limits_{\textbf{p}_{t}^\textup{T}\tilde{\textsf{P}}_{t} \le \textsf{P}_o} \sum\limits_{t = 1}^n {{\textbf{p}^T_t}{\ell_t}(i) - {\textbf{p}^T_t}{\ell_t}({I_t})} } \right]
\!  + \!\mathbb{E}\! \left[ {\frac{{\left[ {\sum\limits_{t = 1}^n {(\textbf{p}_{t}^\textup{T}\tilde{\textsf{P}}_{t}-
 \textsf{P}_o)} } \right]_ + ^2}}{{2\left( {\delta n/2 + {1 \mathord{\left/
 {\vphantom {1 \eta }} \right.
 \kern-\nulldelimiterspace} \eta }} \right)}}} \right] \\
 \hspace{3.3cm}
\mathop   \le
 \limits^{(a)} 2kK\sum\limits_{t = 1}^n {{\eta _t}}  + k\frac{{\ln K}}{{{\eta _n}}}+  \sum\limits_{t = 1}^n  {\eta_t \sqrt{\gamma_{t}}  } .
\end{array}\IEEEyesnumber \label{eq:CstReg4}
\end{IEEEeqnarray*}

Note that our algorithm exhibit a bound in the structure like $ \textup{Regret}_n + \frac{{\textup{Violation}_n^2}}{{O({n^{1 - \alpha }})}} \le {n^{1 - \beta }}$. We can derive a regret bound and the violation of the long-term power budget constraint as
\begin{IEEEeqnarray*}{l}
\begin{array}{l}
 \textup{Regret}_n \le O({n^{1 - \beta }})\\
 \textup{Violation}_n \le \sqrt {O(\left[ {n + {n^{1 - \beta }}} \right]{n^{1 - \alpha }})},
\end{array}\IEEEyesnumber \label{eq:M5}
\end{IEEEeqnarray*}
where the last bound follows the naive fact $-\textup{Regret}_n \le  O(n)$. In practice, we this bound is
 coarse, and we can us the accumulated variance to obtain better violation bound.

Then, according to (\ref{eq:M5}),  we obtain
\begin{IEEEeqnarray*}{l}
\begin{array}{l}
\mathbb{E}\left[ {\mathop {\max }\limits_{\textbf{p}_{t}^\textup{T}\tilde{\textsf{P}}_{t} \le
 \textsf{P}_o} \sum\limits_{t = 1}^n {{\textbf{p}^T_t}{\ell_t}(i) - {\textbf{p}^T_t}{\ell_t}({I_t})} } \right] \mathop  \le \limits^{} 4kKn{\eta _n} + k\frac{{\ln K}}{{{\eta _n}}}\\
 \hspace{3.9cm} +  \sum\limits_{t = 1}^n  {\eta_t \sqrt{\gamma_{t}}  }  \mathop  \le \limits^{} 4k\sqrt {nK\ln K} \\
\mathbb{E}{\left[ {\sum\limits_{t = 1}^n {\textbf{p}_{t}^\textup{T}\tilde{\textsf{P}}_{t}-
 \textsf{P}_o} } \right]_ + } \!\le  \! \sqrt {\!\left( {n + 4k\sqrt {nK\ln K} } \right)\left( {\delta_n n  + {2 \mathord{\left/
 {\vphantom {1 \eta }} \right.
 \kern-\nulldelimiterspace} \eta_n }} \right)} .
\end{array}\IEEEyesnumber \label{eq:CstReg5}
\end{IEEEeqnarray*}
Let ${\eta _n} = {\beta _n}=  \sqrt {\frac{{\ln K}}{{4Kn}}}  = O\left( {{n^{ - 1/2}}} \right),{1 \mathord{\left/{\vphantom {1 \eta_n }} \right. \kern-\nulldelimiterspace} \eta } = O\left( {{n^{1/2}}} \right)$.
 Because $\gamma_{n}= \tilde{O}(\frac{ln(n)}{n})$,  the term
 $ \sum\nolimits_{t = 1}^n  {\eta_t \sqrt{\gamma_{t}}  } = \tilde{O}(ln(n))$. Thus, it can be omitted when
 compared to the first two terms in the r.h.s of (\ref{eq:CstReg4}).   Moreover, in this setting, we set
$\delta_n = 2k\sqrt {\frac{{K\ln K}}{n}} $, such that $ {2kK\eta_n \le \frac{\delta_n}{2}}$. Then,
$\delta_n n = O\left( {{n^{1/2}}} \right)$. We proof the theorem. \end{IEEEproof}

 \emph{Proof of Theorem 3.}
  \begin{IEEEproof}
To defend against the $\theta$-memory-bounded adaptive adversary, we need to adopt the idea of the mini-batch protocol proposed in \cite{Arora12}.
We define a new algorithm by wrapping AOEECC-EXP3++ with a mini-batching loop \cite{Dekel11}. We specify a batch
size $\tau$ and name the new algorithm AOEECC-EXP3++$_\tau$. The idea is to group the overall
timeslots $1,...,n$ into consecutive and disjoint mini-batches of size $\tau$. It can be viewed that one signal mini-batch
as a round (timeslot) and use the average loss suffered during that mini-batch to feed the original AOEECC-EXP3++. Note that our
new algorithm does not need to know $m$, which only appears as a constant as shown in Theorem 2. So our new
AOEECC-EXP3++$_\tau$ algorithm still runs in an adaptive way without any prior about the
environment. If we set the batch $\tau= {(4{k}\sqrt {K\ln K} )^{ - \frac{1}{3}}}{n^{^{\frac{1}{3}}}}$ in Theorem 2 of \cite{Arora12},
we can get the regret upper bound in our Theorem 2.
  \end{IEEEproof}

\subsection{The Stochastic Regime}
 Our proofs are based on the following form of Bernstein's inequality with minor improvement as shown in \cite{Seldin14}.

\textbf{Lemma 2.} (Bernstein's inequality for martingales). Let $X_1,...,X_m$ be martingale difference sequence with
respect to filtration $\mathcal{F}=(\mathcal{F}_i)_{1 \le k \le m}$ and let $Y_k = \sum\nolimits_{j = 1}^k {{X_j}}$ be the
associated martingale. Assume that there exist positive numbers $\nu$ and $c$, such that $X_j \le c$ for all $j$ with probability
$1$ and $\sum\nolimits_{k = 1}^m {\mathbb{E}\left[ {{{\left( {{X_k}} \right)}^2}|{\mathcal{F}_{k - 1}}} \right]}  \le \nu$ with probability 1.
% Then for all $b >0$:}
\begin{IEEEeqnarray*}{l}
\mathbb{P}[{Y_m} > \sqrt {2\nu b}  + \frac{{cb}}{3}] \le {e^{ - b}}.
\end{IEEEeqnarray*}

%We also need to use the following technical lemma, where the proof can be found in \cite{Seldin14}.

%\textbf{Lemma 3. } For any $c > 0$, we have $\sum\nolimits_{n = 0}^\infty  {{e^{ - c\sqrt n }}}  = O\left( {\frac{2}{{{c^2}}}} \right)$.
%
%To obtain the tight regret performance for AOEECC-EXP3++, we need to study and estimate the
%number of times each of channel is selected up to time $n$, i.e., $N_n(f)$. We summarize it in the following lemma.

\textbf{Lemma 3. } Let $\left\{ {{{\underline \varepsilon }_n}(f)} \right\}_{n = 1}^\infty$ be non-increasing deterministic sequences, such that
${{\underline \varepsilon }_n}(f) \le {{ \varepsilon }_n}(f)$ with probability $1$ and ${{\underline \varepsilon }_n}(f) \le {{ \varepsilon }_n}(f^*)$
 for all $n$ and $f$. Define $\nu_n(f)=  \sum\nolimits_{t = 1}^n  \frac{1}{{{k \underline \varepsilon  }_t}(f)} $, and define the event $\mathcal{E}^f_n$
\begin{IEEEeqnarray*}{l}
 {n\Delta (f) - ( {{{\tilde L}_n}(f^*) - {{\tilde L}_n}({f})} )} \\
\hspace{1.3cm} \le {\sqrt {2({\nu _n}(f) + {\nu _n}({f^*})){b_n}}  + \frac{{{(1/k+ 0.25)} {b_n}}}{{3k  {\underline \varepsilon _n}({f^*})}}}\hspace{.3cm}
(\mathcal{E}^f_n).
\end{IEEEeqnarray*}
Then for any positive sequence $b_1, b_2,...,$ and any $n^* \ge 2$ the number of times channel $f$ is played by AOEECC-EXP3++ up to
round $n$ is bounded as:
\begin{IEEEeqnarray*}{l}
\begin{array}{l}
\mathbb{E}[{N_n}(f)] \le \left( {{n^*} - 1} \right) + \sum\limits_{t = {n^*}}^n {{e^{ - {b_t}}}}  + k \sum\limits_{t = {n^*}}^n {{\varepsilon _t}(f){\mathds{1}_{\{ E_n^f\} }}} \\
    \hspace{1.3cm}               + \sum\limits_{t = {n^*}}^n {{e^{ - {\eta _t}{h_{t}}(f)}}}e^{ - \eta_t\lambda_{t}({{\tilde{\Gamma}}_{t}} (f^*)\!-\!{{\tilde{\Gamma}}_{t}} (f))},
\end{array}\IEEEyesnumber \label{eq:LemNum}
\end{IEEEeqnarray*}
where
\begin{IEEEeqnarray*}{l}
\begin{array}{l}
{h_n}(f) = n\Delta (f) - \sqrt {2n{b_n}\left( {\frac{1}{{{k}{{\underline \varepsilon  }_n}(f)}} + \frac{1}{{{k}{{\underline \varepsilon  }_n}({f^*})}}} \right)}  - \frac{{(\frac{1}{4} + \frac{1}{k}){b_n}}}{{3{{\underline \varepsilon  }_n}({f^*})}}.
\end{array}
\end{IEEEeqnarray*}
\begin{IEEEproof}
Note that the elements of the martingale difference sequence $\{ {\Delta (f)
- ({{\tilde \ell}_n}(f) - {{\tilde \ell}_n}({f^*}))} \}_{n = 1}^\infty$ by $\max \{ \Delta (f) +{{\tilde \ell}_n}({f^*}) \}= {\frac{1}{{{k}{{\underline \varepsilon  }_n}({f^*})}}} +1$. Since ${{{\underline \varepsilon  }_n}({f^*})} \le {{{ \varepsilon  }_n}({f^*})} \le 1/(2K) \le 1/4$, we can simplify
the upper bound by using ${\frac{1}{{{{k\underline \varepsilon   }_n}({f^*})}}} +1 \le
    \frac{{(\frac{1}{4} + \frac{1}{k})}}{{{{\underline \varepsilon   }_n}({f^*})}}$.

  We further note that
   \begin{displaymath}
\begin{array}{l}
\sum\limits_{t = 1}^n {{\mathbb{E}_t}\left[{{(\Delta (f) - ({{\tilde \ell}_t}(f) - {{\tilde \ell}_t}({f^*})))}^2}\right]} \\
 \hspace{1.1cm}  \le \sum\limits_{t = 1}^n {{\mathbb{E}_t}\left[{{({{\tilde \ell}_t}(f) - {{\tilde \ell}_t}({f^*}))}^2}\right]} \\
 \hspace{1.1cm}  = \sum\limits_{t = 1}^n {\left( {{\mathbb{E}_t}\left[({{\tilde \ell}_t}{{(f)}^2}\right] + {E_t}\left[({{\tilde \ell}_t}{{({f^*})}^2}\right]} \right)} \\
 \hspace{1.1cm}  \le \sum\limits_{t = 1}^n \left( {\frac{1}{{{\rho_t}(f)}} + \frac{1}{{{\rho_t}({f^*})}}} \right) \\
 \hspace{1.1cm} \mathop \le \limits^{(a)} \sum\limits_{t = 1}^n
 {\left( {\frac{1}{{k{\varepsilon  _t}(f)}} + \frac{1}{{k{\varepsilon  _t}({f^*})}}} \right)} \\
 \hspace{1.1cm}  \le \sum\limits_{t = 1}^n {\left( {\frac{1}{{k{{\underline \varepsilon   }_t}(f)}} +
 \frac{1}{{k{{\underline \varepsilon   }_t}({f^*})}}} \right)}  = {\nu _t}(f) + {\nu _t}({f^*})
\end{array}
\end{displaymath}
with probability $1$.  The above inequality (a) is due to the fact that $q_n(f) \ge \sum\nolimits_{f \in i} {{\varepsilon _n}(f)} \left| {\left\{ {i \in \mathcal{C}:f \in i} \right\}} \right|$. Since each $f$ only belongs to one of the covering strategies $i \in \mathcal{C}$, $\left| {\left\{ {i \in \mathcal{C}:f \in i} \right\}} \right|$ equals to 1 at time slot $n$ if channel $f$ is selected. Thus, $q_n(f) \ge \sum\nolimits_{f \in i} {{\varepsilon _n}(f)}= k{\varepsilon_n}(f)$.

Let $\mathcal{\bar E}_n^f$ denote the complementary of event
 $\mathcal{E}_n^f$. Then by the Bernstein's inequality $\mathbb{P}[\mathcal{\bar E}_n^f] \le e^{-b_n}$. The number of
 times the channel $f$ is selected up to round $n$ is bounded as:
\begin{IEEEeqnarray*}{l}
 \begin{array}{l}
 \mathbb{E}[{N_n}(f)] = \sum\limits_{t = 1}^n {\mathbb{P}[A_t= f]} \\
 \hspace{4em} = \sum\limits_{t = 1}^n {\mathbb{P}[A_t= f|\mathcal{E}_{t - 1}^f]P[\mathcal{E}_{t - 1}^f]} \\
  \hspace{5em}+ \mathbb{P}[A_t= f|\overline {\mathcal{E}_{t - 1}^f} ]P[\overline {\mathcal{E}_{t - 1}^f} ]\\
%     \end{array}
%\end{IEEEeqnarray*}
%\begin{IEEEeqnarray*}{l}
%\begin{array}{l}
 \hspace{4em} \le \sum\limits_{t = 1}^n {\mathbb{P}[A_t= f|\mathcal{E}_{t - 1}^f]} {\mathds{1}_{\{ \mathcal{E}_{t - 1}^f\} }} +
  \mathbb{P}[\overline {\mathcal{E}_{t - 1}^S} ]\\
  \hspace{4em} \le \sum\limits_{t = 1}^n {\mathbb{P}[A_t= f|\mathcal{E}_{t - 1}^f]} {\mathds{1}_{\{ \mathcal{E}_{t - 1}^f\} }} + {e^{ - {b_{t - 1}}}}.
\end{array}
\end{IEEEeqnarray*}
We further upper bound $ {\mathbb{P}[A_t= f|\mathcal{E}_{t - 1}^f]} {\mathds{1}_{\{ \mathcal{E}_{t - 1}^f\} }} $ as follows:
\begin{displaymath}
\begin{array}{l}
\hspace{ -4.5 cm}\mathbb{P} {[A_t= f|{\cal \mathcal{E}}_{t - 1}^f]} {\mathds{1}_{\{ {\cal \mathcal{E}}_{t - 1}^f\} }}
   \end{array}
\end{displaymath}%{{\tilde \Psi}_t}(i) =  {{\tilde{L}}}_{t-1}(i) + \lambda_{t-1} {{\tilde{\Gamma}}_{t-1}} (i)
   \begin{displaymath}
\hspace{ .8em}\begin{array}{l}
= {\rho_t}(f){\mathds{1}_{\{ {\cal \mathcal{E}}_{t - 1}^f\} }}\\
  \hspace{0em} \le ({q_t}(f) + k{\varepsilon_t}(f)){\mathds{1}_{\{ {\cal \mathcal{E}}_{t - 1}^f\} }}\\
 \hspace{0em} =({k\varepsilon  _t}(f) + \frac{{\sum\nolimits_{i:f \in i} {{w_{t - 1}}\left( i \right)} }}{{{W_{t - 1}}}}){\mathds{1}_{\{ {\cal \mathcal{E}}_{t - 1}^f\} }}\\
    \hspace{0em} =({k\varepsilon  _t}(f) + \frac{{\sum\nolimits_{i:f \in i}^{} {{e^{ - {\eta _t}{{\tilde \Psi}_t}(i)}}} }}{{\sum\nolimits_{i = 1}^N {{e^{ - {\eta _t}{{\tilde \Psi}_t}(i)}}} }}){\mathds{1}_{\{ {\cal \mathcal{E}}_{t - 1}^f\} }}\\
  \hspace{0em}\mathop \le \limits^{(a)} ({k\varepsilon  _t}(f) + {e^{ - {\eta _t}\left( {{{\tilde \Psi}_t}(i) - {{\tilde \Psi}_t} ({i^*})} \right)}}){\mathds{1}_{\{ {\cal \mathcal{E}}_{t - 1}^f\} }}\\
    \hspace{.0em}\mathop \le \limits^{(b)} ({k\varepsilon  _t}(f) + {e^{ - {\eta _t}\left( {{\tilde \Psi_{t }}(f) - {{\tilde \Psi}_{t }}({f^*})} \right)}}){\mathds{1}_{\{ {\cal \mathcal{E}}_{t - 1}^f\} }}\\
   \hspace{.0em}   =  ({k\varepsilon  _t}(f) \\
      \hspace{.1em}+ {e^{ - {\eta _t}\left( {{\tilde{L}}}_{t}(f) -{{\tilde{L}}}_{t}(f^*) + \lambda_{t}
   ({{\tilde{\Gamma}}_{t}} (f^*)-{{\tilde{\Gamma}}_{t}} (f)) \right)}}){\mathds{1}_{\{ {\cal \mathcal{E}}_{t - 1}^f\} }} \\
      \end{array}
\end{displaymath}%{{\tilde \Psi}_t}(i) =  {{\tilde{L}}}_{t-1}(i) + \lambda_{t-1} {{\tilde{\Gamma}}_{t-1}} (i)
   \begin{displaymath}
\begin{array}{l}
   \hspace{.0em}\mathop \le \limits^{(c)}  k{\varepsilon  _t}(f){\mathds{1}_{\{ {\cal \mathcal{E}}_{t - 1}^f\} }} \!+ \! {e^{ - {\eta _t}{h_{t}}(f)}}
   e^{ - \eta_t\lambda_{t}({{\tilde{\Gamma}}_{t}} (f^*)\!-\!{{\tilde{\Gamma}}_{t}} (f))}
  %\hspace{.0em}\mathop \le \limits^{(d)}  k{\varepsilon  _t}(f){\mathds{1}_{\{ {\cal \mathcal{E}}_{t - 1}^f\} }} + {e^{ - {\eta _t}{h_{t - 1}}(f)}}.
\end{array}
  \end{displaymath}
The above inequality (a) is due to the fact that channel
$f$ only belongs to one selected strategy $i$ at $t$, inequality (b) is because the  cumulative
regret of each strategy is great than the cumulative regret of each channel that belongs to the
  strategy,  inequality (c) is due to the fact that $ {{{\underline \varepsilon   }_t}(f)}$ is a non-increasing
 sequence ${\upsilon _t}(f) \le \frac{t}{{{{k \underline \varepsilon   }_t}(f)}}$.
  Substitution of this result back into
 the computation of $ \mathbb{E}[{N_n}(f)]$ completes the proof. \end{IEEEproof}

%, and the last inequality (d) is based on the simple fact $e^{-x} \le 1, \forall x \ge 0$
\emph{Proof of Theorem 5.}
\begin{IEEEproof}
The proof is based on Lemma 1 and Lemma 3. Combine  (\ref{eq:Dual1}) and  (\ref{eq:LemNum})
\begin{IEEEeqnarray*}{l}
\begin{array}{l}
\mathbb{E}\left[ {\sum\limits_{t = 1}^n {{\textbf{p}_t^T}{\ell_t}(i) - {\textbf{p}_t^T}{\ell_t}({I_t})} } \right] + \\
\hspace{3cm} \mathbb{E}\left[ {\sum\limits_{t = 1}^n {\lambda (\textbf{p}_{t}^\textup{T}\tilde{\textsf{P}}_{t}- \textsf{P}_o) -  \left( {\frac{{\delta n}}{2} + \frac{1}{\eta }} \right){\lambda ^2}} } \right]\\
\hspace{.3cm} \mathop  \le \sum\limits_{f = 1}^K \mathbb{E}[{N_n}(f)]\Delta(f) +  \sum\limits_{t = 1}^n  {\eta_t \sqrt{\gamma_{t}}  }  \\
 \hspace{3.5cm}\! - \!{ \frac{\delta }{2}} \sum\limits_{t = 1}^n {\lambda _t^2}  + \! \mathbb{E}\left[ {\sum\limits_{t = 1}^n {{\lambda _t}(\textbf{p}_{t}^\textup{T}\tilde{\textsf{P}}_{t}- \textsf{P}_o)} } \right]\!.
\end{array}
\end{IEEEeqnarray*}
Obviously, the last two terms in the r.h.s of the above inequality is negative. By taking maximization over $\lambda$,
we have
\begin{IEEEeqnarray*}{l}
\begin{array}{l}
\!\!\!\!\!\mathbb{E}\left[ {\mathop {\max }\limits_{\textbf{p}_{t}^\textup{T}\tilde{\textsf{P}}_{t} \le \textsf{P}_o}
\sum\limits_{t = 1}^n {{\textbf{p}_t^T}{\ell_t}(i) - {\textbf{p}_t^T}{\ell_t}({I_t})} } \right]
 \!\!+ \!\mathbb{E}\! \left[ \! {\frac{{\left[ {\sum\limits_{t = 1}^n {(\textbf{p}_{t}^\textup{T}\tilde{\textsf{P}}_{t}-
 \textsf{P}_o)} } \right]_ + ^2}}{{2\left( {\delta n/2 + {1 \mathord{\left/
 {\vphantom {1 \eta }} \right.
 \kern-\nulldelimiterspace} \eta }} \right)}}} \right] \\
 \hspace{3.2cm}
\mathop   \le
 \sum\limits_{f = 1}^K \mathbb{E}[{N_n}(f)]\Delta(f) + \sum\limits_{t = 1}^n  {\eta_t \sqrt{\gamma_{t}}  }.
\end{array}\IEEEyesnumber \label{eq:StocCons}
\end{IEEEeqnarray*}

Set $b_n =ln(n \Delta(f)^2)$,  ${{{\underline \varepsilon  }_n}(f)}
={{{ \varepsilon  }_n}(f)}$ and ${{{ \varepsilon  }_n}(f)} = \frac{{{b_n}}}{{n\Delta {{\left( f \right)}^2}}}$.
Thus, $\gamma_n = \sum\nolimits_{f = 1}^K {{\varepsilon _n}\left( f \right)}
 = \sum\nolimits_{f = 1}^K \frac{{\ln (n\Delta {{\left( f \right)}^2})}}{{n\Delta {{\left( f \right)}^2}}} = O(\frac{{\ln (n)}}{n})$.  For any $c \ge 18$ and any $n \ge n^*$,  where $n^*$ is the minimal integer for which
 ${n^*} \ge \frac{{4{c^2}K \ln {{({n^*}\Delta {{(f)}^2})}^2}}}{{\Delta {{(f)}^4}\ln (K)}}$, we have
     \begin{displaymath}
\begin{array}{l}
{h_n}(f) = n\Delta (f) - \sqrt {2n{b_n}\left( {\frac{1}{{k{\varepsilon _n}(f)}} + \frac{1}{{k{\varepsilon _n}({f^*})}}} \right)}  - \frac{{\left( {\frac{1}{4} + \frac{1}{k}} \right){b_n}}}{{3{\varepsilon _n}({f^*})}}\\
   \hspace{2.43em} \ge n\Delta (f) - 2\sqrt {\frac{{n{b_n}}}{{k{\varepsilon _n}(f)}}}  - \frac{{\left( {\frac{1}{4} + \frac{1}{k}} \right){b_n}}}{{3{\varepsilon _n}(f)}}\\
  \hspace{2.43em} = n\Delta (f)(1 - \frac{2}{{\sqrt {k c} }} - \frac{{\left( {\frac{1}{4} + \frac{1}{k}} \right)}}{{3c}})\\
  \hspace{2.43em} \mathop  \ge \limits^{(a)} n\Delta (f)(1 - \frac{2}{{\sqrt c }} - \frac{{1.25}}{{3c}}) \ge \frac{1}{2}n\Delta (f).
\end{array}
 %$k{\underline \varepsilon }_n(f)$
  \end{displaymath}
The above inequality (a) is due to the fact that $(1 - \frac{2}{{\sqrt {k c} }} - \frac{{\left( { \frac{1}{4}+\frac{1}{k}} \right)}}{{3c}}$ is
an increasing function with respect to $k (k \ge 1)$.
%Plus, as indicated in work \cite{Branislav2015}, by a bit more sophisticated  bounding $c$ can be made almost as small as 2 in our case.
 The transmission power  is quasi-concave to the reward such that the accumulated power allocation strategy have $( -h_t(f) + \lambda_{t}({{\tilde{\Gamma}}_{t}} (f)-{{\tilde{\Gamma}}_{t}} (f^*)) ) \le 0$, and by substitution of the lower bound on $h_n(f)$ into Lemma 3. Thus,  we have
\begin{IEEEeqnarray*}{l}
\begin{array}{l}
\!\!\!\mathbb{E}[{N_n}(f)]\! \le \! \frac{{\ln (n)}}{{\Delta {{(f)}^2}}}\! + \!k   \frac{{c\ln {{(n)}^2}}}{{\Delta {{(f)}^2}}} \!+ \!{n^*} \! + \!
\sum\limits_{t = 1}^n \!\!\left(\!{{e^{ - \frac{{\Delta (f)}}{4}\sqrt {\frac{{(t - 1)ln(K)}}{K}} }}}\!\right)\\
 \hspace{3.5em} \le k\frac{{c\ln {{(n)}^2}}}{{\Delta {{(f)}^2}}} + \frac{{\ln (n)}}{{\Delta {{(f)}^2}}}+ {n^*} + O(\frac{{{K
 }}}{{\Delta {{(f)}^2}}}),
\end{array} \IEEEyesnumber \label{eq:StocRegs}
\end{IEEEeqnarray*}
  where we used Lemma 3 to bound the sum of the exponents in the first two terms. In addition, please
  note that $n^*$ is of the order $O(\frac{{k K}}{{\Delta {{(f)}^4}\ln (K)}})$. The last term is bounded by Lemma 10 in \cite{Seldin14}.

The (\ref{eq:StocRegs}) bounds the first item in the r.h.s of  (\ref{eq:StocCons}). For the second term,
since $\eta_n \!\! =\!\! \frac{1}{2}\sqrt {\frac{{\ln K}}{{nK}}}$, $ \sum\nolimits_{t = 1}^n  {\eta_t \sqrt{\gamma_{t}}  }= O(\sqrt{ln(n)})$.
That indicates
\begin{IEEEeqnarray*}{l}
\begin{array}{l}
\!\!\!\!\!\mathbb{E}\!\left[ {\mathop {\max }\limits_{\textbf{p}_t^\textup{T}\tilde{\textsf{P}}_{t} \le \textsf{P}_o}
\sum\limits_{t = 1}^n {{\textbf{p}_t^T}{\ell_t}(i) - {\textbf{p}_t^T}{\ell_t}({I_t})} } \right]
 \!\!+ \!\!\mathbb{E}\!\! \left[ \! {\frac{{\left[ {\sum\limits_{t = 1}^n {(\textbf{p}_{t}^\textup{T}\tilde{\textsf{P}}_{t}-
 \textsf{P}_o)} } \right]_ + ^2}}{{\left( {\delta n + \sqrt {\frac{{nK}}{{\ln K}}}} \right)}}} \right] \\
\mathop   \le k\frac{{c\ln {{(n)}^2}}}{{\Delta {{(f)}}}} + \frac{{\ln (n)}}{{\Delta {{(f)}}}} + O(\frac{{{K
 }}}{{\Delta {{(f)}}}}) + {n^*} + O(\sqrt{ln(n)}).
\end{array}\IEEEyesnumber \label{eq:StocCons2}
\end{IEEEeqnarray*}

Moreover, set $\delta  = 2k\sqrt {\frac{{K\ln K}}{n}} $, we have $ {kK\eta \le \frac{\delta }{2}}$ and ${ {\delta n + \sqrt {\frac{{nK}}{{\ln K}}}} }
 \simeq \left( {2k + 1} \right)\sqrt {2Kn} = O(\sqrt{n})$. Then, we obtain
\begin{IEEEeqnarray*}{l}
 \hspace{-2.43em}\begin{array}{l}
\mathbb{E}\left[ {\mathop {\max }\limits_{\textbf{p}_{t}^\textup{T}\tilde{\textsf{P}}_{t}-
 \textsf{P}_o} \sum\limits_{t = 1}^n {{\textbf{p}_t^T}{\ell_t}(i) - {\textbf{p}_t^T}{\ell_t}({I_t})} } \right] \mathop  \le O(k\frac{{c\ln {{(n)}^2}}}{{\Delta {{(f)}}}}) \\
\mathbb{E}{\left[ {\sum\limits_{t = 1}^n {\textbf{p}_{t}^\textup{T}\tilde{\textsf{P}}_{t}-
 \textsf{P}_o} } \right]_ + } \!\le  \! \sqrt {\!( {n + k\frac{{c\ln {{(n)}^2}}}{{\Delta {{(f)}}}} } ){( {\delta n + \sqrt {\frac{{nK}}{{\ln K}}}} )}} \\
 \hspace{8.5em}   =  O(n^{\frac{3}{4}}).
\end{array}\IEEEyesnumber \label{eq:CstReg5}
\end{IEEEeqnarray*}
 Thus, we proof the theorem.
\end{IEEEproof}

\emph{Proof of Theorem 7.}
\begin{proof} The proof is based on the similar idea of Theorem 5, Lemma 1 and  Lemma 3. Here, we just show the
difference part.  Note that
by our definition ${{\hat \Delta }_n}(f) \le 1$ and the sequence ${\underline \varepsilon _n}(f) = {\underline \varepsilon  _n} =
\min \{ \frac{1}{{2K}},{\beta _n},\frac{{c\ln {{(n)}^2}}}{n}\} $ satisfies the condition of Lemma 10. Note that when ${\beta _n} \ge
\frac{{c\ln {{(n)}^2}}}{n}\}$, i.e., for $n$ large enough such that
$
n \ge \frac{{4{c^2}\ln {{(n)}^4}K }}{{\ln (K)}}
$, we have ${\underline \varepsilon  _n}=\frac{{c\ln {{(n)}^2}}}{n}$. Let $b_n=ln(n)$ and let $n^*$ be
large enough, so that for all $n \ge n^*$ we have $n \ge \frac{{4{c^2}\ln {{(n)}^4}K }}{{\ln (K)}}$ and
$n \ge e^{\frac{1}{\Delta(f)^2}}$. With these parameters and conditions on hand, we are going to bound the
rest of the three terms in the bound on $\mathbb{E}[N_n(f)]$ in Lemma 10. The upper bound of
 $\sum\nolimits_{t = {n^*}}^n {{e^{ - {b_t}}}} $ is easy to obtain. For bounding
 $k\sum\nolimits_{t = {n^*}}^n {{\varepsilon _t}(f){\mathds{1}_{\{ \mathcal{E}_{t - 1}^f\} }}}$,  we note that $\mathcal{E}_{n}^f$ holds and we have
      \begin{displaymath}
\begin{array}{l}
{{\hat \Delta }_n}(f) \ge \frac{1}{n}(\mathop {\max }\limits_k ({{\tilde L}_n}(k)) - {{\tilde L}_n}(f)) \ge \frac{1}{n}({{\tilde L}_n}({f^*}) - {{\tilde L}_n}(f))\\
 \hspace{2.43em} \ge \frac{1}{n}{h_n}(f) = \frac{1}{n}\left( {n\Delta (f) - 2\sqrt {\frac{{n{b_n}}}{{k{{\underline \varepsilon  }_n}}}}  - \frac{{(\frac{1}{4} + \frac{1}{k}){b_n}}}{{3{{\underline \varepsilon  }_n}}}} \right)\\
 \hspace{2.43em} = \frac{1}{n}\left( {n\Delta (f) - \frac{{2n}}{{\sqrt {c k \ln (n)} }} - \frac{{(\frac{1}{4} + \frac{1}{k})n}}{{3c\ln (n)}}} \right)\\
 \hspace{2.43em} \mathop  \ge \limits^{(a)} \frac{1}{n}\left( {n\Delta (f) - \frac{{2n}}{{\sqrt {c\ln (n)} }} - \frac{{1.25n}}{{3c\ln (n)}}} \right)\\
%  \end{array}
%\end{displaymath}
%   \begin{displaymath}
%\begin{array}{l}
%\hspace{2.43em} \mathop  \ge \limits^{(a)} \frac{1}{n}\left( {n\Delta (f) - \frac{{2n}}{{\sqrt {c\ln (n)} }} - \frac{{1.25n}}{{3c\ln (n)}}} \right)\\
\hspace{2.43em} \mathop  \ge \limits^{(b)} \Delta (f)\left( {1 - \frac{2}{{\sqrt c }} - \frac{{1.25}}{{3c}}} \right) \ge \frac{1}{2}\Delta (f),
\end{array}
  \end{displaymath}
  where the inequality (a) is due to the fact that $\frac{1}{n}( n\Delta (f) - \frac{2n}{\sqrt {c k \ln (n)} } -
  \frac{(\frac{1}{4} + \frac{1}{k})n}{3c\ln (n)} )$ is
an increasing function with respect to $k (k \ge 1)$ and the inequality (b) due to the fact that for $n \ge n^*$ we have $\sqrt {ln(n)}
\ge 1/\Delta(f).$ Thus,
\begin{displaymath}
{\varepsilon _K}(f){\mathds{1}_{\{ \mathcal{E}_{K - 1}^f\} }}
\le \frac{{c{{\left( {\ln n} \right)}^2}}}{{n{{\hat \Delta }_n}{{(f)}^2}}} \le
 \frac{{4{c^2}{{\left( {\ln n} \right)}^2}}}{{n\Delta {{(f)}^2}}}
 \end{displaymath}
 and $k\sum\nolimits_{t = {n^*}}^n {{\varepsilon _t}(f){\mathds{1}_{\{ \mathcal{E}_{K - 1}^f\} }}} = O\left(
 {\frac{{k\ln {{\left( n \right)}^3}}}{{\Delta {{(f)}^2}}}} \right)$. Finally, for the last term in Lemma 10, we have
 already get $h_n(f) \ge \frac{1}{2}\Delta(f)$ for $n \ge n^*$ as an intermediate step in the calculation of bound
 on ${{{\hat \Delta }_n}(f)}$. Therefore, the last term
 is bounded in a order of $O(\frac{{{K
 }}}{{\Delta {{(f)}^2}}})$. Use all these results together we obtain the results of the theorem. Note that the
 results holds for any $\eta_n \ge \beta_n$.
 \end{proof}

\subsection{Mixed Adversarial and Stochastic Regime}
\emph{Proof of Theorem 9.}
\begin{proof}
The proof of the regret performance in the mixed adversarial and stochastic regime is simply a combination of the performance of
the AOEECC-EXP3++$^\emph{AVG}$ algorithm in adversarial and stochastic regimes. It is very straightforward from Theorem 1 and Theorem 7.
 \end{proof}
\emph{Proof of Theorem 11.}
\begin{proof}
Similar as above, the proof is very straightforward from Theorem 3 and Theorem 7.
 \end{proof}

\subsection{Contaminated Stochastic Regime}
\emph{Proof of Theorem 13.}
\begin{proof}
The key idea of proving the regret bound under  moderately contaminated stochastic  regime
 relies on how to estimate the performance loss by taking into account the contaminated pairs. The rest of the proof is based on the similar idea of Theorem 7, Lemma 1 and  Lemma 3. Here, we just show the
difference part.  Let $\mathds{1}
 _{n,f}^\star$ denote the indicator functions of the occurrence of contamination at location $(n,f)$, i.e.,
 $\mathds{1} _{n,f}^\star$  takes value $1$ if contamination occurs and $0$ otherwise.
 Let $m_n(f)= \mathds{1}  _{n,f}^\star \tilde \ell_n(f) + (1-\mathds{1}  _{n,f}^\star)\mu(f)$.  If either base arm $f$
 was contaminated on round $n$ then $m_n(f)$ is adversarially assigned a value of loss that is
 arbitrarily affected by some adversary, otherwise we use the expected loss. Let  ${M_n}(f) = \sum\nolimits_{t = 1}^n {{m_n}(f)}$
 then $\left( {{M_n}({f}) - {M_n}(f^*)} \right) - \left( {{{\tilde L}_n}({f}) - {{\tilde L}_n}(f^*)} \right)$ is a martingale.
 After
 $\tau$ steps, for $n \ge \tau$,
  \begin{displaymath}
\begin{array}{l}
\left( {{M_n}({f}) - {M_n}(f^*)} \right) \ge n\min \{ \mathds{1}  _{n,f}^\star ,\mathds{1}  _{n,f^*}^\star \} ({\tilde \ell_n}(f) - {\tilde \ell_n}({f^*}))\\
\hspace{6em} + n\min \{ 1 - \mathds{1}  _{n,f}^\star ,1 - \mathds{1}  _{n,f^*}^\star \} (\mu ({f}) - \mu (f^*))\\
 \hspace{3.6em} \ge  - \zeta n\Delta (f) + (n - \zeta n\Delta (f))\Delta (f) \ge (1-2\zeta){n\Delta (f)}.
\end{array}
  \end{displaymath}

Define the event $\mathcal{Z}_n^f$:
  \begin{displaymath}
(1-2\zeta)n\Delta (f) - \left( {{{\tilde L}_n}({f}) - {{\tilde L}_n}(f^*)} \right)
\le 2\sqrt {{\nu _n}{b_n}}  + \frac{{\left( {\frac{1}{4} + \frac{1}{k}} \right){b_n}}}{{3{{\underline \varepsilon  }_n}}},
  \end{displaymath}
where ${\underline \varepsilon  }_n$ is defined in the proof of Theorem 2 and $\nu _n = \sum\nolimits_{t = 1}^n
 {\frac{1}{{{{k\underline \varepsilon  }_n}}}}$. Then by Bernstein's inequality
  $\mathbb{P}[\mathcal{Z}_n^f] \le e^{-b_n}$. The remanning proof is identical to the proof of Theorem 2.

  For the regret performance in the moderately contaminated stochastic regime, according to our definition with the attacking strength
  $\zeta \in [0,1/4]$, we only need to replace  $\Delta(f)$  by $\Delta(f)/2$ in Theorem 4.
\end{proof}

\section{Proof of Regret for Accelerated AOEECC Algorithm}
  We prove the theorems of the performance results in Section VI in the order they were presented.
\subsection{Accelerated Learning in Adversarial Regime}
The proof the Theorem 16 requires the following Lemma from Lemma 7 \cite{Seldin14lim}. We restate it for completeness.

\textbf{Lemma 4.}  For any probability distribution $\omega$ on $\{1,...,K\}$   and any $m \in [K]$:
\begin{IEEEeqnarray*}{l}
\sum\limits_{f = 1}^n {\frac{{ \omega (f)(K - 1)}}{{ \omega (f)(K - m) + m - 1}} \le \frac{K}{m}} .
\end{IEEEeqnarray*}

 \emph{Proof of Theorem 16.}
  \begin{IEEEproof}
With similar facts and notations as in the proof of Theorem 1,  we have:
Then we have:
\begin{IEEEeqnarray*}{l}
\!\!\begin{array}{l}
{\mathbb{E}_{i \sim {\textbf{p}_t}}}{\tilde{ \Phi}_t}(i) = ( {1 - \sum\nolimits_{f} {{\varepsilon _t}(f)} } ){\mathbb{E}_{i \sim {\textbf{q}_t}}}{\tilde{ \Phi}_t}(i) + {\varepsilon _t}(i){\mathbb{E}_{i \sim \textbf{u}}}{\tilde{ \Phi}_t}(i)\\
\quad\quad \quad \quad \ \  = ( {1 - \sum\nolimits_{f} {{\varepsilon _t}(f)} } )(\frac{1}{{{\eta_t}}}\ln {\mathbb{E}_{i \sim {\textbf{q}_t}}}\exp ( - {\eta_t}({\tilde{ \Phi}_t}(i) \\
\quad\quad \quad \quad \quad \ - {\mathbb{E}_{j \sim {\textbf{q}_t}}} {\tilde{ \Phi}_t}(j))))\\
 \quad\quad \quad \quad \quad \ - \frac{( {1 - \sum\nolimits_{f} {{\varepsilon _t}(f)} } )}{{{\eta_t}}}\ln {\mathbb{E}_{i \sim {\textbf{q}_t}}}\exp ( - {\eta_t}{\tilde{ \Phi}_t}(i))) \\
\quad\quad \quad \quad \quad \  + {\mathbb{E}_{i \sim \textbf{u}}}{\tilde{ \Phi}_t}(i).
\end{array}\IEEEyesnumber \label{eq:AppenA}
\end{IEEEeqnarray*}

In the second step, we use the inequalities $lnx \le x-1$ and $exp(-x) -1 +x \le x^2/2$, for all $x \ge 0$, to obtain:
\begin{IEEEeqnarray*}{l}
\begin{array}{l}
\ln {\mathbb{E}_{i \sim {\textbf{q}_t}}}\exp ( - {\eta_t}({\tilde{ \Phi}_t}(i) - {\mathbb{E}_{j \sim {\textbf{q}_t}}}{\tilde{ \Phi}_t}(j)))\\
\quad\quad \quad = \ln {\mathbb{E}_{i \sim {\textbf{q}_t}}}\exp ( - {\eta_t}{\tilde{ \Phi}_t}(i)) + {\eta_t}{\mathbb{E}_{j \sim {\textbf{q}_t}}}{\tilde{ \Phi}_t}(j)\\
\quad\quad \quad \le {\mathbb{E}_{i \sim {\textbf{q}_t}}}( {\exp ( - {\eta_t}{\tilde{ \Phi}_t}(i)) - 1 + {\eta_t}{\tilde{ \Phi}_t}(j)} )\\
\quad\quad \quad \mathop  \le \limits^{(a)}  {\mathbb{E}_{i \sim {\textbf{q}_t}}}\frac{{\eta_t^2{\tilde{ \Phi}_t}{{(i)}^2}}}{2} \mathop  \le \limits^{(b)}  {\eta_t^2 \mathbb{E}_{i \sim {\textbf{q}_t}}} {{{ \tilde  \ell_t}{{(i)}^2}}} + \lambda_t^2 \eta_t^2 {\mathbb{E}_{i \sim {\textbf{q}_t}}} {{ {{  \tilde{ \textsf{P}}}_{n}}(i) }^2} ,
\end{array}\IEEEyesnumber \label{eq:Apart1}
\end{IEEEeqnarray*}

 Take expectations over all random strategies of losses ${{\tilde \ell}_t}{(i)^2}$, we have
\begin{IEEEeqnarray*}{l}
\begin{array}{l}
%\!\!\!\!\!\!
 \!\!\!{\mathbb{E}_n}\left[{\mathbb{E}_{i \sim {{q}_t}}}{{\tilde \ell}_t}{(i)^2} \right] =
{\mathbb{E}_n}\left[\sum\limits_{i = 1}^N {{{q}_t}(i){{\tilde \ell}_t}{{(i)}^2}} \right]\\
 \!\!\!\!= {\mathbb{E}_n}\!\!\left[\sum\limits_{i = 1}^N {{{q}_t}(i){{(\sum\limits_{f \in i}
  {{{\tilde \ell}_t}(f)} )}^2}}
  \right]
  \le {\mathbb{E}_n}\!\!\left[ \sum\limits_{i = 1}^N {{{q}_t}(i){k}\!\!\sum\limits_{f
   \in i} {{{\tilde \ell}_t}{{(f)}^2}} }\right]
 \\
  \!\!\!\!= \!{\mathbb{E}_n}k \left[\! \sum\limits_{f = 1}^K {{{\tilde \ell}_t}{{(f)}^2}} \!\!\!\!\sum\limits_{i
  \in {\mathcal{P}}:f \in i} \!\!\!{{{q}_t}(i)} \right]\! = \!
 {k}{\mathbb{E}_n}\!\! \left[\!\sum\limits_{f' = 1}^K \!{{{\tilde \ell}_t}{{(f')}^2}{q_{t,f}}(f')} \right]\\
 \!\!\!\! = {k}{\mathbb{E}_n}\left[ {\sum\limits_{f' = 1}^K¡¡{{{\left( {\frac{{{\ell_t}(f)}}{{{{{\tilde \varrho }_t}}(f)}}
 {\mathds{1}_t}(f)} \right)}^2}} {q_{t,f}}(f')} \right]\\
       \end{array}
\end{IEEEeqnarray*}%{{\tilde \Psi}_t}(i) =  {{\tilde{L}}}_{t-1}(i) + \lambda_{t-1} {{\tilde{\Gamma}}_{t-1}} (i)
\begin{IEEEeqnarray*}{l}
\begin{array}{l}
 \!\!\!\! \le {k}{\mathbb{E}_n}\left[ {\sum\limits_{f = 1}^K {\frac{{{q_{t,f}}(f)}}{{{{{\tilde \varrho }_t}}{{(f)}^2}}}{\mathds{1}_t}(f
 )} } \right] = {k}\sum\limits_{f = 1}^K¡¡{\frac{{q_{t,f}(f)}}{{{{{\tilde \varrho }_t}}(f)}}} \\
 \!\!\!\!= {k}\!\!\sum\limits_{f = 1}^K \!\! {\frac{{q_{t-1,f}(f)}}{{
{\rho _t}(f)
 + (1-{\rho _t}(f))\frac{{{m_t} - 1}}{{{K} - 1}}}}} \mathop  \le \limits^{(a)}  {k}\!\!\sum\limits_{f = 1}^K\!\! {\frac{{2\rho _t}(f)}{{
{\rho _n}(f)
 + (1-{\rho _t}(f))\frac{{{m_t} - 1}}{{{K} - 1}}}}} \\
 \mathop  \le \limits^{(b)}  2k \frac{K}{m},
\end{array}\IEEEyesnumber \label{eq:Apart1t}
\end{IEEEeqnarray*}
where the above inequality $(a)$ follows the fact that $( {1 - \sum\nolimits_{f} {{\varepsilon _t}(f)} }) \ge \frac{1}{2}$ by the definition of
${{\varepsilon _t}(f)}$ and the equality   (\ref{eq:Lim2}) and the above inequality $(b)$ follows the Lemma 4.  Similarly,
\begin{IEEEeqnarray*}{l}
\begin{array}{l}
%\!\!\!\!\!\!
{\mathbb{E}_n}\left[{\mathbb{E}_{i \sim {\textbf{q}_t}}} {{ {{  \tilde{ \textsf{P}}}_{n}}(i) }^2}  \right]  \le 2k  \frac{K}{m},
\end{array}\IEEEyesnumber \label{eq:ApartConst2}
\end{IEEEeqnarray*}

 Note that ${\varphi_{t-1}}(f)= \sum\nolimits_{f \in i} {{\varepsilon _n}(f)} \left| {\left\{ {i \in \mathcal{C}:f
\in i} \right\}} \right|, \forall f \in [1,n]$ Take expectations over all random strategies of losses
 ${{\tilde \ell}_t}{(i)}$ with respective
to distribution $u$, we have
\begin{IEEEeqnarray*}{l}
\begin{array}{l}
%\!\!\!\!\!\!
{\mathbb{E}_n}\left[{\mathbb{E}_{i \sim {\varphi_t}}}{{\tilde \ell}_t}{(i)} \right] =
{\mathbb{E}_n}\left[\sum\limits_{i = 1}^N {{\varphi_t}(i){{\tilde \ell}_t}{{(i)}}} \right]\\
= {\mathbb{E}_n}\!\!\left[\sum\limits_{i = 1}^N {{\varphi_t}(i){{(\sum\limits_{f \in i} {{{\tilde \ell}_t}(f)} )}}}
  \right]    \le  {\mathbb{E}_n}\!\!\left[ \sum\limits_{i = 1}^N {{\varphi_t}(i) (
 \sum\limits_{f \in i} {{{\tilde \ell}_t}{{(f)})}} }\right]
 \\= \!{\mathbb{E}_n}  \left[\! \sum\limits_{f = 1}^K {{{\tilde \ell}_t}{{(f)}}} \!\!\!\!\sum\limits_{i \in {\mathcal{P}}:f \in
  i} \!\!\!{{\varphi_t}(i)} \right]\! = \!
 { }{\mathbb{E}_n}\!\! \left[\!\sum\limits_{f' = 1}^K \!{{{\tilde \ell}_t}{{(f')}}{\varphi_{t}}(f')} \right]\\
 %\end{array}
%\end{IEEEeqnarray*}
%\begin{IEEEeqnarray*}{l}
%\begin{array}{l}
\le {k}{\mathbb{E}_n}\left[ {\sum\limits_{f' = 1}^K {\frac{{\varphi_{t}}(f')}{{{{{\tilde \rho }_t}}{{(f')}
}}}{\mathds{1}_t}(f')} } \right] = {k}\sum\limits_{f' = 1}^K {\frac{{\varphi_{t}}(f')}{{{{{\tilde \rho }_t}}(f')}}} \\
 = {k}\sum\limits_{f' = 1}^K {\frac{{\varphi_{t}}(f')}{{\tilde\rho _t}(f)
 + (1-{\tilde\rho _t}(f))\frac{{{m_t} - 1}}{{{K} - 1}}}}\\
  \mathop  \le \limits^{(a)}  {k}\sum\limits_{f' = 1}^K {\frac{{\tilde\rho _t}(f)}{{\tilde\rho _t}(f)
 + (1-{\tilde\rho _t}(f))\frac{{{m_t} - 1}}{{{K} - 1}}}}
  \mathop  \le \limits^{(b)}  2k \frac{K}{m},
\end{array}\IEEEyesnumber \label{eq:Aparts4t}
\end{IEEEeqnarray*}
where the above inequality $(a)$ is due to the fact that ${\tilde\rho _n}(f) \ge {\varphi_{t-1}}(f)$ and the above inequality $(b)$ follows the Lemma 4.

In the third step,  take expectation over all random strategies of losses up to time $n$, we obtain
\begin{IEEEeqnarray*}{l}
\begin{array}{l}%\!\! \sum\limits_{t = 1}^n  \!{\mathbb{E}_{i \sim u}}{{\tilde \Phi}_t}(i)
\hspace{-.1cm}{\mathbb{E}_n}\left[ \sum\limits_{t = 1}^n {\mathbb{E}_{i \sim {\textbf{p}_t}}}{{\tilde \Phi}_t}(i) \right]
 \le \frac{2kK}{m} \!\sum\limits_{t = 1}^n \eta_t +  \frac{2kK}{m} \!\sum\limits_{t = 1}^n \eta_t \lambda_t^2 +  \frac{{\ln N}}{\eta_n } \\
\hspace{.3cm} + {
\mathbb{E}_n}\left[\sum\limits_{t = 1}^n {\mathbb{E}_{{I_t} \sim {\textbf{p}_t}}} {{{\tilde \Phi}_t}(j)} \right] + {\mathbb{E}_n}\left[ \sum\limits_{t = 1}^{n - 1} ({{\Upsilon _t}({\eta _{t + 1}}) - {\Upsilon _t}({\eta _t}))})\right]\!.
\end{array}
\end{IEEEeqnarray*}
The last term in the r.h.s of the inequality is less than or equals to zero as indicated in \cite{Bubeck12}.  Then, we get
\begin{IEEEeqnarray*}{l}
R_{\tilde{ \Phi}_n(i)}(n)=  \mathbb{E}_n \left[\sum\limits_{t = 1}^n {{\mathbb{E}_{i \sim {\textbf{p}_t}}}\tilde{ \Phi}_n(i)}  - \sum\limits_{t = 1}^n
{{\mathbb{E}_{{I_t} \sim {\textbf{p}_t}}}\tilde{ \Phi}_n(i)} \right] \\
\hspace{1.4cm} \le \frac{2kK}{m} \!\sum\limits_{t = 1}^n \eta_t + \frac{{\ln N}}{\eta_n } +  \frac{2kK}{m} \!\sum\limits_{t = 1}^n \eta_t \lambda_t^2 \\
\end{IEEEeqnarray*}
\begin{IEEEeqnarray*}{l}
%\hspace{.7cm}\mathop  \le \limits^{(a)} k K \!\sum\limits_{t = 1}^n \eta_t + \frac{{\ln N}}{\eta_n }
%+ k \sum\limits_{t = 1}^n {\sum\limits_{f = 1}^K {{\varepsilon _t}(f)} } \\
%\hspace{.7cm}\mathop  \le \limits^{(b)} 2k K \!\sum\limits_{t = 1}^n \eta_t + \frac{{\ln N}}{\eta_n } \\
\hspace{.7cm}\mathop  \le \limits^{(a)}\frac{2kK}{m} \!\sum\limits_{t = 1}^n \eta_t + k\frac{{\ln K}}{\eta_n } +   \frac{2kK}{m} \!\sum\limits_{t = 1}^n \eta_t \lambda_t^2.
%\le k K \!\sum\limits_{t = 1}^n \eta_t + \frac{{\ln N}}{\eta_n }
\IEEEyesnumber \label{eq:CstResg1}
\end{IEEEeqnarray*}
Note that, the
 inequality $(a)$ is due to the fact that $N \le K^{k}$. %Setting $\eta_n=b_n$, we prove the theorem.

Combine  (\ref{eq:Dual1}) and  (\ref{eq:CstReg1}) gives that
\begin{IEEEeqnarray*}{l}
\begin{array}{l}
\mathbb{E}\left[ {\sum\limits_{t = 1}^n {{\textbf{p}^T_t}{\ell_t}(i) - {\textbf{p}^T_t}{\ell_t}({I_t})} } \right] + \\
\hspace{2.4cm} \mathbb{E}\left[ {\sum\limits_{t = 1}^n {\lambda (\textbf{p}_{t}^\textup{T}\tilde{\textsf{P}}_{t}- \textsf{P}_o)
-  \left( {\frac{{\delta_n n}}{2} + \frac{1}{\eta  \sqrt{\gamma_{t}} }} \right){\lambda ^2}} } \right]\\
\mathop  \le \limits^{(a)} \frac{2kK}{m} \sum\limits_{t = 1}^n {{\eta _t}}
 + k\frac{{\ln K}}{{{\eta _n}}}+  \sum\limits_{t = 1}^n  {\eta_t \sqrt{\gamma_{t}}  } + \left( {\frac{kK}{m}\eta_n  -
 \frac{\delta_n }{2}} \right)\sum\limits_{t = 1}^n {\lambda _t^2}   \\
\hspace{3.3cm}+ \mathbb{E}\left[ {\sum\limits_{t = 1}^n {{\lambda _t}
(\textbf{p}_{t}^\textup{T}\tilde{\textsf{P}}_{t}- \textsf{P}_o)} } \right].
\end{array}
\end{IEEEeqnarray*}

  For the above inequality $(a)$, we use the trick by letting $\eta_n = \eta_t$ indicated in [16] (page 25) again to extract the $\eta_t$ from the
sum of $\lambda_t^2$ over $n$ and the inequality in (\ref{eq:ApartConst2}).   Let $ {kK\eta_n \le \frac{m\delta_n}{2}}$ by setting properly the values  such that  $\eta_n = O(\delta_n)$ (shown
in the next). Thus, the last two terms in the r.h.s of the above inequality is non-positive.  By taking maximization over $\lambda$,
we have
\begin{IEEEeqnarray*}{l}
\begin{array}{l}
\!\!\!\mathbb{E}\! \left[ {\mathop {\max }\limits_{\textbf{p}_{t}^\textup{T}\tilde{\textsf{P}}_{t} \le \textsf{P}_o} \sum\limits_{t = 1}^n {{\textbf{p}^T_t}{\ell_t}(i) - {\textbf{p}^T_t}{\ell_t}({I_t})} } \right]
\! \! + \!\mathbb{E}\! \!\left[ {\frac{{\left[ {\sum\limits_{t = 1}^n {(\textbf{p}_{t}^\textup{T}\tilde{\textsf{P}}_{t}-
 \textsf{P}_o)} } \right]_ + ^2}}{{2\left( {\delta n/2 + {1 \mathord{\left/
 {\vphantom {1 \eta }} \right.
 \kern-\nulldelimiterspace} \eta }} \right)}}} \right] \\
 \hspace{3.3cm}
\mathop   \le
 \limits^{(a)} \frac{2kK}{m}\sum\limits_{t = 1}^n {{\eta _t}}  + k\frac{{\ln K}}{{{\eta _n}}}+  \sum\limits_{t = 1}^n  {\eta_t \sqrt{\gamma_{t}}  } .
\end{array}\IEEEyesnumber \label{eq:CstReg4}
\end{IEEEeqnarray*}

Then,  we obtain
\begin{IEEEeqnarray*}{l}
\begin{array}{l}
\mathbb{E}\left[ {\mathop {\max }\limits_{\textbf{p}_{t}^\textup{T}\tilde{\textsf{P}}_{t} \le
 \textsf{P}_o} \sum\limits_{t = 1}^n {{\textbf{p}^T_t}{\ell_t}(i) - {\textbf{p}^T_t}{\ell_t}({I_t})} } \right] \mathop  \le \limits^{} \frac{4kK}{m}
 n{\eta _n} + k\frac{{\ln K}}{{{\eta _n}}}\\
 \hspace{3.9cm} +  \sum\limits_{t = 1}^n  {\eta_t \sqrt{\gamma_{t}}  }  \mathop  \le \limits^{} 4k\sqrt {n\frac{K}{m}\ln K} \\
\mathbb{E}{\left[ {\sum\limits_{t = 1}^n {\textbf{p}_{t}^\textup{T}\tilde{\textsf{P}}_{t}-
 \textsf{P}_o} } \right]_ + } \!\le  \! \sqrt {\!\left( {n + 4k\sqrt {nK\ln K} } \right)\left( {\delta_n n  + {2 \mathord{\left/
 {\vphantom {1 \eta }} \right.
 \kern-\nulldelimiterspace} \eta_n }} \right)} .
\end{array}\IEEEyesnumber \label{eq:CstReg5}
\end{IEEEeqnarray*}
Let ${\eta _n} = {\beta _n}=  \sqrt {\frac{{\ln K}}{{4Kn}}}  = O\left( {{n^{ - 1/2}}} \right),{1 \mathord{\left/{\vphantom {1 \eta_n }} \right. \kern-\nulldelimiterspace} \eta } = O\left( {{n^{1/2}}} \right)$.
 Because $\gamma_{n}= \tilde{O}(\frac{ln(n)}{n})$,  the term
 $ \sum\nolimits_{t = 1}^n  {\eta_t \sqrt{\gamma_{t}}  } = \tilde{O}(ln(n))$. Thus, it can be omitted when
 compared to the first two terms in the r.h.s of (\ref{eq:CstReg4}).   Moreover, in this setting, we set
$\delta_n = 2\frac{k}{m}\sqrt {\frac{{K\ln K}}{n}} $, such that $ {2kK\eta_n \le \frac{m\delta_n}{2}}$. Then,
$\delta_n n = O\left( {{n^{1/2}}} \right)$. This completes the proof.
  \end{IEEEproof}

 \emph{Proof of Theorem 17.}
 \begin{IEEEproof}
 The proof of Theorem 17 for adaptive adversary   uses the same idea as in the proof
 of Theorem 2. Here, if we set the batch $\tau= {(4{k}\sqrt {\frac{n}{m}\ln n} )^{ - \frac{1}{3}}}{n^{^{\frac{1}{3}}}}$ in Theorem 2 of \cite{Arora12},
we can get the regret upper bound in our Theorem 17.
 \end{IEEEproof}

\subsection{Cooperative Learning of AOEECC in  Stochastic Regime}
To obtain the tight regret performance for cooperative learning of  AOEECC-EXP3++, we need to study and estimate the
number of times each of channel is selected up to time $n$, {i}.f., $N_n(f)$. We summarize it in the following lemma.

\textbf{Lemma 5. } In the multipath probing case, let $\left\{ {{{\underline \varepsilon }_n}(f)} \right\}_{n = 1}^\infty$ be non-increasing deterministic sequences, such that
${{\underline \varepsilon }_n}(f) \le {{ \varepsilon }_n}(f)$ with probability $1$ and ${{\underline \varepsilon }_n}(f) \le {{ \varepsilon }_n}(f^*)$
 for all $n$ and $f$. Define $\nu_n(f)= \sum\nolimits_{t = {1}}^n  \frac{1}{{{k \underline \varepsilon  }_t}(f)} $, and define the event $\Xi^f_n$
\begin{IEEEeqnarray*}{l}
 {mt\Delta (f) - ( {{{\tilde L}_{n}}(f^*) - {{\tilde L}_{n}}({f})} )} \\
\hspace{.3cm} \le {\sqrt {2({\nu _{n}}(f) + {\nu _n}({f^*})){b_{n}}}  + \frac{{{(1/k+ 0.25)} {b_{n}}}}{{3k
{\underline \varepsilon _{n}}({f^*})}}}\hspace{.3cm}
(\Xi^f_{n}).
\end{IEEEeqnarray*}
Then for any positive sequence $b_1, b_2,...,$ and any $n^* \ge 2$ the number of times channel $f$ is played by AOEECC-EXP3++ up to
round $n$ is bounded as:
\begin{IEEEeqnarray*}{l}
\begin{array}{l}
\mathbb{E}[{N_n}(f)] \le \left( {{n^*} - 1} \right) + \sum\limits_{t = {n^*}}^{n} {{f^{ - {b_t}}}}  +
 k \sum\limits_{t = {n^*}}^{n} {{\varepsilon _t}(f){\mathds{1}_{\{ \Xi_{n}^f\} }}} \\
    \hspace{1.6cm}               + \sum\limits_{t = {n^*}}^{n} {{f^{ - {\eta _t}{\hslash_{t - 1}}(f)}}}
    e^{ - \eta_t\lambda_{t}({{\tilde{\Gamma}}_{t}} (f^*)\!-\!{{\tilde{\Gamma}}_{t}} (f))},
\end{array}
\end{IEEEeqnarray*}
where
\begin{IEEEeqnarray*}{l}
\begin{array}{l}
{\hslash_n}(f) = {mt}\Delta (f) - \sqrt {2{mt}{b_n}\left( {\frac{1}{{{k}{{\underline \varepsilon
 }_n}(f)}} + \frac{1}{{{k}{{\underline \varepsilon  }_n}({f^*})}}} \right)}  - \frac{{(\frac{1}{4} +
 \frac{1}{k}){b_n}}}{{3{{\underline \varepsilon  }_n}({f^*})}}.
\end{array}
\end{IEEEeqnarray*}

\begin{IEEEproof}
Note that AOEECC-EXP3++ probes $L_n$ strategies rather than $1$ strategy each timeslot $n$. Let $\# \left\{  \cdot  \right\}$ stands for the number of elements
 in the set $\left\{  \cdot  \right\}$. Hence,
 \begin{IEEEeqnarray*}{l}
\begin{array}{l}
\mathbb{E}[{N_n}(f)] = \mathbb{E}[\# \left\{ {1 \le t \le n:{A_t} = f,\mathcal{E}_n^f} \right\} + \\
\hspace{4cm}\# \left\{ {1 \le t \le n:{A_t} = f,\overline {\mathcal{E}_n^f} } \right\}],
\end{array}
\end{IEEEeqnarray*}
where $A_t$ denotes the action of channel selection at timeslot $t$.
By the following  simple trick, we have
 \begin{IEEEeqnarray*}{l}
\begin{array}{l}
\!\!\!\!\!\!\!\!\mathbb{E}[{N_n}(f)] = \mathbb{E}[\# \left\{ {1 \le t \le n:{A_t} = f,\mathcal{E}_n^f} \right\}] + \\
\hspace{3.1cm}\mathbb{E}[\# \left\{ {1 \le t \le n:{A_t} = f,\overline {\mathcal{E}_n^f} } \right\}]] \\
 \hspace{.4cm}\le \mathbb{E}[ \sum\limits_{t = 1}^n \mathds{1}_{ \left\{ {1 \le t \le n:{A_t} = f} \right\}}\mathbb{P}[\# \{\mathcal{E}_n^f\}]] + \\
\hspace{3.1cm}\mathbb{E}[ \sum\limits_{t = 1}^n \mathds{1}_{ \left\{ {1 \le t \le n:{A_t} = f} \right\}}\mathbb{P}[\# \{\overline {\mathcal{E}_n^f}\}]] \\
 \hspace{.4cm}\le \mathbb{E}[\sum\limits_{t = 1}^n  \mathds{1}_{ \left\{ {1 \le t \le n:{A_t} = f} \right\}}\mathbb{P}[\Xi _{mt}^f]] + \\
\hspace{3.1cm}\mathbb{E}[ \sum\limits_{t = 1}^n \mathds{1}_{ \left\{ {1 \le t \le n:{A_t} = f} \right\}}\mathbb{P}[\Xi _{mt}^f]].
\end{array}\IEEEyesnumber \label{eq:Apart23t}
\end{IEEEeqnarray*}

%Because, $ \mathbb{E}[ \# \left\{ {1 \le t \le n:{A_t} = f,\mathcal{E}_n^f} \right\}]= \mathbb{E}\{
%\# \left\{ {1 \le t \le n:{A_t} = f} \right\}\mathbb{P}[\mathcal{E}_n^f]\}$

% \hspace{.4cm}\le \mathbb{E}[ \mathds{1}_{ \left\{ {1 \le t \le mt:{A_t} = f,\Xi _{mt}^f} \right\}}\# \{\mathcal{E}_n^f\} + \\
%\hspace{4.cm}\# \left\{ {1 \le t \le n: {A_t} = f,\overline {\mathcal{E}_n^f} } \right\}] .

%$\mathbb{E}[ \# \left\{ {1 \le t \le n:{A_t} = f,\mathcal{E}_n^f} \right\}] =

Note that the elements of the martingale difference sequence in the  $\{ {\Delta (f)
- ({{\tilde \ell}_n}(f) - {{\tilde \ell}_n}({f^*}))} \}_{n = 1}^\infty$ by $\max \{ \Delta (f) +{{\tilde \ell}_n}({f^*}) \}= {\frac{1}{{{k}{{\underline \varepsilon  }_n}({f^*})}}} +1$. Since ${{{\underline \varepsilon  }_n}({f^*})} \le {{{ \varepsilon  }_n}({f^*})} \le 1/(2n) \le 1/4$, we can simplify
the upper bound by using ${\frac{1}{{{{k\underline \varepsilon   }_n}({f^*})}}} +1 \le
    \frac{{(\frac{1}{4} + \frac{1}{k})}}{{{{\underline \varepsilon   }_n}({f^*})}}$.

We further note that
   \begin{displaymath}
\begin{array}{l}
{\mathbb{E}_t}  \left\{  \# \left\{
\sum\limits_{t = 1}^{n} {\left[{{(\Delta (f) - ({{\tilde \ell}_t}(f) - {{\tilde \ell}_t}({f^*})))}^2}\right]} \right\} \right\}\\
 \hspace{1.1cm}  \mathop \le \limits^{(a)}   {\mathbb{E}_t}  \left\{ m
\sum\limits_{t = 1}^{n} {\left[{{(\Delta (f) - ({{\tilde \ell}_t}(f) - {{\tilde \ell}_t}({f^*})))}^2}\right]}\right\} \\
 \hspace{1.1cm}  \le m\sum\limits_{t = 1}^{n}  {{\mathbb{E}_t}\left[{{({{\tilde \ell}_t}(f) - {{\tilde \ell}_t}({f^*}))}^2}\right]} \\
 \hspace{1.1cm}  = m\sum\limits_{t = 1}^{n}  {\left( {{\mathbb{E}_t}\left[({{\tilde \ell}_t}{{(f)}^2}\right] + {E_t}\left[({{\tilde \ell}_t}{{({f^*})}^2}\right]} \right)} \\
 \hspace{1.1cm}  \le m\sum\limits_{t = 1}^{n}  \left( {\frac{1}{{{\tilde \varrho_t}(f)}} + \frac{1}{{{\tilde \varrho_t}({f^*})}}} \right) \\
 \hspace{1.1cm} \mathop \le \limits^{(b)} m\sum\limits_{t = 1}^{n}  {\left( {\frac{1}{{k{\varepsilon  _t}(f)}} + \frac{1}{{k{\varepsilon  _t}({f^*})}}} \right)} \\
 \hspace{1.1cm}  \le m\sum\limits_{t = 1}^{n} {\left( {\frac{1}{{k{{\underline \varepsilon   }_t}(f)}} +
 \frac{1}{{k{{\underline \varepsilon   }_t}({f^*})}}} \right)} = m{\nu _n}(f) + m{\nu _n}({f^*})
\end{array}
\end{displaymath}
with probability $1$. The above inequality $(a)$ is because the number of probes for each channel
$f$  at timeslot $t$ is at most $m$ times, so does the accumulated value of the variance ${{(\Delta (f) - ({{\tilde \ell}_t}(f)
- {{\tilde \ell}_t}({f^*})))}^2}$. The above inequality (b) is due to the fact
that ${{\tilde \varrho }_n}(f) \ge
{{\tilde \rho }_n}(f) \ge \sum\nolimits_{f \in i}
 {{\varepsilon _n}(f)} \left| {\left\{ {i \in \mathcal{C}:f \in i} \right\}} \right|$. Since each $f$ only belongs to one of the covering strategies $i \in \mathcal{C}$, $\left| {\left\{ {i \in \mathcal{C}:f \in i} \right\}} \right|$ equals to 1 at time slot $n$ if channel $f$ is selected. Thus, ${{\tilde \rho }_n}(f) \ge \sum\nolimits_{f \in i} {{\varepsilon _n}(f)}= k{\varepsilon_n}(f)$.

Let $\mathcal{\bar E}_n^f$ denote the complementary of event
 $\mathcal{E}_n^f$. Then by the Bernstein't inequality $\mathbb{P}[\mathcal{\bar E}_n^f] \le f^{-b_n}$. According to (\ref{eq:Apart23t}), the number of
 times the channel $f$ is selected up to round $n$ is bounded as:
\begin{IEEEeqnarray*}{l}
 \begin{array}{l}%\mathds{1}_{[A_t= f]}
  \mathbb{E}[{N_n}(f)] \le  \sum\limits_{t = 1}^n {\mathbb{P}[A_t= f|\Xi_{t - 1}^f]P[\Xi_{t - 1}^f]} \\
  \hspace{5em}+ \mathbb{P}[A_t= f|\overline {\Xi_{t - 1}^f} ]P[\overline {\Xi_{t - 1}^f} ]\\
 \hspace{4em} \le \sum\limits_{t = 1}^n {\mathbb{P}[A_t= f|\Xi_{t - 1}^f]} {\mathds{1}_{\{ \Xi_{t - 1}^f\} }} +
  \mathbb{P}[\overline {\Xi_{t - 1}^S} ]\\
  \hspace{4em} \le \sum\limits_{t = 1}^n {\mathbb{P}[A_t= f|\Xi_{t - 1}^f]} {\mathds{1}_{\{ \Xi_{t - 1}^f\} }} + {f^{ - {b_{t - 1}}}}.
\end{array}
\end{IEEEeqnarray*}
We further upper bound $ {\mathbb{P}[A_t= f|\Xi_{t - 1}^f]} {\mathds{1}_{\{ \Xi_{t - 1}^f\} }} $ as follows:
    \begin{displaymath}
\begin{array}{l}
\mathbb{P} {[A_t= f|{ \Xi}_{t - 1}^f]} {\mathds{1}_{\{ { \Xi}_{t - 1}^f\} }}
= {{{\tilde \rho }_t}}(f){\mathds{1}_{\{ { \Xi}_{t - 1}^f\} }}\\
  \hspace{4em} \le ({ {q}_t}(f) + k{\varepsilon_t}(f)){\mathds{1}_{\{ { \Xi}_{t - 1}^f\} }}\\
 \hspace{4em} =({k\varepsilon  _t}(f) + \frac{{\sum\nolimits_{i:f \in i} {{w_{t - 1}}\left( i \right)} }}
 {{{W_{t - 1}}}}){\mathds{1}_{\{ {\Xi}_{t - 1}^f\} }}\\
    \hspace{4em} =({k\varepsilon  _t}(f) + \frac{{\sum\nolimits_{i:f \in i}^{} {{f^{ - {\eta _t}{
    \tilde L_{t - 1}}(i)}}} }}{{\sum\nolimits_{i = 1}^N {{f^{ - {\eta _t}{ \tilde L_{t - 1}}(i)}}} }})
    {\mathds{1}_{\{ { \Xi}_{t - 1}^f\} }}\\
  \hspace{4em}\mathop \le \limits^{(a)} ({k\varepsilon  _t}(f) + {e^{ - {\eta _t}\left( {{{\tilde \Psi}_t}(i) - {{\tilde \Psi}_t} ({i^*})} \right)}}){
  {\mathds{1}_{\{ { \Xi}_{t - 1}^f\} }}}\\
    \hspace{4.0em}\mathop \le \limits^{(b)} ({k\varepsilon  _t}(f) + {e^{ - {\eta _t}\left( {{\tilde
    \Psi_{t }}(f) - {{\tilde \Psi}_{t }}({f^*})} \right)}}){{\mathds{1}_{\{ { \Xi}_{t - 1}^f\} }}}\\
   \hspace{4.0em}   =  ({k\varepsilon  _t}(f) \\
      \hspace{4.1em}+ {e^{ - {\eta _t}\left( {{\tilde{L}}}_{t}(f) -{{\tilde{L}}}_{t}(f^*) + \lambda_{t}
   ({{\tilde{\Gamma}}_{t}} (f^*)-{{\tilde{\Gamma}}_{t}} (f)) \right)}}){\mathds{1}_{\{ { \Xi}_{t - 1}^f\} }} \\
   \hspace{4.0em}\mathop \le \limits^{(c)}  k{\varepsilon  _t}(f){\mathds{1}_{\{ { \Xi}_{t - 1}^f\} }} \!+ \! {e^{ - {\eta _t}{\hbar_{t}}(f)}}
   e^{ - \eta_t\lambda_{t}({{\tilde{\Gamma}}_{t}} (f^*)\!-\!{{\tilde{\Gamma}}_{t}} (f))}
\end{array}
%\vspace{-0.4cm}
  \end{displaymath}
The above inequality (a) is due to the fact that channel
$f$ only belongs to one selected strategy $i$ at $t$, inequality (b) is because the  cumulative
regret of each strategy is great than the cumulative regret of each channel that belongs to the
  strategy,  inequality (c) is due to the fact that $ {{{\underline \varepsilon   }_t}(f)}$ is a non-increasing
 sequence ${\upsilon _t}(f) \le \frac{t}{{{{k \underline \varepsilon   }_t}(f)}}$.
  Substitution of this result back into
 the computation of $ \mathbb{E}[{N_n}(f)]$ completes the proof.
 \end{IEEEproof}

 \emph{Proof of Theorem 18.}
\begin{IEEEproof}
The proof is based on Lemma 5. Let $b_n =ln(n \Delta(f)^2)$ and ${{{\underline \varepsilon  }_n}(f)}
={{{ \varepsilon  }_n}(f)}$. For any $c \ge 18$ and any $n \ge n^*$, where $n^*$ is the minimal integer for which
 ${n^*} \ge \frac{{4{c^2}n \ln {{({n^*}\Delta {{(f)}^2})}^2}}}{{m^2\Delta {{(f)}^4}\ln (n)}}$, we have
     \begin{displaymath}
\begin{array}{l}
{\hslash_n}(f) = mt\Delta (f) - \sqrt {2mt{b_n}\left( {\frac{1}{{k{\varepsilon _n}(f)}} + \frac{1}{{k{\varepsilon _n}({f^*})}}} \right)}  - \frac{{\left( {\frac{1}{4} + \frac{1}{k}} \right){b_n}}}{{3{\varepsilon _n}({f^*})}}\\
   \hspace{2.43em} \ge mt\Delta (f) - 2\sqrt {\frac{{mt{b_n}}}{{k{\varepsilon _n}(f)}}}  - \frac{{\left( {\frac{1}{4} + \frac{1}{k}} \right){b_n}}}{{3{\varepsilon _n}(f)}}\\
  \hspace{2.43em} = mt\Delta (f)(1 - \frac{2}{{\sqrt {k c} }} - \frac{{\left( {\frac{1}{4} + \frac{1}{k}} \right)}}{{3c}})\\
  \hspace{2.43em} \mathop  \ge \limits^{(a)}m n\Delta (f)(1 - \frac{2}{{\sqrt c }} - \frac{{1.25}}{{3c}}) \ge \frac{1}{2}mt\Delta (f),
\end{array}
 %$k{\underline \varepsilon }_n(f)$
  \end{displaymath}
where ${\varepsilon _n}(f) = \frac{{c\ln (n\Delta {{(f)}^2})}}{{tm\Delta {{(f)}^2}}}$.  The transmission power  is quasi-concave to the reward such
that the cooperative learning strategy has $( -\hbar_t(f) + \lambda_{t}({{\tilde{\Gamma}}_{t}} (f)-{{\tilde{\Gamma}}_{t}} (f^*)) ) \le 0$. By substitution of the lower bound on $\hbar_n(f)$ into Lemma 5, we have
\begin{IEEEeqnarray*}{l}
\begin{array}{l}
\!\!\!\!\!\mathbb{E}[{N_n}(f)] \le {n^*} \!+  \! \frac{{\ln (n)}}{{\Delta {{(f)}^2}}} \!+ \! k \frac{{c\ln {{(n)}^2}}}{{m\Delta {{(f)}^2}}}  +  \!\!
\sum\limits_{t = 1}^n \!(\!{{f^{ - \frac{{m\Delta (f)}}{4}\sqrt {\frac{{(t - 1)ln(n)}}{n}} }}}\!)\\
 \hspace{3.2em} \le k\frac{{c\ln {{(n)}^2}}}{{m\Delta {{(f)}^2}}} + \frac{{\ln (n)}}{{\Delta {{(f)}^2}}} + O(\frac{{{n
 }}}{{m^2\Delta {{(f)}^2}}}) + {n^*},
\end{array}\IEEEyesnumber \label{eq:Apart13}
\end{IEEEeqnarray*}
  where lemma 3 is used to bound the sum of the exponents. In addition, please
  note that $n^*$ is of the order $O(\frac{{k n}}{{m^2\Delta {{(f)}^4}\ln (n)}})$.

The rest of the proof follows the same line in the proof of the Theorem 3. Thus, we complete the proof.
\end{IEEEproof}

 \emph{Proof of Theorem 19-Theorem 21.}
 The proofs of Theorem 19-Theorem 21 use similar idea as in previous proofs. We omitted here for brevity.

\begin{figure*}
 \subfigure[]{
 \label{fig:subfig:a}
\includegraphics[width=1.823in]{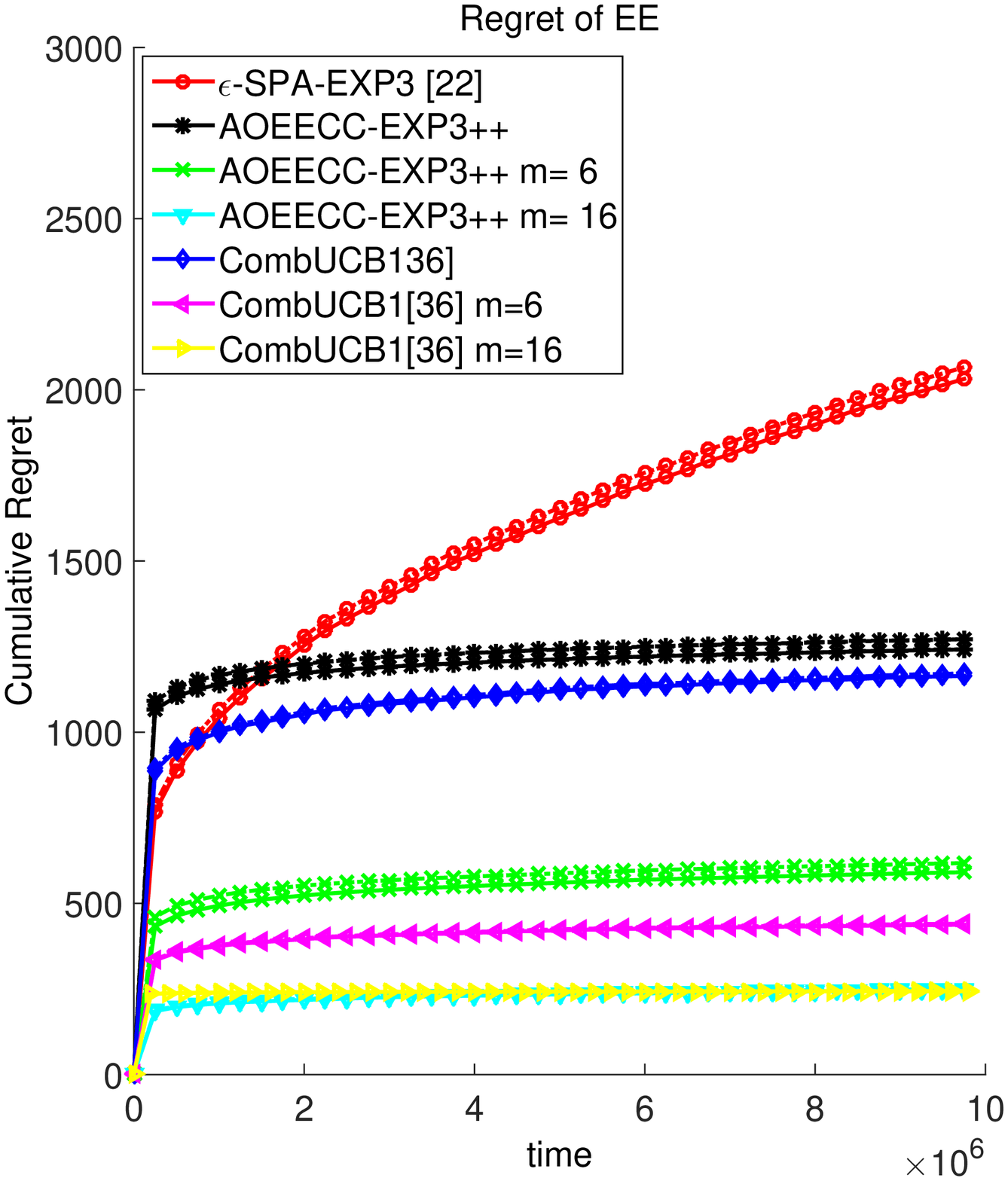}
\includegraphics[width=1.823in]{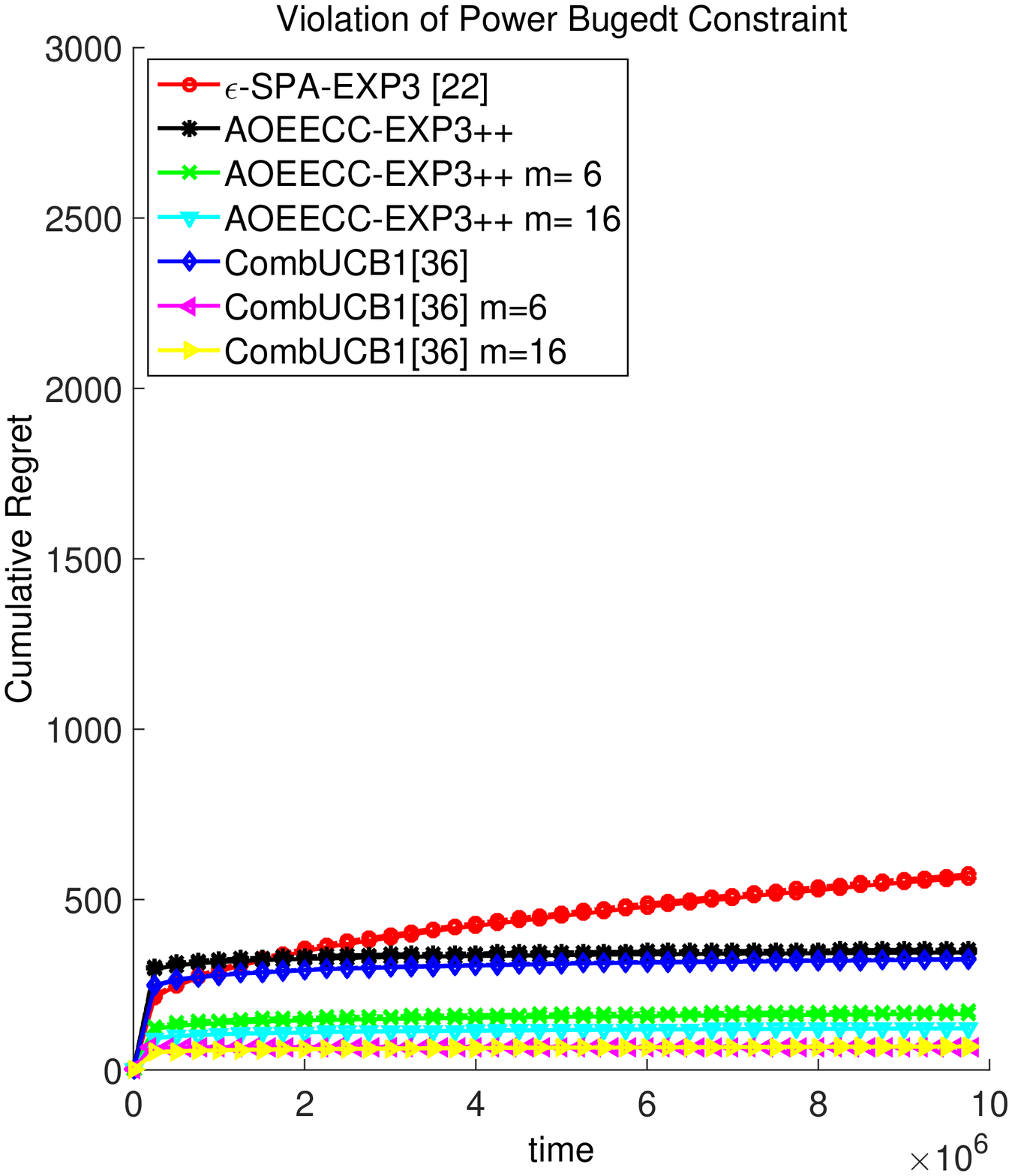}}
\hspace{-0.182in}
 \subfigure[]{
  \label{fig:subfig:b}
  \includegraphics[width=1.823in]{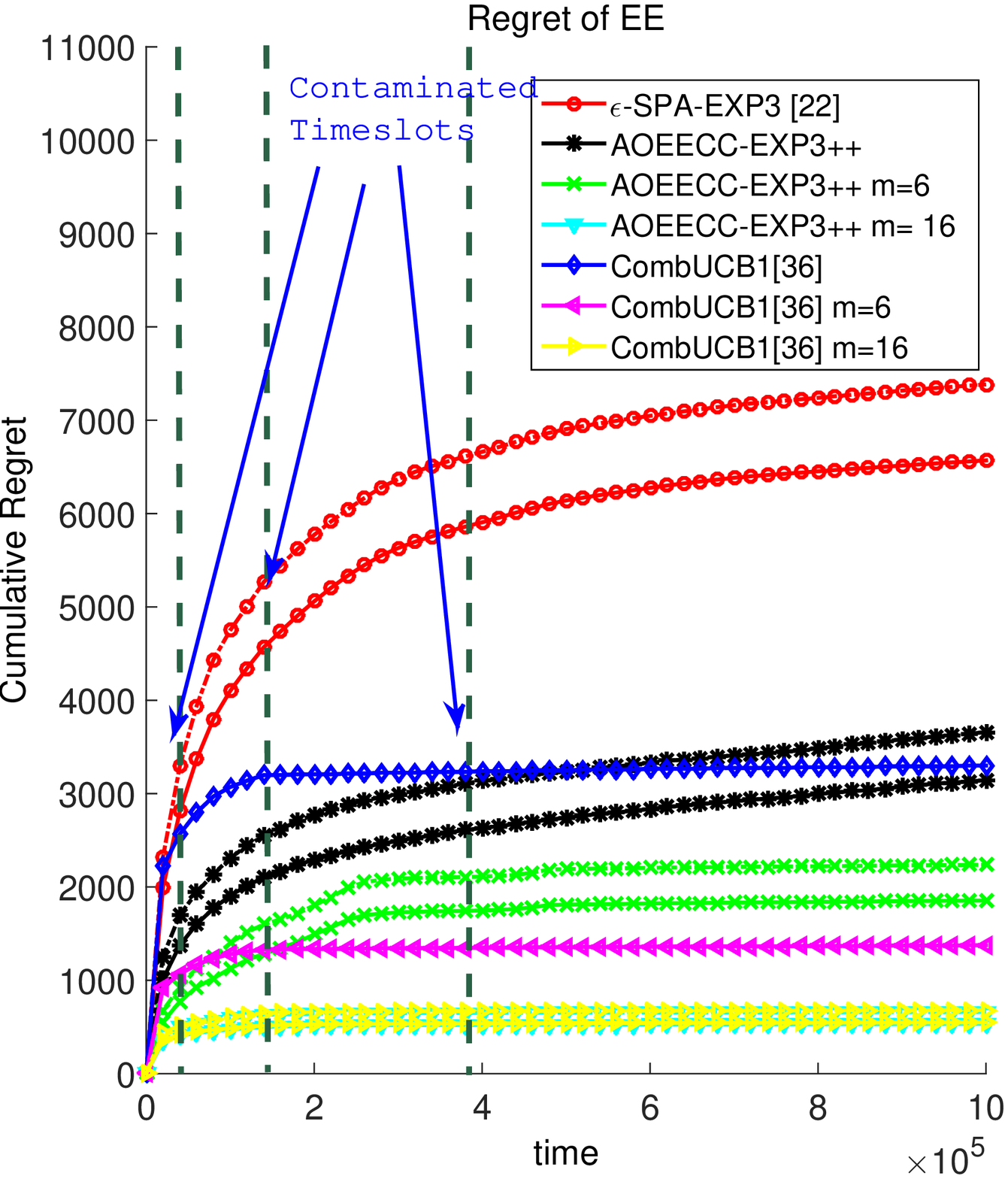}
  \includegraphics[width=1.823in]{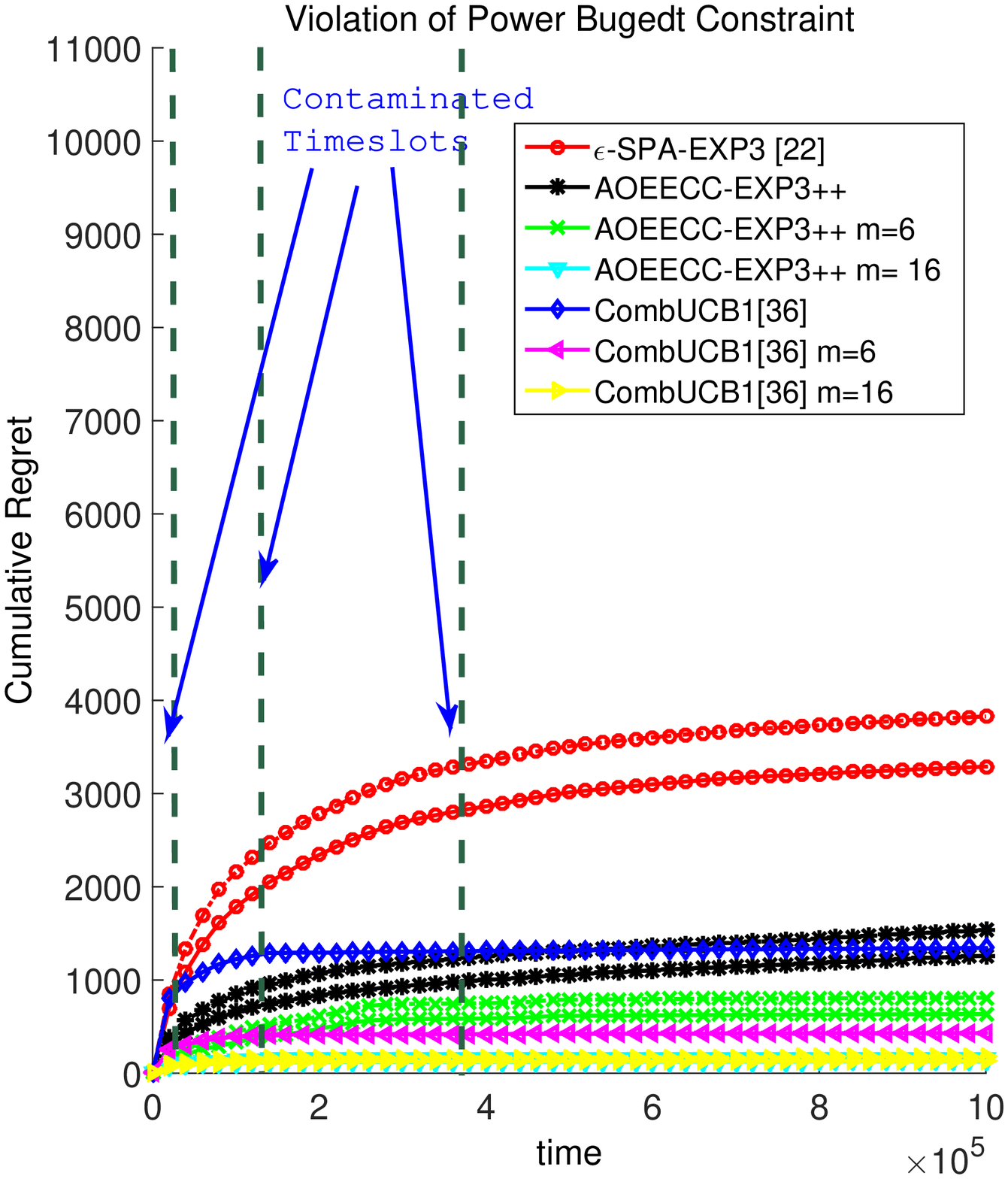}}
 \subfigure[]{
 \label{fig:subfig:a}
\includegraphics[width=1.823in]{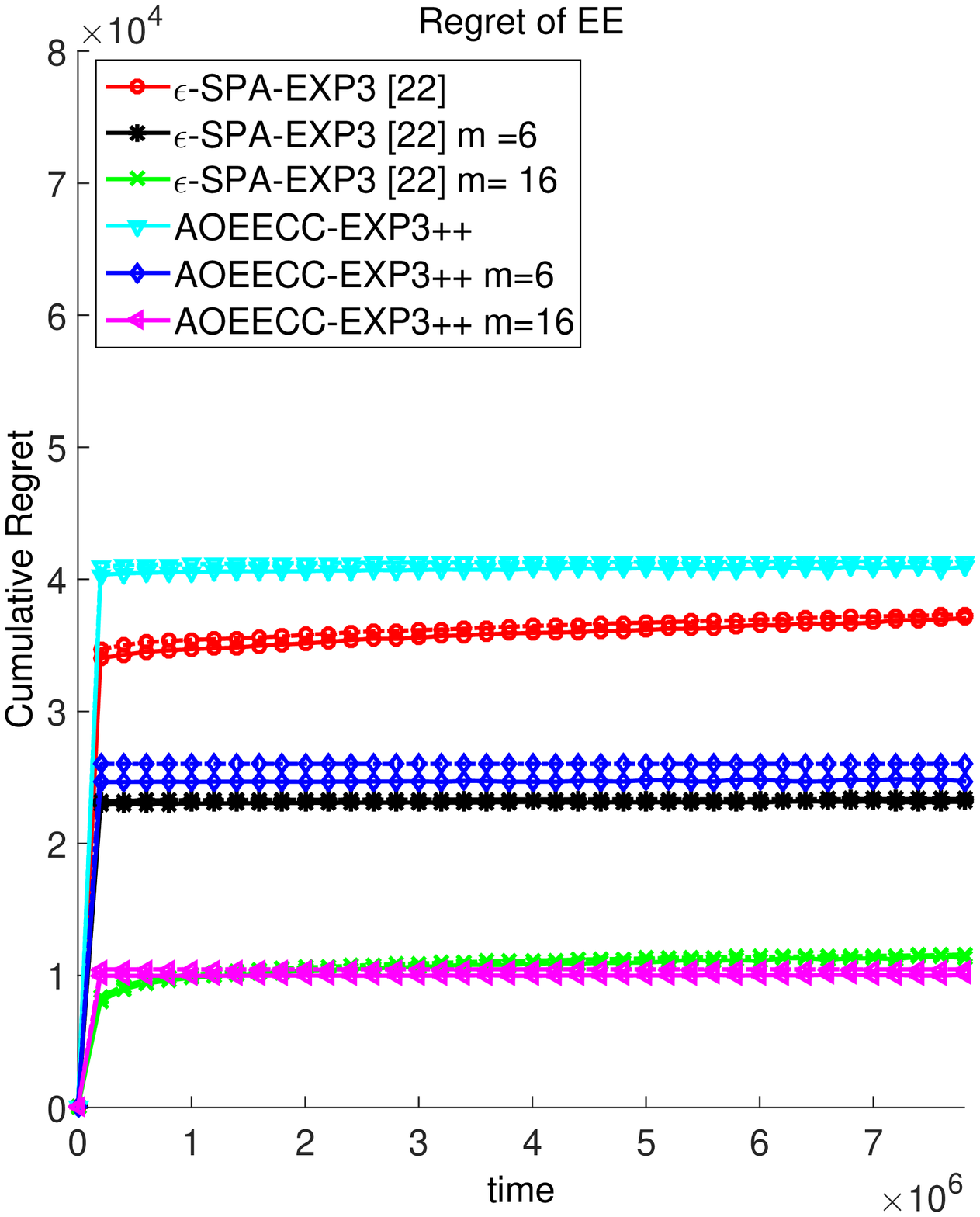}
\includegraphics[width=1.823in]{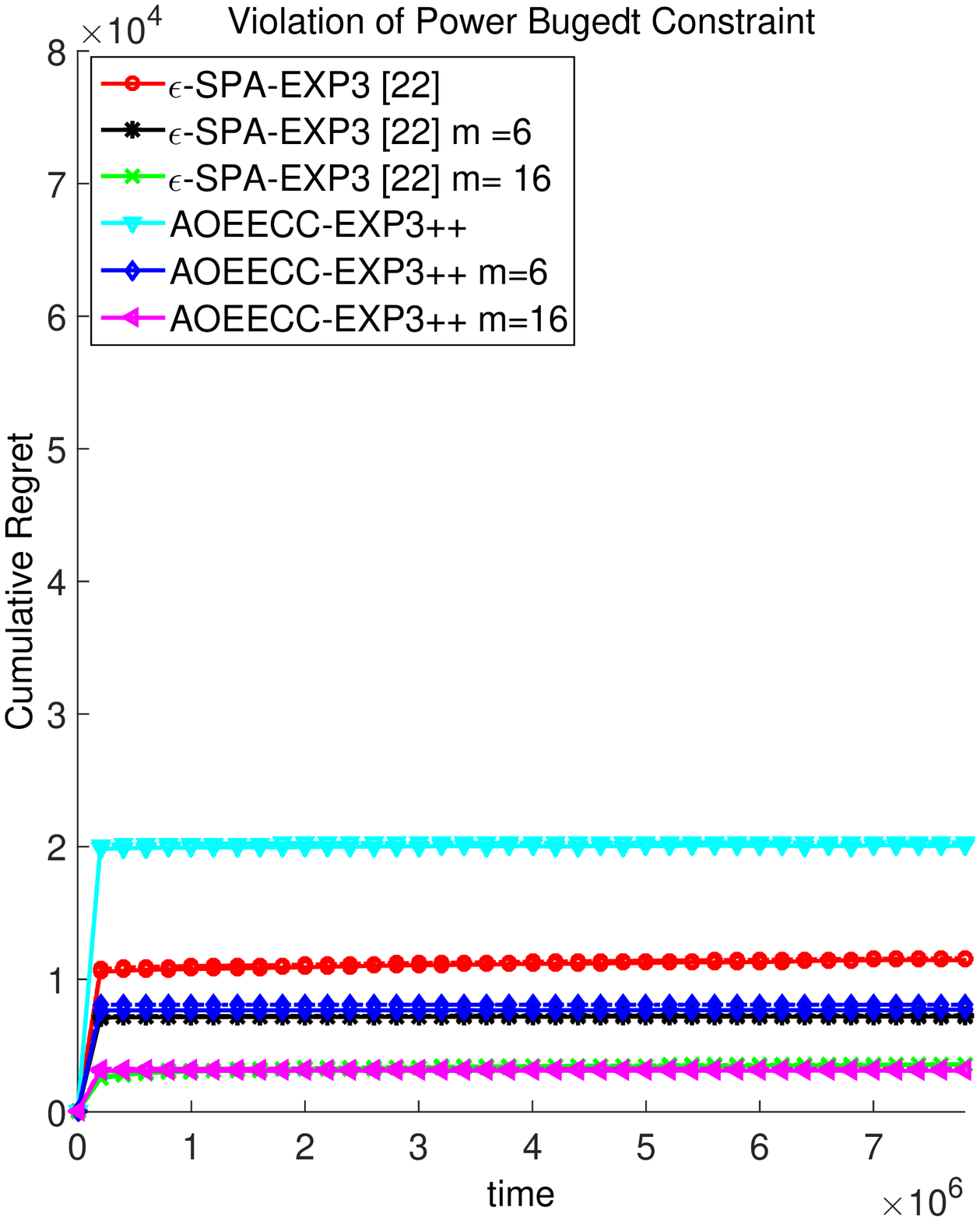}}
\hspace{-0.182in}
 \subfigure[]{
  \label{fig:subfig:b}
  \includegraphics[width=1.823in]{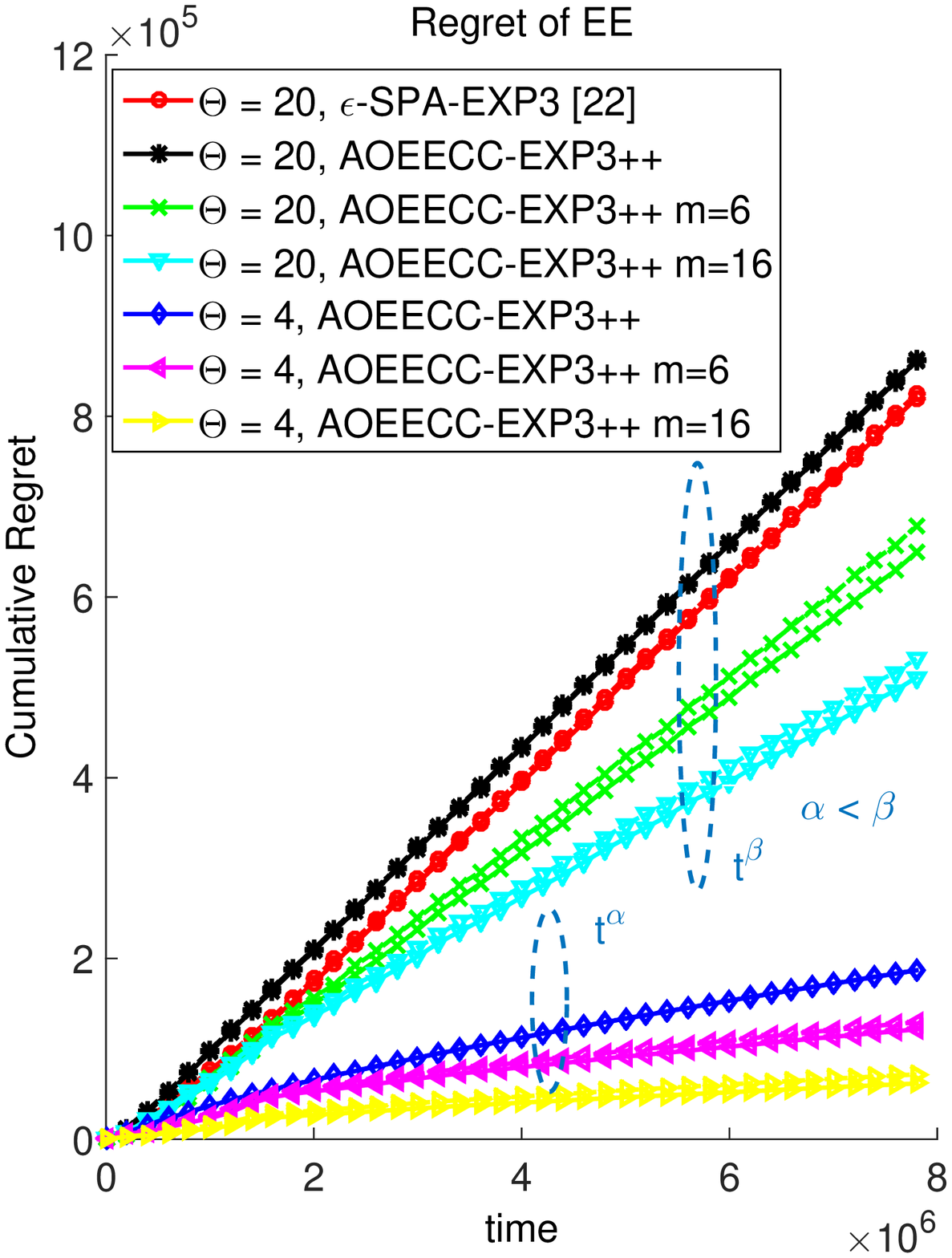}
  \includegraphics[width=1.823in]{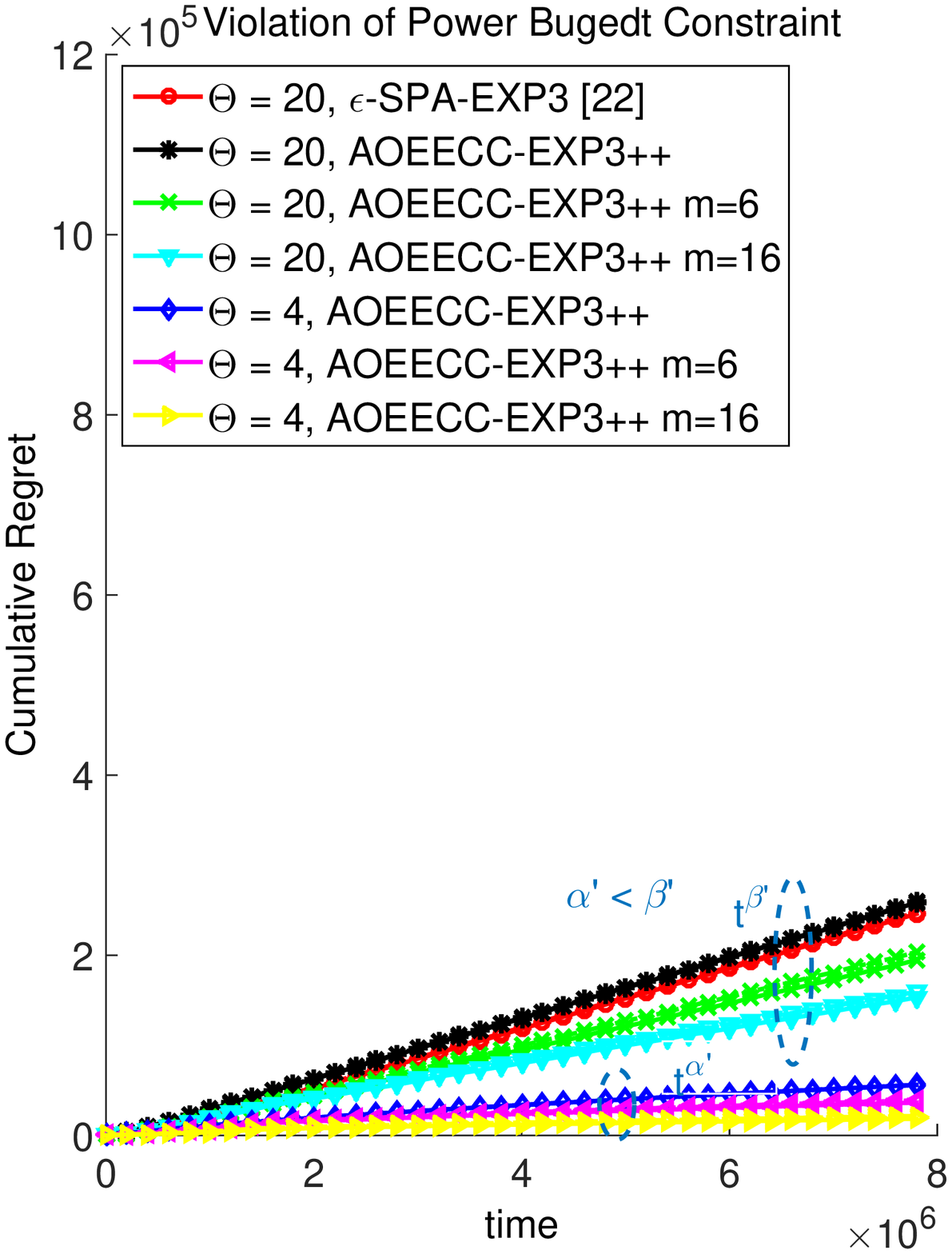}}
  \caption{Performance Comparison in Different Regimes.}
\vspace{-.2cm}
\end{figure*}

%\section{Numerical and Simulation Results}
\section{Implementation Issues and Simulation Results}
\subsection{Computational Efficient Implementation of the AOEECC-EXP3++ Algorithm}
The implementation of Algorithm $1$ requires the computation of probability distributions and storage of $N$ strategies, which has a time and space complexity $O(K^{k})$. As the number of channels increase, the strategy will become
exponentially large, which is very hard to be scalable and results in low efficiency. To address this important problem,
 a computational efficient enhanced algorithm is proposed by utilizing the dynamic programming techniques. The key idea
of the enhanced algorithm is to select the transmitting  channels one by one until $k$ channels are chosen, instead of choosing a strategy
from the large strategy space in each timeslot. Interesting readers can find details in \cite{Pan15}
\cite{QianJSAC12}. The linear time and space complexity are achievable for AOEECC-EXP3++, which
 is highly efficient and can  be easily implemented in practice.

\subsection{Simulation Results}
We evaluate the performance of our  AOEECC-EXP3++ Algorithm on a cognitive radio system which contains
 $16$ nodes and 8 USRP devices. There is  line-of-sight path between the
two nodes of a path at a specific distance, which was varied for different
experiments ranging from $10$ meters to $60$ meters with fixed topology. We conduct all our experiments on our own built system. The
maximum transmission rate for each sensor node ranges from 20bps to 240kbps. We use the
USRPs as CR nodes and the sensor nodes as the PUs. There are 32 channels available for the
PUs in PC. The transmission bandwidth of PUs are 4 $M$Hz, while the bandwidth of each USRP with 4 SPA
radios (channel) is $350kHz$. The RF performance of a single channel is operating at $3.5Ghz$ with the receive
noise figure less than $8$dB, and the the maximum output power of each USRP device is 11.5$dBM$, and the average
transmission power is about $8.63dBM$. We only count
the average measured circuit   and processing power that
is related to data transmission, which is about $46.7dBm$. We set the $\textsf{P}_o $ of each
SU to be $9.24dBM$. We implement our
SPA models and algorithms that builds up on the software suit built upon GNU radio. We assume
that all the SU will agree upon a common control
channel (CCC), where the channel $17$ is used as the CCC. We take $\epsilon=1$ to get the
maximum achievable EE.

In Fig. 3, W.l.o.g., we normalize the  EE into unitary value in every timeslot $n$. Then, we have $M=1$, $k=4$ and $K=32$. All computations of the
collected datasets were conducted on an
 off-the-shelf desktop with dual
$6$-core Intel i7 CPUs clocked at $2.66$Ghz.  To show the advantages of our AOEECC-EXP3++ algorithms, we
compare their performance to other existing MAB based algorithms, which includes:
the EXP3 based  combinatorial version (implemented by ourselves)
of the $\epsilon$-SPA for non-stochastic MABs in CC \cite{XYinfocom11}, and we named it as ``$\epsilon$-SPA-EXP3"; The combinatorial stochastic
MAB algorithm, i.e.,
``CombUCB1",  with the tight regret bound as proved in \cite{Branislav2015}, and the cooperative learning versions of algorithms of ours and others.
  In Fig. 3, the solid lines in the graphs  represent the mean performance over the experiments and the dashed lines represent the mean plus on standard deviation (std) over the ten repetitions of the corresponding experiments. For a given optimal channel access strategy, small regret values indicate the large value of EE. We set all versions of our AOEECC-EXP3++ algorithms
parameterized by ${\xi _t}(f) = \frac{{\ln (t{{\hat \Delta }_t}{{(f)}^2})}}{{32t{{\hat \Delta }_t}{{(f)}^2}}}$, where ${{{\hat \Delta }_t}(f)}$ is the
empirical estimate of ${{{ \Delta }_t}(f)}$, and parameters $\eta_t$ and $\delta_t$ according to the theorems.

In our first group of experiments  in the  stochastic regime (environment) as shown in Fig. 1(a), it is clear to see that AOEECC-EXP3++
 enjoys almost the same (cumulative) regrets as  CombUCB1 and has much lower regrets over time than the
 adversarial $\epsilon$-SPA-EXP3. We also see the significantly regrets reduction when accelerated learning ($m=6, 16$)
  is employed for both  AOEECC-EXP3++ and  CombUCB1. For the subplot of the violation of budgeted
  constraint, we also see very similar behaviors among all algorithms for a fixed setting of the CC topology.

In our second group of experiments in the moderately contaminated stochastic environment, there are several contaminated
timeslots as labeled in Fig. 1(b), which is made by irregular jamming behaviors at some rounds.
In this case, the contamination does not make the whole dataset be fully adversarial, but drawn
from a different stochastic model. Despite the corrupted rounds the AOEECC-EXP3++
algorithm successfully returns to the stochastic operation
mode and achieves better results than $\epsilon$-SPA-EXP3 and has very close and comparable performance as CombUCB1. We also
see the cooperative learning is highly efficient for all algorithms.
\begin{figure}
%\vspace{-.3cm}
\centering%.6
\includegraphics[scale=.42]{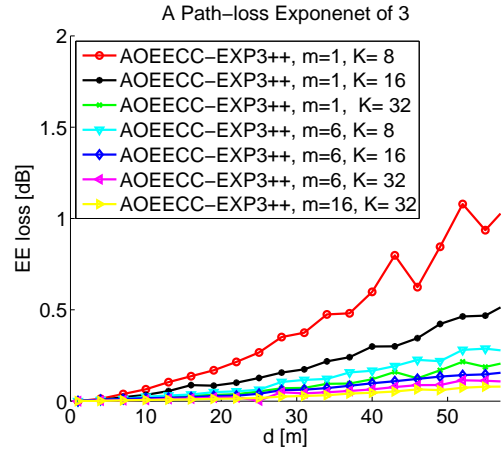}
\caption{EE loss  of AOEECC-EXP3++  in a Path-loss Model}
\label{fig:digraph}
\vspace{-.5cm}
\end{figure}

We conducted the third group of experiments in the
adversarial regimes.  We present the oblivious adversary case in Fig. 1(c). Due to the strong  interference effect on each channel and
the arbitrarily changing feature of the jamming behavior, all algorithms experience very high accumulated
regrets.  It can be find that our AOEECC-EXP3++
algorithm will have close and slightly worst learning  performance when compared to $\epsilon$-SPA-EXP3, which confirms our theoretical
analysis. Note that we do not implement stochastic MAB algorithms such as CombUCB1, since it is not applicable in this regime.

\begin{figure}
%\vspace{-.3cm}
\centering%.6
\includegraphics[scale=.42]{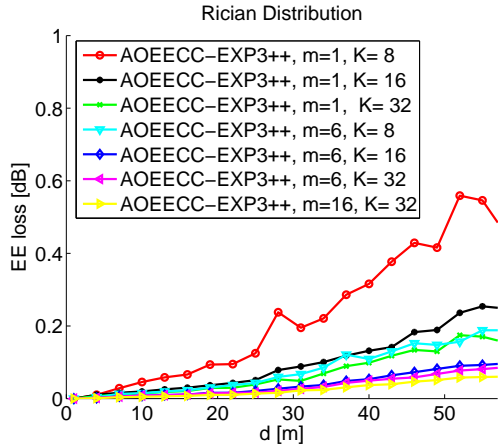}
\caption{EE loss of AOEECC-EXP3++  in  Rician Distribution}
\label{fig:digraph}
\vspace{-.5cm}
\end{figure}

In our fourth set of experiments shown in Fig. 1(d), we simulate the adaptive jamming attack case in the adversarial regime with a typical large
memory  $\Theta=20$. We can see large performance degradations for all algorithms when compared to the oblivious jammer case. The multiplicative effect of
 $\Theta$ makes the  AOEECC-EXP3++  and $\epsilon$-SPA-EXP3 very hard to combat this type of jamming attack, although the regret curve is
 still sublinear after normalization.

\begin{figure*}
%\vspace{-.12cm}
%\vspace{-1em}
\centering
\includegraphics[scale=.4 ]{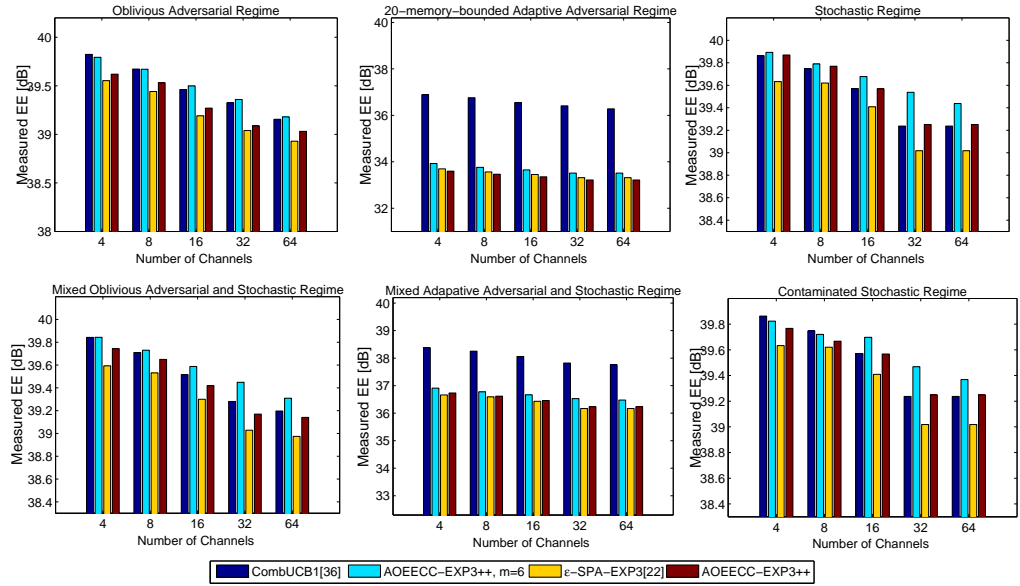}
\caption{Measured EE in Different Regimes. }
\label{fig:digraph}
\vspace{-1em}
\vspace{-.3cm}
\end{figure*}

We also compare the average EE loss of the proposed  AOEECC-EXP3++ algorithm (after a run of $10^7$ rounds) with respect to the
optimal solution for 100 random channel realizations with  a path-loss exponent of
3, a noise figure of 7 dB, a carrier frequency of 3.5 GHz, a noise bandwidth of 10 MHz, and the average circuit power $P_c^t(i_f)$ = $29.2dBm$
 for each transmitting channel $f$. The
result is shown in Fig. 4. We can find that with the increasing of the number of available channels, the EE loss
is decreasing. This confirms the well-known ``multi-channel" diversity in wireless communications. In addition,
increasing  $m$ also reduces the average EE loss

Moreoever, we conduct another group of experiment to verify the performance of our algorithms
in the fading environments. We consider the Rician fading has a direct-to-scattered
signal path ratio of $\bar K = 6 dB$, which is expected to dominate mobile communications.  Fig. 5 shows the gaps between ours and optimal EE solutions, where
we see similar phenomena but with
a larger variance when compared with Fig. 4. Nevertheless, the figure shows that the results are reasonable for
this typical conditions. It can be shown that the methodology presented in this paper can
be applied to find the power allocation for any channel distribution for EECC.

%In our fourth experiments shown in , summarize the EE of our algorithms over time $8*10^6$ by counting the number of received data packets in the receive side for a typical scenarios, $k=4$ and $K=12$ and $k=16$ and $K=100$, in the moderately
%contaminated regimes. We can see that our AOEECC-EXP3++$^{EMP}$ always outperforms Anti-Jam-EXP3 and CombUCB1, which shows the advantage of our algorithms
%in different wireless communication scenarios with scalability.

%, and it is interesting to see
%that the cumulative regret is not sensitive to the size of $K$.
%Interestingly, for different size of channel set $K=4, 8, 16$, the cumulative regret is not sensitive
% to $K$ for the $\theta$-memory-bounded adaptive adversary,  which indicates the

For brevity, we do not plot the regret performance figures for the mixed adversarial and stochastic regime. However, in our last experiments, we
compare the measured EE (dB) for all the four different regimes after a relative long period of learning rounds $n=2*10^7$.  We plot our results
in Fig. 6. It is easy to find that our algorithm AOEECC-EXP3++  attains almost all the advantages of the stochastic MAB algorithms CombUCB1, and
has better EE performance than $\epsilon$-SPA-EXP3.

\section{Conclusion}
In this paper, we proposed the first adaptive multi-channel SPA algorithm for EECC without the knowledge about the nature of environments.
At first, we captured  features of  general CC environments and divided them into four regimes, and then provided solid theoretical analysis for each of them.
We find that our formulated constrained regret
minimization problem requires joint control of learning rate and exploration parameters to achieve best performance.  We
have also found and verified that cooperative learning is an effective approach to improve the performance of EECC.
 Extensive simulations were conducted  to verify the learning performance. The proposed algorithm could be implemented
 efficiently in practical CC  with different sizes. We believe that the idea and algorithms of this paper can be
 applied to other wireless communications problems in unknown environments.

 %In the future, we plan to
%extend our idea to general distributed, multi-hop multi-user wireless networks. % and other combinatorial MAB problems in other type of networks.
%Obviously, they are more interesting for networking researchers, but it might not be
% easy to  obtain computational efficient algorithms. %as indicated in \cite{OR2014}.

%\bibliographystyle{IEEEtran}
%\bibliography{IEEEabrv,PanZhou-Ref}

 \begin{IEEEbiography}[{\includegraphics[width=2.6in,height=1.25in,clip,keepaspectratio]{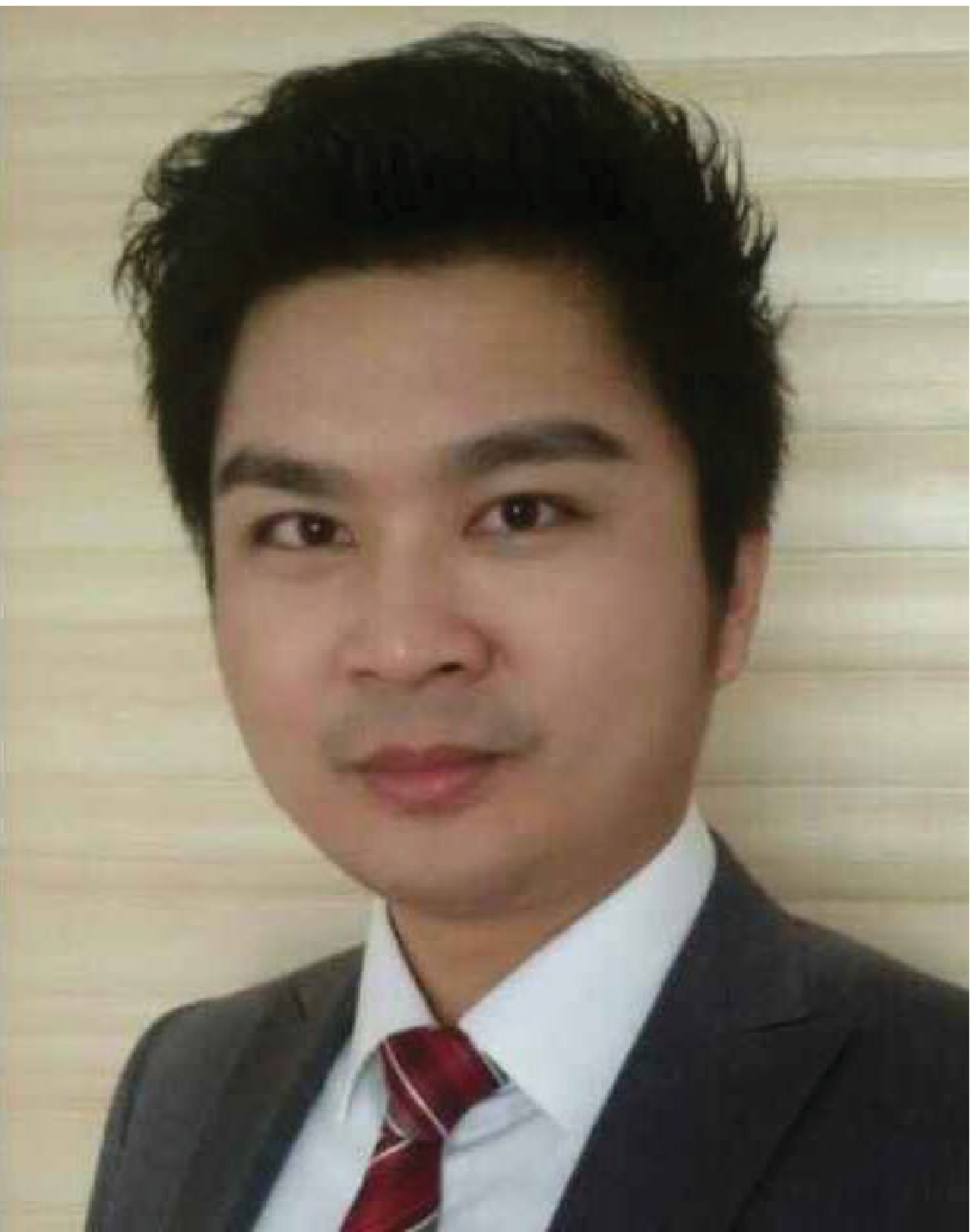}}]{Pan Zhou(S'07--M'14)} is currently an associate professor
 with School of Electronic Information and Communications, Huazhong University of Science and Technology, Wuhan, P.R. China. He received his Ph.D. in the School of Electrical and Computer Engineering at the Georgia
Institute of Technology (Georgia Tech) in 2011, Atlanta, USA. He
received his B.S. degree in the \emph{Advanced Class} of
HUST, and a M.S. degree in the Department of Electronics and Information Engineering
from HUST, Wuhan, China, in 2006 and 2008, respectively.
He held honorary degree in his bachelor and merit research award
of HUST in his master study. He was a
senior technical member at Oracle Inc, America during 2011 to 2013, Boston, MA, USA,  and worked on hadoop and distributed storage system for big data
analytics at Oracle cloud Platform.  His current research interest includes:  wireless communication and networks, security and privacy,  machine learning and big data.
\end{IEEEbiography}

 \begin{IEEEbiography}[{\includegraphics[width=2.6in,height=1.25in,clip,keepaspectratio]{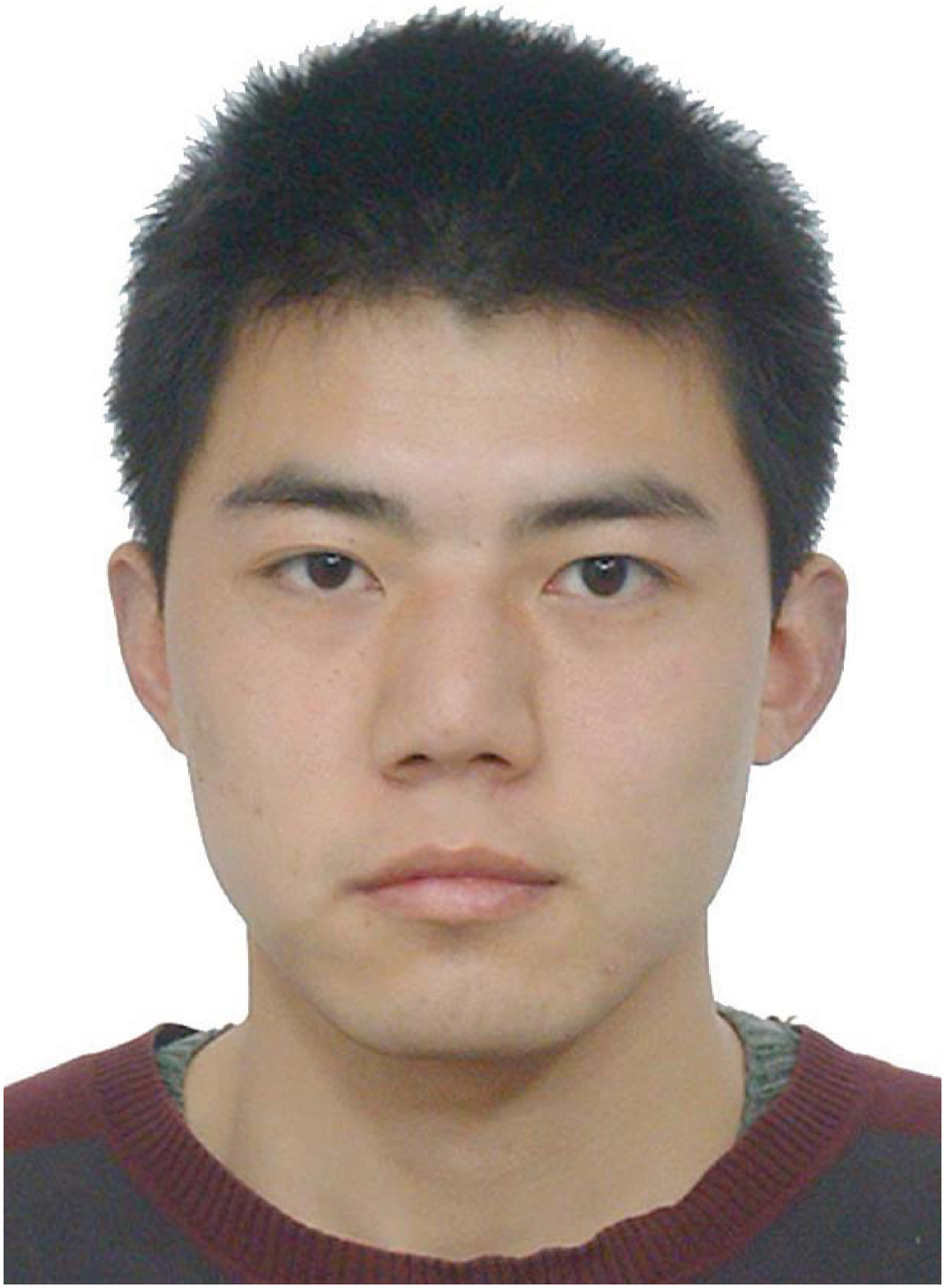}}]
{Chenhui Hu(S'08)}  is currently pursuing his Ph.D. degree in engineering and applied sciences at Harvard University, Cambridge, MA. He is also a research assistant in the Center for Advanced Medical Imaging Sciences (CAMIS) of Massachusetts General Hospital. He received his B.S. and M.S. degrees in electronic engineering from Shanghai Jiao Tong University, Shanghai, China, in 2007 and 2010, respectively. He was a recipient of the Outstanding Graduates of Shanghai in 2010 and the 3rd IEEE ComSoc Asia-Pacific Outstanding Paper Award in 2014. His current research interests include mobile wireless communication, machine learning, statistical signal processing, and brain network analysis.
\end{IEEEbiography}

\begin{IEEEbiographynophoto}{Tao Jiang (M'06--SM'10)}
 is currently a full Professor in the School of Electronic Information and Communications, Huazhong University of Science and Technology, Wuhan, P. R. China. He received the B.S. and M.S. degrees in applied geophysics from China University of Geosciences, Wuhan, P. R. China, in 1997 and 2000, respectively, and the Ph.D. degree in information and communication engineering from Huazhong University of Science and Technology, Wuhan, P. R. China, in April 2004. From Aug. 2004 to Dec. 2007, he worked in some universities, such as Brunel University and University of Michigan-Dearborn, respectively. He has authored or co-authored over 160 technical papers in major journals and conferences and six books/chapters in the areas of communications and networks. He served or is serving as symposium technical program committee membership of some major IEEE conferences, including INFOCOM, GLOBECOM, and ICC, etc.. He is invited to serve as TPC Symposium Chair for the IEEE GLOBECOM 2013 and IEEEE WCNC 2013. He is served or serving as associate editor of some technical journals in communications, including in IEEE Communications Surveys and Tutorials, IEEE Transactions on Vehicular Technology, and IEEE Internet of Things Journal, etc.. He is a recipient of the NSFC for Distinguished Young Scholars Award in P. R. China. He is a senior member of IEEE.
\end{IEEEbiographynophoto}

 \begin{IEEEbiographynophoto}{Dapeng Oliver Wu (S'98--M'04--SM¡¯06--F'13) }
 received B.E. in Electrical Engineering from Huazhong University of Science and Technology, Wuhan, China, in 1990, M.E. in Electrical Engineering from Beijing University of Posts and Telecommunications, Beijing, China, in 1997, and Ph.D. in Electrical and Computer Engineering from Carnegie Mellon University, Pittsburgh, PA, in 2003.

He is a professor at the Department of Electrical and Computer Engineering, University of Florida, Gainesville, FL.  His research interests are in the areas of networking, communications, signal processing, computer vision, machine learning, smart grid, and information and network security. He received University of Florida Research Foundation Professorship Award in 2009, AFOSR Young Investigator Program (YIP) Award in 2009, ONR Young Investigator Program (YIP) Award in 2008, NSF CAREER award in 2007, the IEEE Circuits and Systems for Video Technology (CSVT) Transactions Best Paper Award for Year 2001, and the Best Paper Awards in IEEE GLOBECOM 2011 and International Conference on Quality of Service in Heterogeneous Wired/Wireless Networks (QShine) 2006.

Currently, he serves as an Associate Editor for IEEE Transactions on Circuits and Systems for Video Technology, Journal of Visual Communication and Image Representation, and International Journal of Ad Hoc and Ubiquitous Computing.  He is the founder of IEEE Transactions on Network Science and Engineering.  He was the founding Editor-in-Chief of Journal of Advances in Multimedia between 2006 and 2008, and an Associate Editor for IEEE Transactions on Wireless Communications and IEEE Transactions on Vehicular Technology between 2004 and 2007. He is also a guest-editor for IEEE Journal on Selected Areas in Communications (JSAC), Special Issue on Cross-layer Optimized Wireless Multimedia Communications.  He has served as Technical Program Committee (TPC) Chair for IEEE INFOCOM 2012, and TPC chair for IEEE International Conference on Communications (ICC 2008), Signal Processing for Communications Symposium, and as a member of executive committee and/or technical program committee of over 80 conferences. He has served as Chair for the Award Committee, and Chair of Mobile and wireless multimedia Interest Group (MobIG), Technical Committee on Multimedia Communications, IEEE Communications Society. He was a member of Multimedia Signal Processing Technical Committee, IEEE Signal Processing Society from Jan. 1, 2009 to Dec. 31, 2012.  He is an IEEE Fellow.
\end{IEEEbiographynophoto}

\end{document}